\documentclass[twocolumn]{aastex7}

\usepackage[flushleft]{threeparttable}

\received{September 11, 2026}

\submitjournal{ApJ}

\shorttitle{New PLZs and Metallicity Distributions in the MCs}
\shortauthors{Mart\'inez-V\'azquez et al.}

\graphicspath{{./}{figures/}}

\begin{document}

\title{New Optical and Mid-Infrared RR Lyrae Period-Luminosity-Metallicity Relations for Precision Distances and Metallicity Distributions in the Magellanic Clouds}

\correspondingauthor{C.~E.~Mart\'inez-V\'azquez}

\author[0000-0002-9144-7726]{Clara E. Mart\'inez-V\'azquez}
\affiliation{NSF NOIRLab, 670 N. A'ohoku Place, Hilo, HI 96720, USA}
\email{}

\author[0000-0002-1650-2764]{Joseph P. Mullen}
\affiliation{Department of Physics, University of North Florida, 1 UNF Dr., Jacksonville, FL 32224, USA}
\affiliation{Department of Physics and Astronomy, Vanderbilt University, VU Station 1807, Nashville, TN 37235, USA}
\email{}

\author[0000-0001-9910-9230]{Massimo Marengo}
\affiliation{Department of Physics, Florida State University, 77 Chieftain Way, Tallahassee, FL 32306, USA}
\email{}

\author[0000-0002-4896-8841]{Giuseppe Bono}
\affiliation{Dipartimento di Fisica, Universit\`a di Roma Tor Vergata, via della Ricerca Scientifica 1, 00133 Roma, Italy}
\affiliation{INAF -- Osservatorio Astronomico di Roma, via Frascati 33, 00078 Monte Porzio Catone, Italy}
\affiliation{INFN Sezione di Roma, Università di Roma Tor Vergata, Via della Ricerca Scientifica, 1 , Roma, 00133, Italy}
\email{}

\author[0000-0001-7511-2830]{Vittorio F. Braga}
 \affiliation{INAF -- Osservatorio Astronomico di Roma, via Frascati 33, 00078 Monte Porzio Catone, Italy}
\email{}

\author[0000-0001-5870-3735]{Santi Cassisi}
\affiliation{INAF – Osservatorio Astronomico di Abruzzo, via M. Maggini, sn, 64100 Teramo, Italy}
\affiliation{INFN, Sezione di Pisa, Largo Pontecorvo 3, 56127 Pisa, Italy}
\email{}

\author[0000-0003-3096-4161]{Brian Chaboyer}
\affiliation{Department of Physics and Astronomy, Dartmouth College, 6127 Wilder Laboratory, Hanover, NH 03755, USA}
\email{}

\author[0000-0001-8209-0449]{Massimo Dall'Ora}
\affiliation{INAF -- Osservatorio Astronomico di Capodimonte, Salita Moiariello 16, 80131 Napoli, Italy}
\email{}

\author[0000-0002-2662-3762]{Valentina D'Orazi}
\affiliation{Dipartimento di Fisica, Universit\`a di Roma Tor Vergata, via della Ricerca Scientifica 1, 00133 Roma, Italy}
\affiliation{INAF -- Osservatorio Astronomico di Padova, Vicolo dell’Osservatorio 5, 35122, Padova, Italy}
\email{}

\author[0000-0001-5829-111X]{Michele Fabrizio}
\affiliation{INAF -- Osservatorio Astronomico di Roma, via Frascati 33, 00078 Monte Porzio Catone, Italy}
\affiliation{Space Science Data Center -- ASI, via del Politecnico snc, 00133 Roma, Italy}
\email{}

\author[0000-0003-0376-6928]{Giuliana Fiorentino}
\affiliation{INAF -- Osservatorio Astronomico di Roma, via Frascati 33, 00078 Monte Porzio Catone, Italy}
\email{}

\author[0000-0001-5292-6380]{Matteo Monelli}
\affiliation{INAF -- Osservatorio Astronomico d’Abruzzo, via Mentore Maggini s.n.c., 64100 Teramo, Italy}
\affiliation{ IAC -- Instituto de Astrof\'isica de Canarias, Calle V\'ia Lactea s/n, E-38205 La Laguna, Tenerife, Spai}
\affiliation{Departmento de Astrof\'isica, Universidad de La Laguna, E-38206 La Laguna, Tenerife, Spain}
\email{}

\author[0000-0001-6074-6830]{Peter B. Stetson}
\affiliation{Herzberg Astronomy and Astrophysics, National Research Council, 5071 West Saanich Road, Victoria, British Columbia V9E 2E7, Canada}
\email{}

\author[0000-0002-5032-2476]{Fr\'ed\'eric Th\'evenin}
\affiliation{Universit\'e de Nice Sophia--antipolis, CNRS, Observatoire de la C\^ote d'Azur, Laboratoire Lagrange, BP 4229, F-06304 Nice, France}
\email{}

\begin{abstract}

RR Lyrae (RRL) stars, thanks to the existence of their period-luminosity relations, have been widely used as primary standard candles within the Milky Way and the Local Group of galaxies. We present in this work the latest empirical RRL period–luminosity-metallicity (PLZ) and period–Wesenheit–metallicity (PWZ) relations combining optical ($V, I$) and mid-infrared data ($W1$) using OGLE and WISE photometry, anchored to Gaia DR3 parallaxes. We calibrate these relations with the largest, most complete, and homogeneous catalog of Galactic halo field RRL spectroscopic metallicities. These new PLZ/PWZ relations provide distance measurements in remarkable agreement with those derived from eclipsing binaries in the Large and Small Magellanic Clouds (MC): $(m-M)_{0, LMC} = 18.47 \pm 0.07 (stat) \pm 0.05 (sys)$~mag and $(m-M)_{0, SMC} = 18.87 \pm 0.09 (stat) \pm 0.06 (sys)$~mag (within 1\% and 10\% of eclipsing binary distances, respectively). We have also re-calibrated photometric metallicity relations to derive individual [Fe/H] abundances of LMC and SMC RRL stars. This allowed us to investigate the metallicity gradient of the MCs and discover two distinct populations of fundamental mode RRL stars in the LMC (only one is instead found in the SMC). Our newly derived mid-infrared period-luminosity relations are poised to become an essential tool to derive precise distances of older stellar populations in the Local Group of galaxies, and can easily be adapted to observations obtained in JWST NIRCam photometric bands.
\end{abstract}

\keywords{RR Lyrae variable stars (1410), Distance indicators (394), Metallicity (1031), Large Magellanic Cloud (903), Small Magellanic Cloud (1468), Dwarf galaxies (416), Local Group (929), Galaxy distances (590)}

\section{Introduction} \label{sec:intro}

The accurate determination of astronomical distances is a fundamental challenge in astrophysics, directly impacting our understanding of the structure, formation, and evolution of galaxies. The Magellanic Clouds (MCs), the nearest satellite galaxies to the Milky Way, serve as crucial laboratories for studying stellar populations, galaxy interactions, and as anchors for the cosmic distance scale. Over the past decades, numerous studies have sought to refine distance measurements to the Large Magellanic Cloud (LMC) and Small Magellanic Cloud (SMC), utilizing various standard candles such as Cepheids, eclipsing binaries, RR Lyrae stars (RRL), Type II Cepheids and the RGB Tip \citep[e.g.,][]{Pietrzynski2013, Graczyk2014, Moretti2014, Muraveva2018_vmc, Ripepi2017, Cusano2021, Hoyt2023, Sicignano2024, Udalski2025}. Among these, RRL stars are particularly valuable for tracing the old stellar population ($>$ 10 Gyr), as they are distributed throughout the entire extent of the LMC and SMC, unlike eclipsing binaries, which are primarily confined to the bar and inner disk of the LMC \citep{Pietrzynski2019} and the SMC \citep{Graczyk2014}.

Historically, Cepheid variables have been extensively used as standard candles, given their relative high brightness and well-established period-luminosity relation \citep[Leavitt Law;][]{Leavitt1912}, which has been pivotal in extragalactic distance measurements \citep{FeastWalter1987, Freedman2009, Freedman2012, Walker2012, Riess2019}. However, Cepheids trace a younger stellar population, whereas RRL stars provide insight into the ancient structural properties of their host galaxies. RRL stars exhibit a period-luminosity-metallicity (PLZ) relation that becomes significantly more precise in the infrared, as evolutionary and extinction effects are minimized and the intrinsic scatter in the relation decreases. 
The advent of mid-infrared (MIR) observations from Spitzer Space Telescope and the Wide-field Infrared Survey Explorer (WISE, \citealt{2010AJ....140.1868W,2011ApJ...731...53M}) has further strengthened RRL-based distance determinations, as demonstrated by \citet{Neeley2017}, The Carnegie RR Lyrae Program \citep{Muraveva2018_carnegie}, SMASH \citep{Garofalo2018}, \cite{Gilligan2021}
and \cite{Mullen2023}, who employed $\sim$3.5, 4.5 microns bands. These wavelengths mitigate extinction effects and sample the Raileigh-Jeans tail of the star's SEDs, thus decreasing the dependence from temperature and evolutionary effects, improving the precision of the PLZ relation.

The importance of accurate distance measurements extends beyond mapping galactic morphology; it also plays a critical role in determining the Hubble constant ($H_{0}$) and refining the cosmic distance scale \citep{Freedman2021, Pietrzynski2019, Riess2023}. Systematic uncertainties in RRL PLZ relations directly affect measurements of extragalactic distances and, consequently, the tension between local and cosmic microwave background-derived values of $H_{0}$ \citep{Freedman2021, Riess2024}. By refining RRL PLZ relations using high-precision datasets, we can provide an independent and robust distance estimate to the MCs, which serve as anchors for the extragalactic distance scale \citep{Riess2019}. We can also pave the way for adding new anchor galaxies, within the Local Group, with precision distances derived from their RRL-rich old stellar populations.

LMC and SMC are known for their complex shape \citep[e.g.,][]{ElYoussoufi2019, Nidever2024} and they have been the target of numerous photometric surveys: OGLE \citep{Udalski1992, Udalski2015}, SuperMACHO \citep{Rest2005}, SMASH \citep{Nidever2017}, and the VMC survey \citep{Cioni2011}. Recent works have significantly improved RRL distance determinations to the LMC and SMC by integrating multi-wavelength data. \citet{JacyszynDobrzeniecka2017} used OGLE-IV RRL distributions across the MCs to derive their three-dimensional structure, finding that the LMC RRL component is regular and well described by triaxial ellipsoids, with no clear evidence for a pronounced bar-like substructure. \citet{Muraveva2018_vmc} used near-infrared time-series photometry from the VMC survey to derive individual distances to SMC RRL stars. They found that the old SMC population traced by RRL has an ellipsoidal three-dimensional distribution with a significant line-of-sight depth, and that RRL stars in the eastern SMC show signatures of the interaction between the MCs. \cite{Cusano2021} further characterized the spatial distribution of RRL stars and provided insights into the three-dimensional structure of the LMC. These studies have collectively demonstrated that the LMC is not a simple, thin disk system but is rather dynamically evolved, while the SMC appears highly disturbed, likely due to repeated interactions with the LMC and the Milky Way.

In this work, we build upon these efforts to derive new, high-precision PLZ relations for RRL stars by combining optical and mid-infrared data from the OGLE survey \citep{Udalski1992} and the WISE mission \citep{2010AJ....140.1868W}. By addressing systematic biases in previous calibrations and leveraging the improved sensitivity of mid-infrared wavelengths, we derive a new robust distance of the MCs based on old-age stellar population standard candles (RRL). Our measurement is independent from young-age distance indicators such as Cepheids and eclipsing binaries, and aims to reach a comparable accuracy (a few percent). We achieve this goal by relying entirely on photometric data, in order to derive, self-consistently, the metallicity of individual RRL from their light curves, without the need of time-consuming spectral observations. By operating in a key mid-infrared wavelength ($\sim 3.4 \ \mu$m), our method can be easily adapted to observations obtained with the JWST NIRCam camera \citep{Rieke2023}. This will open the possibility of deriving similar high precision distances for other galaxies in the Local Group and beyond, thus extending the available set of anchors for the cosmological distance scale, ultimately aiding the ongoing efforts to reconcile the discrepancies in $H_{0}$ measurements and enhance our understanding of cosmic expansion.

This paper is structured as follows. Section~\ref{sec:dataset} describes the data used in this work. Section~\ref{sec:membership} evaluates the LMC/SMC membership of each RRL based on the latest Gaia proper motion information.
Section~\ref{sec:metallicity} derives and analyze the new metallicity distributions for the LMC and SMC obtained from the best calibrated $\phi_{31}$-$P$-[Fe/H] relations to date.
Section~\ref{sec:plz} describes a new methodology to redefine PLZ and PWZ and reports the coefficients of this work's empirically derived relations.
These relations are used to obtain accurate and precise distances to both the LMC and SMC (Section~\ref{sec:distances}) and to reassess the spatial distribution of the LMC and SMC old-age stellar component. Section~\ref{sec:enrichment} analyzes the chemical enrichment and metallicity gradients of both the LMC and SMC, and compares the photometric vs. spectroscopic metallicities of the oldest LMC and SMC globular clusters. The conclusions and final remarks of this work are given in Section~\ref{sec:Conclusions}.

\section{Dataset} \label{sec:dataset}

\subsection{Merging Optical and Mid-infrared Catalogs}

We downloaded the full catalog of 48,603 RRL in the MCs from the \textit{OGLE Catalogue of Variable Stars} database\footnote{\url{https://ogledb.astrouw.edu.pl/~ogle/OCVS/rrlyr_query.php}} \citep{Soszynski2016}. We then cross-matched the OGLE-IV catalog with the Gaia DR3 database \citep{GaiaDR3} to incorporate proper motion information. We used the algorithm described in \cite{Marrese2022} for sparse catalogs to cross-match the literature RRLs with Gaia DR3 data, keeping only those stars with a Gaia counterpart. A total of 1,709 sources do not include proper motion information, and therefore, they were excluded from our analysis (see next subsection).

The resulting OGLE-Gaia catalog was finally cross-matched with CatWISE2020\footnote{Using a matching radius of 1.0 arcsec.} \citep{Eisenhardt2020, Marocco2021} to obtain deeper mean photometry at 3.4 and 4.6 microns (also known as band $W1$ and W2) and a total of 10,439 stars were found. The CatWISE catalog contains almost 9 years of data and an average of $\sim$ 178 epochs per source. Given the small amplitude that RRL show in the mid-infrared (see e.g., Figure 3 in \citealt{Mullen2021}) and the amount of epochs collected by WISE, we are confident that the profile-fit photometry magnitude in $W1$ (\verb|w1mpro|, essentially the flux average of all epochs) will be nearly as accurate as the intensity average magnitude extracted from $W1$ light curve fitting.  

At first, we included $W2$ in our analysis; however, when analyzing the data carefully, we realized some inconsistency in the W1-W2 color information (greater than a few tenths of magnitude). $W2$ CatWISE photometry is in general shallower than $W1$, and more subject to blending in crowded areas. Therefore, we decided not to include $W2$ in our current analysis. 

In the following, we will refer to the matched OGLE-Gaia catalog as the \textit{Optical} catalog and to the matched OGLE-Gaia-CatWISE as the \textit{MIR} catalog.

\begin{table*}[]
\begin{scriptsize}
\caption{Final catalog of selected LMC and SMC RRL stars and their derived parameters.}
\label{tab:catalog}
\addtolength{\tabcolsep}{-0.4em}
\hspace{-2cm}
\begin{tabular}{ccccccccccccc}
\hline
OGLE ID &  Gaia DR3 Source ID  & CatWISE ID &         RA &        DEC & System & Type &  $\log{P_{F}}$ &  E(V-I) &       [Fe/H] &    $\mu_{0}^{(a)}$ &  $\mu_{0}^{(b)}$ &  $\mu_{0}^{(c)}$\\

&  &  & (deg) & (deg) & \\
\hline
OGLE-LMC-RRLYR-00001 &        4654159362336311296 & J042745.55-704311.9 &  66.939375 & -70.720000 &     LMC & RRab &   -0.197396 &      0.138 & -1.61 &  18.650 &      18.659 &    18.448 \\
OGLE-LMC-RRLYR-00002 &        4655689126613147648 &                     &  66.993540 & -70.183167 &     LMC & RRab &   -0.203760 &      0.120 & -1.07 &         &             &    18.499 \\
OGLE-LMC-RRLYR-00003 &        4654183379793888256 & J042808.48-702123.2 &  67.035420 & -70.356333 &     LMC & RRab &   -0.182767 &      0.124 & -1.55 &  18.324 &      18.319 &    18.387 \\
OGLE-LMC-RRLYR-00004 &        4654182623879626752 & J042819.19-702212.0 &  67.080465 & -70.370000 &     LMC & RRab &   -0.290404 &      0.131 & -1.41 &  18.142 &      18.131 &    18.434 \\
OGLE-LMC-RRLYR-00005 &        4655690711456452608 & J042820.97-700854.7 &  67.087755 & -70.148472 &     LMC & RRab &   -0.191551 &      0.120 & -1.28 &  18.881 &      18.898 &    18.599 \\

 ... & \\
 OGLE-SMC-RRLYR-6670 &        4687687221373818880 &                     &  23.081160 & -72.203167 &     SMC & RRab &   -0.248091 &      0.043 & -1.25 &         &             &           \\
 OGLE-SMC-RRLYR-6671 &        4637646175155200000 & J013302.82-755425.6 &  23.261415 & -75.907000 &     SMC & RRab &   -0.259242 &      0.089 & -2.49 &  18.747 &             &           \\
 OGLE-SMC-RRLYR-6672 &        4630765603186673664 &                     &  23.297370 & -80.580111 &     SMC & RRab &   -0.230903 &      0.109 & -1.88 &         &             &    18.832 \\
 OGLE-SMC-RRLYR-6675 &        4687823831400950784 &                     &  24.224505 & -70.976833 &     SMC & RRab &   -0.197817 &      0.050 & -1.14 &         &             &           \\
 OGLE-SMC-RRLYR-6681 &        4633137249767534592 &                     &  27.389370 & -78.969944 &     SMC & RRab &   -0.215493 &      0.097 & -1.12 &         &             &           \\
\hline
\end{tabular}
\end{scriptsize}
\begin{tablenotes}
    \item RA and DEC are in J2000 (extracted from OGLE).
    \item System tag is based on our membership classification.
    \item $\log{P_{F}}$ is the logarithm of the fundamentalized period, P (obtained from OGLE).
    \item (a) Distance derived using P$W1$Z.
    \item (b) Distance derived using P$W(W1, V-W1)$Z.
    \item (c) Distance derived using P$W(I, V-I)$Z.
\end{tablenotes}
\end{table*}

In this work, we use the periodic properties (period, amplitude, Fourier decomposition components) derived from the OGLE-IV dataset of RRL in the MCs as this survey has a higher cadence, and therefore offer more accurate light curve parameters.

\subsection{Ancillary Field \& Stellar Association Datasets}\label{sec:ancillary}

To address the systematic uncertainties and parameters correlation present in most RRL PLZ relations, we developed a novel calibration approach using a combination of LMC RRLs and field stars (see Section~\ref{sec:plz}). The sample of field RRLs is chosen from an extensive catalog of 9169 field RRLs, from which we have either [Fe/H] abundances derived from HR spectra ($R \ga 20$,000) or an estimate of their metallicity based on the $\Delta$S method \citep{1959ApJ...130..507P}, applied to medium resolution spectra ($R \sim 2$,000) from the Large Scale Area Multi-Object Spectroscopic Telescope (LAMOST) DR2 survey \citep{2012RAA....12..735D, 2014IAUS..298..310L} and Sloan Extension for Galactic Understanding and Exploration \citep[SEGUE,][]{2009AJ....137.4377Y}. For a complete and detailed description of the homogeneous metallicity scale, the HR metallicity catalog's demographics, the $\Delta$S calibration, and the spectrum selection criteria adopted in our combined metallicity catalog, we refer the reader to the \cite{Crestani2021} paper. 

The resulting HR+$\Delta$S metallicity catalog was cross-matched with the $Gaia$ DR3 database \citep{GaiaDR3} to provide astrometric data.  

Infrared photometry for all stars in the metallicity catalog is obtained by cross-matching all stars with those in the CATWISE catalog in order to provide $W1$-band magnitudes. Similarly, visible time-series photometric measurements in the $V$ band were taken from the All-Sky Automated Survey for Supernovae (ASAS-SN, \citealt{2014ApJ...788...48S, 2018MNRAS.477.3145J}). For a more detailed description of the ASAS-SN survey and how we have integrated such measurements into our calibration sample, we refer the reader to  \cite{Mullen2021} and \cite{Mullen2022}.

Lastly, as an external independent validation of our derived relations in Section~\ref{sec:plz_results} we introduce additional V, I, and $W1$ photometry datasets to derive the distance to three stellar systems (M4, Reticulum, and Sculptor). 

\begin{figure}
    \centering
    \includegraphics[width=1.0\linewidth]{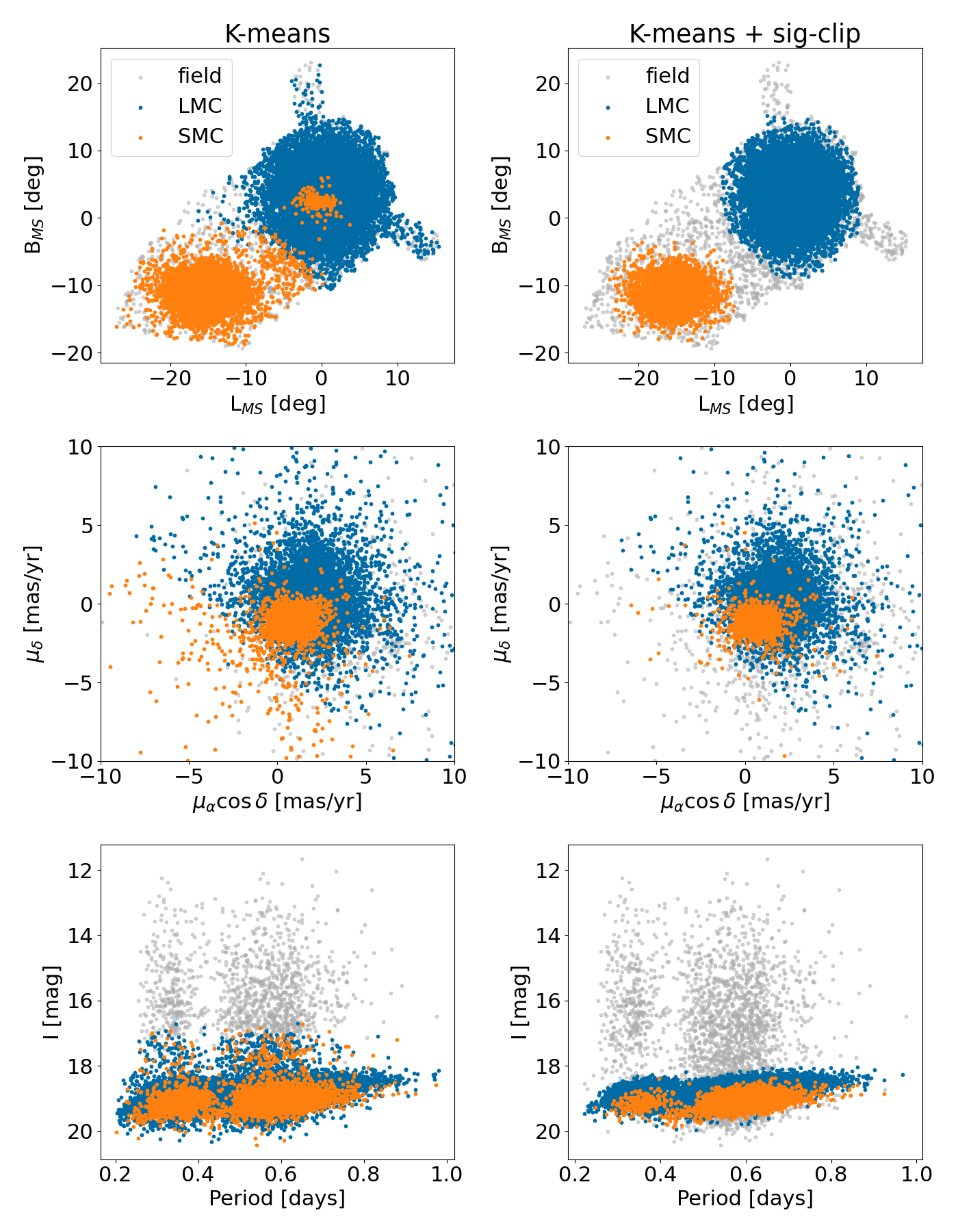}
    \caption{Separation of LMC (blue), SMC (orange) and field (gray) member RRL stars using k-means clustering (leftmost panels) and the final selection (rightmost panels) refined with sigma-clipping performed in z direction and elliptical distance from center of the ellipsoid to each star.}
    \label{fig:clustering}
\end{figure}

\subsection{Reddening}\label{sec:reddening}
To correct the MC RRL for reddening we adopted \citet{Skowron2021}'s $E(V-I)$\footnote{Specifically, we use \texttt{E(V-I) median}.} reddening map based on OGLE-IV red clump stars, when available \footnote{\url{https://ogle.astrouw.edu.pl/cgi-ogle/get_ms_ext.py}}. Not all the area covered by this work, however, has OGLE $E(V-I)$ values. In those cases, and for the field RRL in our metallicity catalog, we used the $E(V-I)$ values derived from \citet{Schlegel1998}'s $E(B-V)$ maps with the proper \citet{Schlafly2011} correction applied. According to the extinction law provided by \cite{Cardelli1989} calculated for the effective mean wavelengths of Landolt’s (Landolt 1992) filter/photocathode combination for stars of spectral type A0III–F5III, we have $A_V = 3.06\, E(B-V)$ and $A_I = 1.80\, E(B-V)$, and therefore $E(V-I) = E(B-V) / 1.25$.

\section{LMC and SMC membership}\label{sec:membership}

A two step selection process is performed in order to obtain a clean sample of LMC and SMC members. 

In the first step, thanks to Gaia DR3 and the proper motion information, we follow a different separation than the on-sky categorization provided by OGLE by taking advantage of the 2D velocity information of the OGLE-IV MC RRL stars. We separated the LMC from SMC RRL and those from the MW field RRL by fitting a \textit{k-means} model to our data, assuming three clusters in our sample. The information used to identify the clusters are the position of the stars in the sky (RA, Dec), their proper motions, and their magnitudes in the $I$ band. To optimize the process, we standardize each parameter involved in the clustering detection by changing its reference to its central values and normalizing by its standard deviation. Both optical and MIR catalogs were jointly cleaned in this process. The result of this exercise is shown in the leftmost panels of Figure~\ref{fig:clustering}.
With this sample we derive a PLZ in I-band (following the procedure described in \S~\ref{sec:plz_non-iterative}).

For the second step, we use the first iteration of the PLZ relation in I-band to obtain individual distances so we get the cleanest separation possible, especially from those stars in front of the MC that belong to the halo of the Milky Way. In this step, we excluded the RRd stars since individual distances are only provided for RRab and RRc stars. Information on how distances are obtained is explained later in \S~\ref{sec:distances}. With the equatorial coordinates and the distances, we derive Cartesian (x,y,z) coordinates for each star. As described in \citet{JacyszynDobrzeniecka2017}, there is an extension of stars along the line-of-sight (z) direction that we would like to avoid, which is a blend artifact. Similar to what those authors did, we performed an iterative 3-sigma clipping selection in the z-direction. Furthermore, to select a pure sample of LMC and SMC members, we derived the elliptical distance (r') from each star to the center of each spheroid. We performed a spatial selection of members by fitting ellipses in the three Cartesian planes using the \verb|astroML| \citep{Ivezic2014} function \verb|fit_bivariate_normal| in an iterative way to remove outliers farther than 5-sigma for the LMC and 4-sigma for the SMC. Figure~\ref{fig:clustering} (right columns) shows the final member selection of LMC (blue), SMC (orange) and field (gray) RRL. This selection excluded a total of 1,722 stars for which the k-means method was not able to accurately determine membership. We ended up with a sample of 28,271 LMC and 4,624 SMC RRL members. Table~\ref{tab:catalog} shows the IDs of the final LMC and SMC members in the different catalogs (OGLE, Gaia, CatWISE), their coordinates, membership to each system, together with pulsation properties (such the logarithm of the fundamentalized period and the pulsation type), reddening (\S~\ref{sec:reddening}), and the newly derived metallicity and distances -- which will be described in \S~\ref{sec:metallicity} and \ref{sec:distances}.

The center of the LMC RRL star distributions based on this new membership selection is at RA = $5^{h} 22^{m} 02.8^{s}$ and Dec = $-69\degr32\arcmin31.8\arcsec$ (J2000), which is slightly offset North-East form the one obtained by \citet{JacyszynDobrzeniecka2017} (RA = $5^{h} 21^{m} 31.2^{s}$, Dec =$-69\degr36\arcmin36\arcsec$). It is worth noting that the derived center of the RRL stars (i.e., old population of the galaxy) is slightly offset to the North-West from the one adopted by 
\cite{Pietrzynski2019} for the young stellar population in the LMC (RA = $5^{h} 20^{m} 12^{s}$ and Dec = $-69\degr18\arcmin00\arcsec$). On the other hand, for the SMC -- also based on our membership selection-- the center is at RA = $0^{h} 56^{m} 46.2^{s}$ and Dec = $-72\degr47\arcmin02.4\arcsec$, slightly South-West from the one obtained by \citet{JacyszynDobrzeniecka2017} (RA = 0:55:48.0, Dec =-72:46:48). Table~\ref{tab:coordinates} also lists the center of both distributions in the cartesian and Magellanic Stream (MS) reference frame \citep{Nidever2008} converted using \verb|gala| \citep{Price-Whelan/gala, Price-Whelan/gala-soft}. The center of both galaxies in the different coordinate systems were calculated by fitting each pair of coordinates to a bivariate normal distribution, with the \verb|astroML| function \verb|fit_bivariate_normal|, using a ranked-based statistics which is robust to outliers. 

\begin{table}[]
\begin{threeparttable}
\caption{Center of the Magellanic Clouds in different coordinate frames (Equatorial, Magellanic Stream, and Cartesian) based on their RRL population.}\label{tab:coordinates}
\begin{tabular}{cccc}
\toprule
Coordinate & \multicolumn{2}{c}{System} & Units \\
\cline{2-3}
&    LMC      &    SMC      &    \\
\hline
RA  &  80.511660   &   14.192393  & deg \\ 
DEC & --69.542167  &  --72.784000 & deg  \\
\hline
L$_{MS}$ & --0.278143 &  --15.358386 & deg \\ 
B$_{MS}$ & 2.538484   &  --11.382220 & deg \\
\hline
x   &   2.813946  &    16.920120 & kpc \\  
y   &  16.977628  &     4.165105 & kpc \\
z   & --46.076418 &  --56.102905 & kpc \\
\hline
\end{tabular}
\begin{tablenotes}
\item Cartesian coordinates $x, y, z$ of the center of the LMC and SMC were obtained from the new P$W(I, (V-I)$Z relation derived in this work (see \S~\ref{sec:distances} for details). 
\end{tablenotes}
\end{threeparttable}
\end{table}

\section{Metallicities of the LMC and SMC} \label{sec:metallicity}

Photometric metallicities from OGLE-IV LMC and SMC RRLs have been derived by \cite{2016AcA....66..269S}, but only for RRab stars. In order to fill this gap, we derive our own set of [Fe/H] abundances for all RRab and RRc variables in the OGLE-IV LMC and SMC samples. In doing so, we also ensure that our derived metallicities are consistent with the spectroscopic abundances we have adopted for our HR+$\Delta$S metallicity catalog described in section~\ref{sec:ancillary}.

Many relations exist for photometric metallicities (e.g., \citealt{1996A&A...312..111J,2005AcA....55...59S,2007MNRAS.374.1421M,2013ApJ...773..181N,2014IAUS..301..461M,Martinez-Vazquez2016,2016ApJS..227...30N,2020arXiv200802280I,Mullen2021,Mullen2022,2021ApJ...920...33D,2023ApJ...944...88L,2025MNRAS.536.2749M}), using modern datasets in a variety of wavelengths. Among the available calibrations, Mullen's $\phi_{31}$-P-[Fe/H] relations are particularly compelling, having been derived from the largest homogeneous spectroscopic [Fe/H] datasets to date, with metallicities extending farther into lower and higher metallicity regimes than many of the prior works. Additionally, they are available for both fundamental and first overtone modes. We refer the reader to \cite{Mullen2021,Mullen2022}, for a more detailed discussion of the strengths and limitations of the various relations. In comparison to more recent calibrations, metallicity estimates of \cite{2025MNRAS.536.2749M}, tend to appear systematically less metal-rich at the high-metallicity end and more metal-rich at the low-metallicity end for the RRab, when compared to \cite{Mullen2021}. This trend is apparent in other relations quoted in \cite{Mullen2021}, where the peak of the metallicity distribution tends to dominate the fit at the expense of the tails.

Out of the aforementioned listed relations, only \cite{2005AcA....55...59S,2021ApJ...920...33D} are direct $I$-band calibrations. While convenient for this work, given the availability of $I$-band $\phi_{31}$ from OGLE, we are not adopting these two relations. This is due to a number factors such as \cite{2005AcA....55...59S} only having published relations for RRab, and \cite{2021ApJ...920...33D} having a significantly smaller calibration size of 80 RRab and 24 RRc.

The calibration of Mullen's relations, although only available in $V$ and some mid-infrared bands, is instead based on 1980 RRab and 594 RRc Milky Way field RRL stars that have either [Fe/H] abundances derived from high-resolution (HR) spectra or an estimate of their metallicity based on the $\Delta S$ method \citep{Crestani2021, Fabrizio2021} and light curves $V$-band photometry from ASAS-SN. The reliability of these relations was tested with globular clusters, which showed a scatter of less than 0.1 dex for the systems\footnote{Individual errors are considered to be within 0.4 dex.}. 

\citet{Mullen2022} found a small offset between the metallicities derived with their relations for RRab and RRc variables, likely due to differences in calibration between the high resolution and $\Delta S$ metallicities for RRL in the two pulsation modes. To ensure that all metallicities are homogeneous and consistent with the $\Delta S$ metallicities from \citet{Crestani2021}, we re-measured this offset with the help of mono-metallic Galactic globular clusters, rich in both RRab and RRc variables. This process is explained in Appendix~\ref{ap:metallicity_check}, that reports the revised form of Mullen's relations that correct for this issue, and also the prescription to convert between our metallicities and the common metallicity scale of \citet{Carretta2009}.

Since Mullen's relations are in Johnson $V$-band, we transformed the $\phi_{31}$ values from OGLE-IV, derived from $I$-band light curves, and used the \cite{1999A&A...348..815D} (DF99 hereafter) coefficients of the Fourier inter-relations for RRab and RRc stars to transform between $\phi_{31}^{I}$ and $\phi_{31}^{V}$ (see their Table 4). Several transformations exist in the literature \citep{2010MNRAS.402..691D,2016AcA....66..269S} but DF99 inter-relations give the most consistent result between our spectroscopic metallicities and are available for both the RRab and RRc pulsation modes. Due to the fact the OGLE performed cosine decomposition to the light curves and Mullen's RRab $\phi_{31}$-P-[Fe/H] relation uses a sine decomposition, we also added a factor $\pi$ to the resulting $\phi_{31}$ values of RRab stars. Therefore the conversion was done using the following relations:

\begin{equation}
\phi_{31}^{s,V} (\rm{RRab}) = -0.039 + \phi_{31}^{c,I}\cdot0.788 + \pi
\end{equation}

\begin{equation}
\phi_{31}^{c,V} (\rm{RRc}) = -0.249 + \phi_{31}^{c,I}\cdot0.995
\end{equation}

\noindent
where $\phi_{31}^{c,\lambda}$ and $\phi_{31}^{s,\lambda}$ are derived from the cosine and sine form of the Fourier decomposition, respectively, and $\lambda$ indicates the band. Our complete catalog LMC and SMC RRL metallicities, derived with the Mullen's relations as described above, is listed in Table~\ref{tab:catalog}.

\begin{table*}
 \centering
  \caption{PLZ and PWZ relations coefficients defined as $M=a(\log P + 0.26) +b([Fe/H]+1.5)+c$. The number of stars in a given fit is denoted by $N$, $\alpha$ is the color coefficient used in the Wesenheit magnitude, and the fitting methodology is given in the comments, as explained in Section~\ref{sec:plz_iterative} and ~\ref{sec:plz_non-iterative}.}\label{tab:fits}
 \begin{tabular}{lcccccccc}
 Band & Type & $N_{Field}$ &$N_{LMC}$  & $\alpha$ & $a$ & $b$ & $c$ & Comments\\
 \hline
 $W1$& ab + c & 1268+677 & 3992+407 & \nodata & $-1.97\pm 0.08$ & $0.20 \pm 0.027$ & $-0.440 \pm 0.008$ & Iterative\\
  I & ab & \nodata & 3993 & \nodata & $-1.497\pm 0.100$ & $0.170 \pm 0.003$ & $0.207 \pm 0.006$ & Fixed b\\
 W(W1, V-W1)& ab + c & 1097+426 & 3864+401 & 0.068 & $-2.063\pm 0.091$ & $0.195 \pm 0.039$ & $-0.514 \pm 0.009$ & Iterative\\
 W(W1, V-W1)& ab & & 3864 & 0.068 & $-2.075\pm 0.042$ & $0.183 \pm 0.003$ & $-0.518 \pm 0.002$ & Fixed b\\
 W(W1, I-W1)& ab & \nodata  & 3992 & 0.124 & $-2.053\pm 0.042$ & $0.185 \pm 0.003$ & $-0.524 \pm 0.002$ & Fixed b\\
 W(I, V-I)& ab & \nodata  & 3864 & 1.385 & $-2.461\pm 0.115$ & $0.150 \pm 0.003$ & $-0.451 \pm 0.007$ & Fixed b\\
  \hline
 \end{tabular}
\end{table*}

\section{Refining Period-Luminosity/Wesehheit-Metallicity relations}\label{sec:plz}

In general, throughout the literature, stellar system distances from RRL PLZ/PWZ relations have steadily decreased with newer calibrations (as demonstrated by M4, Reticulum, Sculptor in \cite{Mullen2023} Figure 4). This trend is partly attributed to gradual zero-point changes between HST, Gaia DR2, and the most recent Gaia DR3 (Figure 5 in \citealt{Mullen2023}). With more recent astrometric information, the LMC and SMC can serve as important, well-studied benchmarks to derive new and updated PLZ/PWZ relations based on OGLE and mid-IR bands. The MCs have the benefits of possessing a large, diverse population of RRL (ideal for aiding in calibration) and very precise distances coming from eclipsing binaries \citep{Graczyk2014,Pietrzynski2019} ideal for validating any fits. 

In addition to this work providing new empirical fits, we introduce new fitting methodologies to overcome widespread degeneracies in published PLZ/PWZ relations. There exists an overwhelming number of different PLZ/PWZ relations in literature, even within the same band, making it often hard to choose what relation to use for distances. One can make arguments for favored relations based on differences in the sample of sources or their attributed parameters ([Fe/H], extinction, parallax, period, magnitude, pulsation mode, etc.). However, it is often not until one applies the relation(s) and compares it to more definitive geometrical distances (i.e., parallax, EBs) that one sees how a relation performs. It is inadvisable to intuit the fit performance by simply comparing the fit coefficients of the standard bi-linear PLZ form (see eqn. in Table~\ref{tab:fits}). A strong correlation exists between period and metallicity slope; see Figure 3 in \cite{Mullen2023}. Therefore, based upon differences in fitting techniques, one can trade off between period and [Fe/H] dependencies and still get the same derived distances.

By decomposing the fits as demonstrated in the following sections, degeneracies between period and [Fe/H] coefficients can be minimized, and a future of self-consistent non-degenerate RRL relations can be developed. The methodologies can be categorized as either relying on an initial fit of the period slope for samples of stars sharing a common distance (Section~\ref{sec:plz_iterative}) or those relying on fixing the [Fe/H] slope based on theoretical models (Section~\ref{sec:plz_non-iterative}). 

The fundamentalization required to include RRc type is explored in Section~\ref{subsec:fundamentalization}. Details of these fitting techniques, including a discussion of their benefits and limitations, are explored in Sections~\ref{sec:plz_iterative}, and \ref{sec:plz_non-iterative}. In Section~\ref{sec:plz_results} we validate the results with an astrophysical application of our relations.

\subsection{Fundamentalization}\label{subsec:fundamentalization}

In order to simultaneously fit both fundamental (RRab) and first-overtone (RRc) RRL, the RRc stars have their period fundamentalized in this work. For the field RRc stars fundamentalization, we add 0.127 to $\log P$ \citep[following e.g.][]{1971A&A....14..293I,2022MNRAS.tmp.2792B}. To ensure that this relationship is applicable for the MC RRL, we have derived the relation between the fundamental mode and first-overtone period from the double-mode RRL (RRd) available in OGLE-IV MC RRL database\footnote{We exclude in this analysis the anomalous RRd identified by the OGLE team.}. We performed a least squares polynomial fit to the fundamental period (P$_{F}$) and first-overtone period (P$_{FO}$) in the sample of 2786 RRd in the MC. The fit is shown in eq.~\ref{eq:fundamentalization}.

\begin{equation}\label{eq:fundamentalization}
    P_{F} = 1.3122 (\pm 0.0003) P_{FO} + 0.0113 (\pm 0.0001); \sigma = 0.0003
\end{equation}

We use Equation~\ref{eq:fundamentalization} to fundamentalize the MC RRc; however, we note that the fundamentalization relation used for either the field or MC produces nearly identical results, and the selection of either relation does not affect the PLZ/PWZ fits.

\subsection{Iterative method} \label{sec:plz_iterative}

\begin{figure*}
\centering
\includegraphics[width=.49\textwidth]{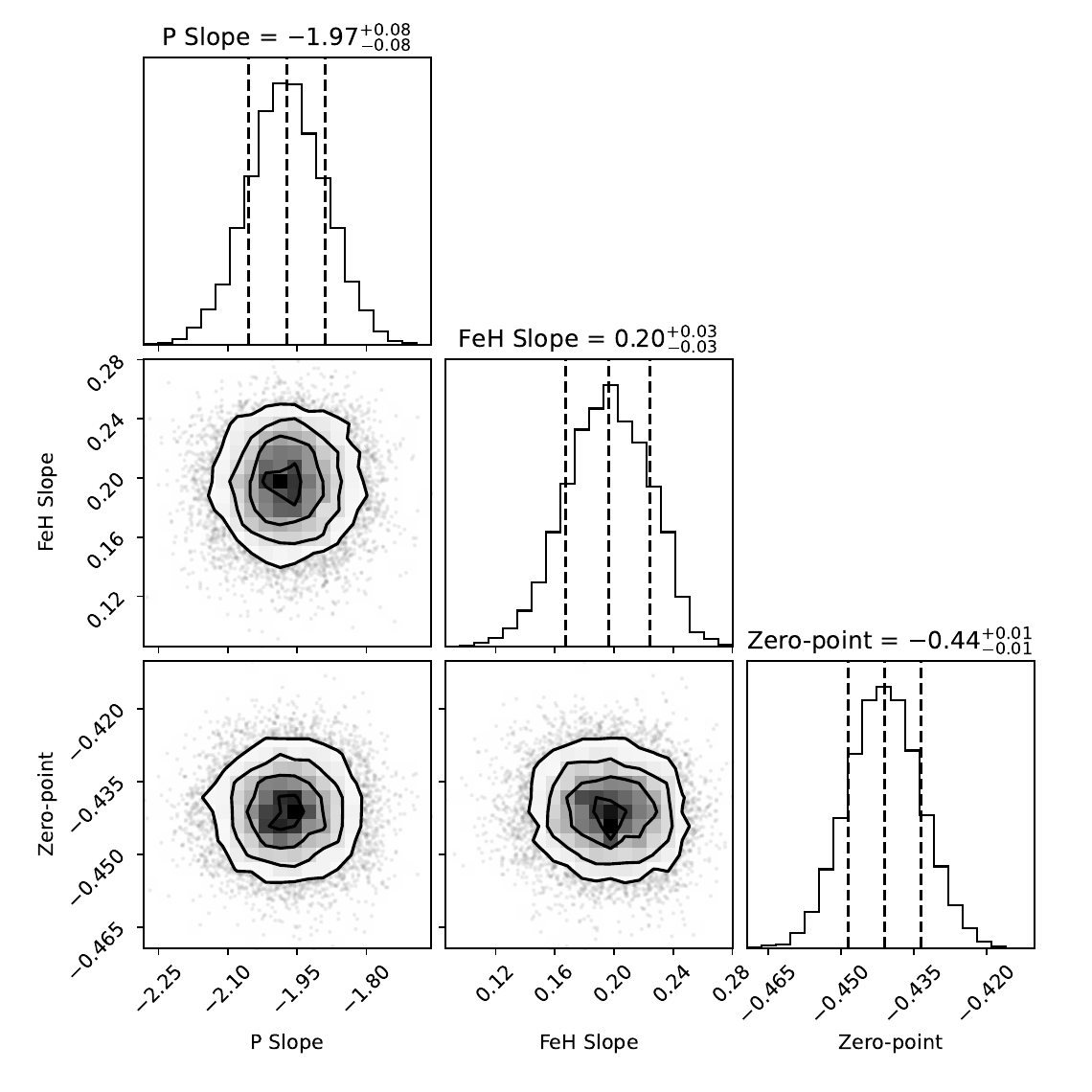}
\includegraphics[width=.49\textwidth]{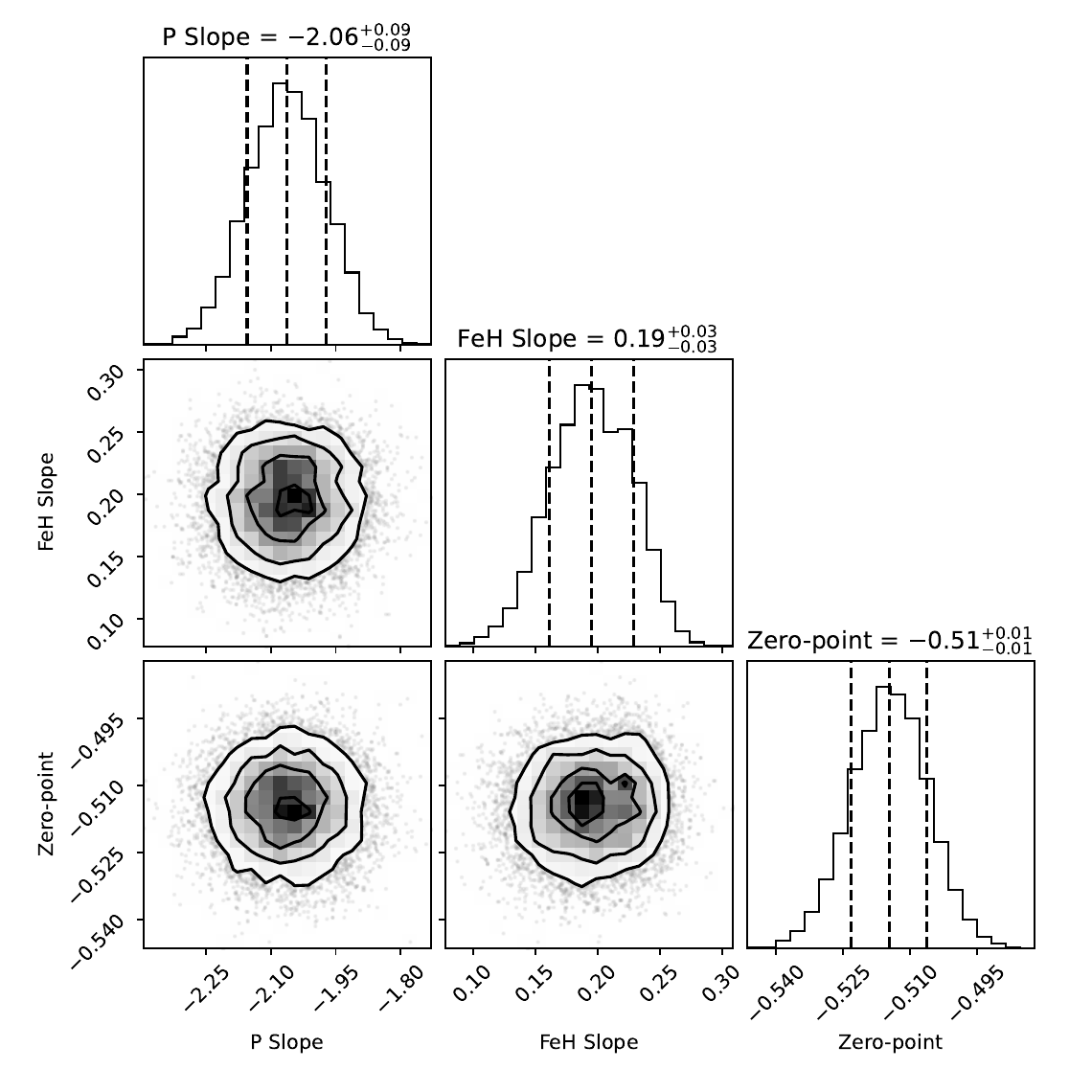}
\caption{Posterior distribution of the P-$W1$-[Fe/H] fit (left) coefficients and of the P-($W1, V-W1$)-[Fe/H] fit coefficients (right) determined by a Monte Carlo simulation with bootstrap resampling. In the figure, zero-point stands for the coefficient $c$ in Table~\ref{tab:fits}; ``P Slope'' is the log period slope (coefficient $a$), and the [Fe/H] slope (coefficient $b$) is ``[Fe/H] Slope.''}.\label{fig:PLZ}
\end{figure*}

Our approach in fitting RRL PLZ relations draws on the strengths of both MC and field star datasets. We deconstruct the standard 3-component linear PLZ relation by separately fitting the period slope using the LMC RRL. The [Fe/H] slope and zero point of the period-luminosity relation utilize instead the field RRL (leveraging their accurate parallaxes and homogeneous spectroscopic metallicities). The fitting process follows this general procedure:
\begin{enumerate}
\item We fit a linear PL relation to the LMC data. From this step, we get an initial approximation of the period slope (temporarily ignoring their metallicity and dispersion along the line of sight). 
\item    \begin{enumerate}
    \item We fit a PLZ relation to the field MW Halo RRLs, where we fix the period slope to that from the prior step, and fit the metallicity slope and zero point.
    \item We return to the LMC data and fit a PLZ relation, holding the metallicity slope and zero point constant to those from the prior step, and retrieve an updated period slope (now taking into account the photometric metallicity of the individual MC RRL).
    \end{enumerate}
\item We repeat steps 2a and 2b until convergence.
\end{enumerate}

The only purpose of step~1 is to derive a prior for the period slope, to be replaced with  fitted values through the iterative procedure. Step 2a anchors the PLZ zero point entirely to Gaia DR3 parallaxes: note that no LMC/SMC distances are used or assumed anywhere in the fitting procedure, not just because of a non-negligible spread in the LMC RRL distances (Section~\ref{sec:distances}), but also because we intend to use EBs distances to validate our PLZ relations. Similarly, the metallicity slope is entirely constrained on the spectroscopic metallicities of the halo stars fit in step 2a: the advantage in this case rests on the broad extent and homogeneity of the spectroscopic catalog ($-2.5 <$ [Fe/H] $< 0.0$), providing maximum leverage in the fit. By fitting the period slope separately on an independent dataset based on LMC stars (step 2b) we break the correlation between period and metallicity slope, and exploit the broad range of periods of the RRL in the Large Magellanic Cloud. Even though the metallicity slope is not changed in step 2b, we do take into account the individual metallicities of the LMC stars, which are derived for each star using the OGLE $I$-band Fourier parameters, as described in section~\ref{sec:metallicity}. Conversely, the line-of-sight depth of the LMC contributes additional scatter to the PL relation. Nevertheless, the large size of the RRL sample should prevent this from significantly affecting the precision of the period-slope determination in step 2b, unless period is correlated with line-of-sight position. No such correlation, however, is expected, since we do not find any dependence of period with respect to the projected radial distance across the plane of the sky.

We start with the $W1$ band fit according to the above procedure. For the LMC MIR sample, we utilize individual OGLE reddening and extinction conversions as laid out in Section~\ref{sec:reddening}, transformed to the $W1$ band using the Galactic ISM ratio $A_{W1}/A_V = 0.061$. In the $W1$ sample, 97\% of the stars have relatively low reddenings, with $E(B-V) < 0.15$, which corresponds to $A_{W1} < 0.028\,$mag.  The maximum reddening in the sample is $E(B-V) = 0.43$, corresponding to $A_{W1} = 0.08\,$mag. Thus, uncertainties in the reddening values, generally estimated to be $\sim$10\%, will not significantly impact the PLZ fit. 

For the field sample, we utilize the methods from \cite{Mullen2023}. Reddenings were determined from the Lallement et al. (2019) 3-D maps. Gaia EDR3 parallaxes were processed according to the recommendations in \citet{GaiaDR3}. In particular, the parallaxes were corrected for the Gaia zero-point systematic error using the calibration of \cite{EDR3bias}. Additionally, in order to obtain a clean sample of RRLs, only RRL with $RUWE < 1.4$ and \texttt{astrometric\_excess\_noise} $< 0.2$ were allowed in the final catalog, prior to fitting the $PLZ$ relation.

As we apply the methodology described above, the explicit fits, which take into account the uncertainties in all of the observed quantities, were performed using a bootstrap-resampled Monte Carlo orthogonal distance regression \citep{Boggs, Zwolak} fitting technique. The LMC data are fit using the regular form of the PL/PLZ fit (see Table~\ref{tab:fits} caption), while the field stars utilize the Astrometric Based Luminosity (ABL), following the procedure outlined in \citealt{Layden19,Gilligan2021,Mullen2023}:

 \begin{equation}
\varpi 10^{0.2m_o - 2} = 10^{0.2 [a (\log P + 0.26) + b ([\mathrm{Fe/H}] + 1.5) + c ] }
 \label{eqnfitfunctionE}
\end{equation}
 where $\varpi$ is the Gaia DR3 parallax in mas, $m_o$ is the absorption corrected apparent magnitude, and the $a,b$ and $c$ are determined in the fit. When RRc stars are utilized, $P$ refers to the fundamentalized period (see Section~\ref{subsec:fundamentalization}).

In order to accurately estimate the uncertainties in the fitted coefficients, bootstrap resampling was used to generate 10,000 datasets, i.e. each individual step of the fit was performed 10,000 times. The bootstrap samples have the same size as the original dataset, and were generated by random resampling with replacement. For each realization, the properties of every selected datapoint were perturbed according to their associated uncertainty distributions, assuming Gaussian uncertainties in [Fe/H], $\varpi$, period, magnitude, and reddening. We assumed a uniform period error of $10^{-4}$ days and a 10\% uncertainty in the reddening for both the LMC and field sources. All LMC photometric metallicities are given uncertainties of 0.3 dex \cite[from][]{Mullen2021, Mullen2022}. The parameters that are kept fixed in each iteration (i.e. the period slope in step 2a, and the metallicity slope and zero point in step 2b) are drawn from the posterior distributions of the previous step in the iteration.
 
We found that the fits usually converge within the first half-dozen rounds (i.e. the median slopes and zero point change by much less than the dispersion around the fit). To ensure stability, we however iterate in all cases for a total of 20 rounds.
 
The initial fits revealed several significant outliers ($>4\,\sigma$) in both the field and LMC samples, including stars incorrectly classified as LMC members by the initial clustering algorithm. Rather than applying additional clustering procedures or magnitude cuts based on assumed relations, we relied on an iterative outlier-rejection procedure. After each iteration, we computed the RMS scatter and temporarily excluded stars lying more than $3\sigma$ from the fit. In subsequent iterations, the fit and RMS were recalculated using the cleaned subset, while the outlier criterion continued to be evaluated against the full original sample. This approach allowed stars rejected in earlier iterations to be re-included as the fitted relation evolved, producing a progressively cleaner and more self-consistent sample. Provided that the initial fit is reasonably accurate and the rejection threshold is not overly restrictive, the procedure typically converges within five iterations, with only minor changes to the sample membership. We applied this procedure only once, in order to finalize a final clean sample that was then adopted to fit all the subsequent relations.
 
The resulting fit for the $W1$-band PLZ is summarized in Table~\ref{tab:fits}. Figure~~\ref{fig:PLZ} shows the corner plot for the three fitting parameters: as expected the round shape of the posterior distributions confirms that independently fitting the period and metallicity terms largely removes the correlation between period and [Fe/H] in the RRL sample.
 
In addition to fitting iterative PLZ relations in the $W1$ band, we also fit the PWZ relation in a Wesenheit band constructed using the $W1$ and Johnson $V$ bands, defined as  $W (W1,V) = W1 - (V - W1) * 0.068$ \citep[following][]{Neeley2017}. For this fit we adopted OGLE $V$-band photometry (whenever available) for LMC sources; for the field RRL we instead used flux-averaged magnitudes calculated from the ASAS-SN survey (see section~\ref{sec:ancillary}). In this fit, we assume a uniform uncertainty of 0.05 mag for the $V$-band photometry, corresponding to the approximate error attributed at the mean magnitude of LMC sources (see \citealt{2016AcA....66....1S}). The best fit coefficients are reported in Table~\ref{tab:fits}.

\subsection{Non-iterative method} \label{sec:plz_non-iterative}

Inversion of the $I$-band PLZ relation has been widely used to measure the metallicity distributions of RRL populations in clusters and nearby galaxies \citep[see e.g.][]{MartinezVazquez2016a, Braga2016}. For this reason we decided to include a new calibration of PLZ and PWZ relations that include the $I$ band, designed to be consistent with the relations we have derived for the $W1$ band. The iterative method described above is however not practical for these relations, due to the lack of an homogeneous set of $I$ band photometry for our field halo stars sample. As a consequence, our new $I$-band PLZ relation will be entirely based on the LMC photometric data from OGLE.

Having to rely on a single dataset to fit all three PLZ coefficients, however, has one significant drawback: the correlation between period and metallicity slope, as described above, limits the robustness of the fit. To avoid this issue, we fix the metallicity slope to the theoretical value from \cite{Neeley2017}, and only perform a two-parameter fit (period slope and zero point) in the $I$-band relation. While this choice is not ideal, it is justified by the excellent agreement we found between the metallicity slopes derived in section~\ref{sec:plz_iterative}, and the corresponding theoretical values in \cite{Neeley2017}. The adopted values are reported in Table~\ref{tab:fits}. Furthermore, the absolute value of the [Fe/H] slope is small, so a slight deviation in its exact value has a smaller impact than the period slope. Furthermore, the approximate [Fe/H] slope is generally well known, and has a minimal wavelengths dependence in the spectral range of interest (see Table 6 in \citealt{2015ApJ...808...50M}).

With this assumptions, we fit the two remaining coefficients with the same bootstrap Monte Carlo technique as in the previous section (but the fit is performed only once, with no iterations). Even when fitting relations that do not involve WISE bands (e.g. the $I$-band PLZ and the $W(I, V - I)$ PWZ, we restrict the fit on sources that have both OGLE and WISE photometry. This is because, in order to fit the zero point of the relations, we need the distance of each individual RRL in the sample. These distances (and their uncertainties) are derived using the $W1$-band derived in section~\ref{sec:plz_iterative}, adopting a Monte Carlo procedure (10,000 samples derived assuming Gaussian uncertainties in period, metallicity, reddening, and using the posterior distributions obtained in section~\ref{sec:plz_iterative} for the PLZ coefficients). This choice allows us to anchor the zero point of the non-iterative PLZ and PWZ relations to the same Gaia parallaxes as in the iterative method, which are entirely independent on the LMC detached eclipsing binary distance. The metallicity of the individual stars used for this fit is the same photometric metallicity derived in section~\ref{sec:metallicity}, that were adopted for the iterative method.

Table~\ref{tab:fits} shows the coefficients of the individual fits. We only provide an RRab type fit for the non-iterative technique, due to the scarcity of RRc in the LMC sample compared to the field source. 

As a test for our overall fitting strategy, we provide the $W(W1, V - W1)$ PWZ relation using both the iterative and the non-iterative method. The resulting fitting parameters, reported in Table~\ref{tab:fits}, are in excellent agreement, well within their respective $1 \sigma$ uncertainties.

\begin{table}[!t]

    \begin{center}
    \caption{Distance of known systems}\label{tab:validation}
    \begin{tabular}{llc}
    \tableline
    \tableline
    System & Bands & $\mu_{0}$ [mag] \\
    \tableline
    Reticulum & $W1$         & 18.23$\pm$0.04(stat)$\pm$0.10(syst)\\
    Reticulum & $W(W1,V-W1)$ & 18.22$\pm$0.04(stat)$\pm$0.11(syst) \\
    \tableline
    M4 & $W1$                & 11.20$\pm$0.01(stat)$\pm$0.09(syst) \\
    M4 & $W(W1,V-W1)$        & 11.21$\pm$0.01(stat)$\pm$0.10(syst) \\
    \tableline
    Sculptor & $W1$          & 19.51$\pm$0.03(stat)$\pm$0.10(syst) \\
    Sculptor & $W(W1,V-W1)$  & 19.51$\pm$0.03(stat)$\pm$0.11(syst) \\
    \tableline
    \end{tabular}  
    \end{center}
\end{table}

\subsection{Independent validation} \label{sec:plz_results}
To provide an independent validation for our new iterative relations, we follow \citet{Mullen2023}, by deriving the distance modulus of three stellar systems: two globular clusters (M4 and Reticulum) and the Sculptor dSph galaxy. For these calculations, we utilize the same sources, photometry, extinction, and metallicity as presented in Section 4 of \cite{Mullen2023}, with their metallicity converted into the $\Delta S$ scale using the prescription in \ref{ap:metallicity_check} (equation~\ref{eq:C09toC21}). The results are presented in Table~\ref{tab:validation}. We do not derive distances based on the non-iterative relations, since these are anchored on the distances of LMC stars from the iterative relations, and as such they would not be independent determinations.

For each system, we found an almost perfect agreement between the distance derived from the PLZ and the PWZ relation. The values posted in Table~\ref{tab:validation} are also in excellent agreement with other determinations anchored to Gaia DR3 distances. As previously noted, all DR3-based distance tend to be smaller (but still within 1 to 2$\sigma$) than values measured with other methods (although this bias is reduced in this work, with respect to the results in \citealt{Mullen2023}).

\begin{figure*}
    \centering
    \includegraphics[width=0.49\linewidth]{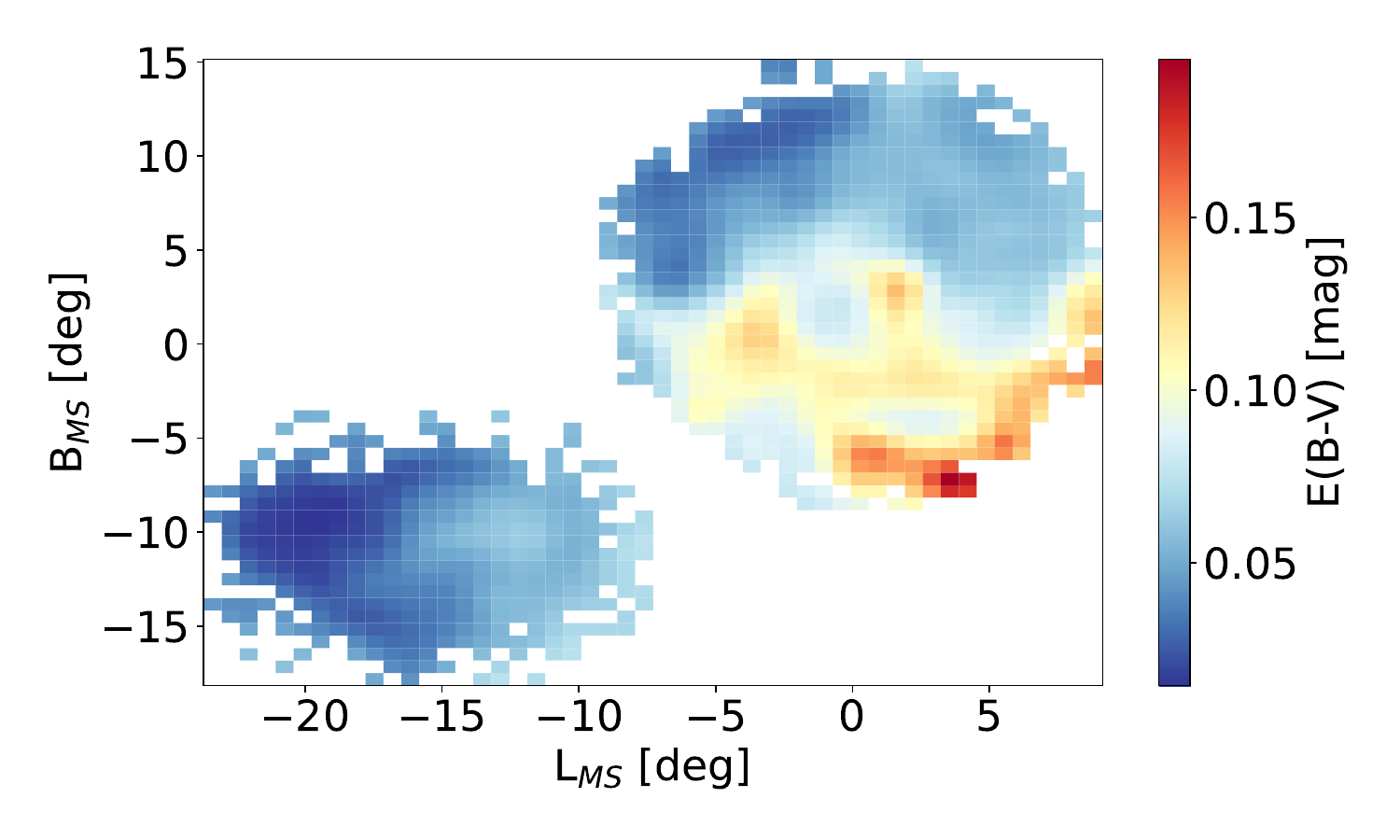}
    \includegraphics[width=0.49\linewidth]{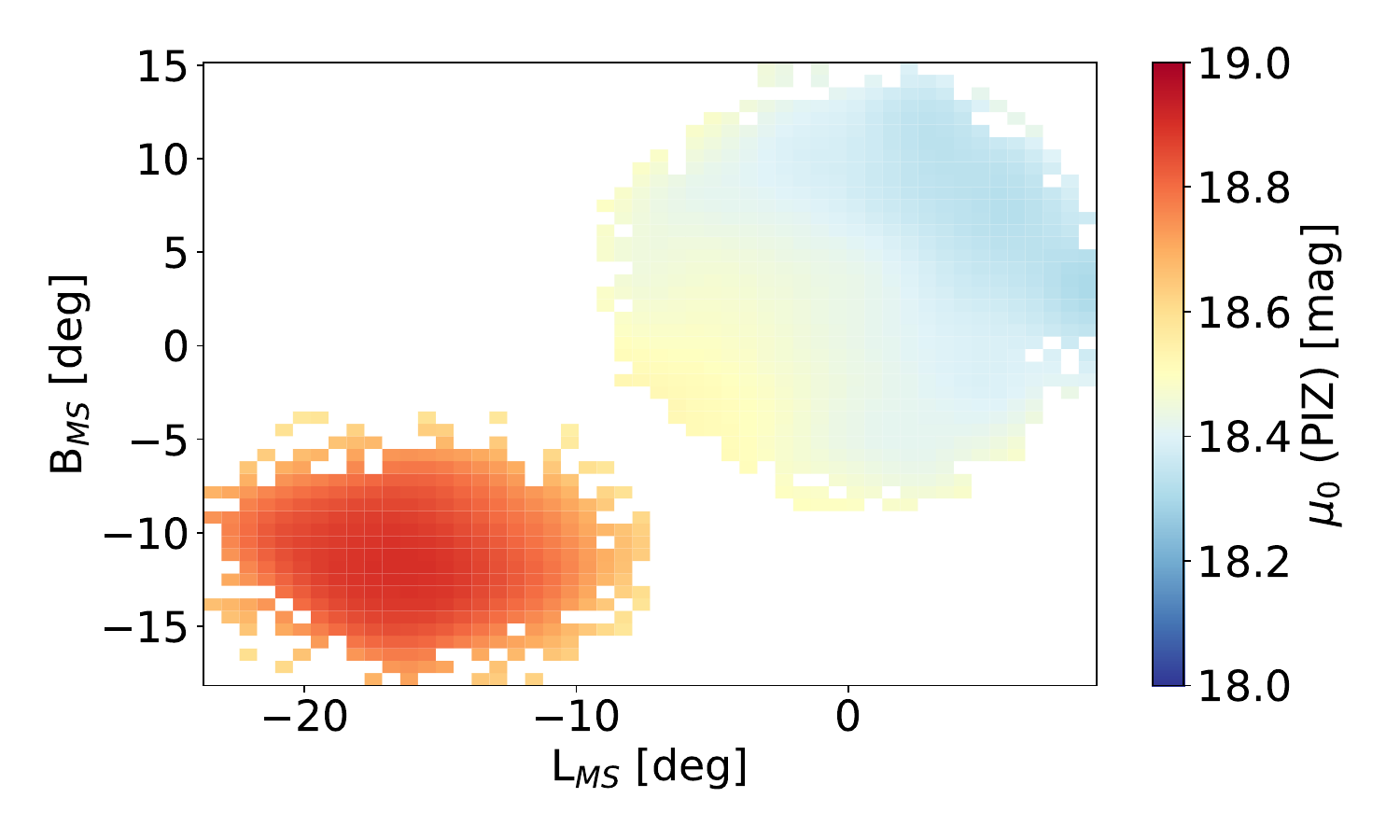}
    \includegraphics[width=0.49\linewidth]{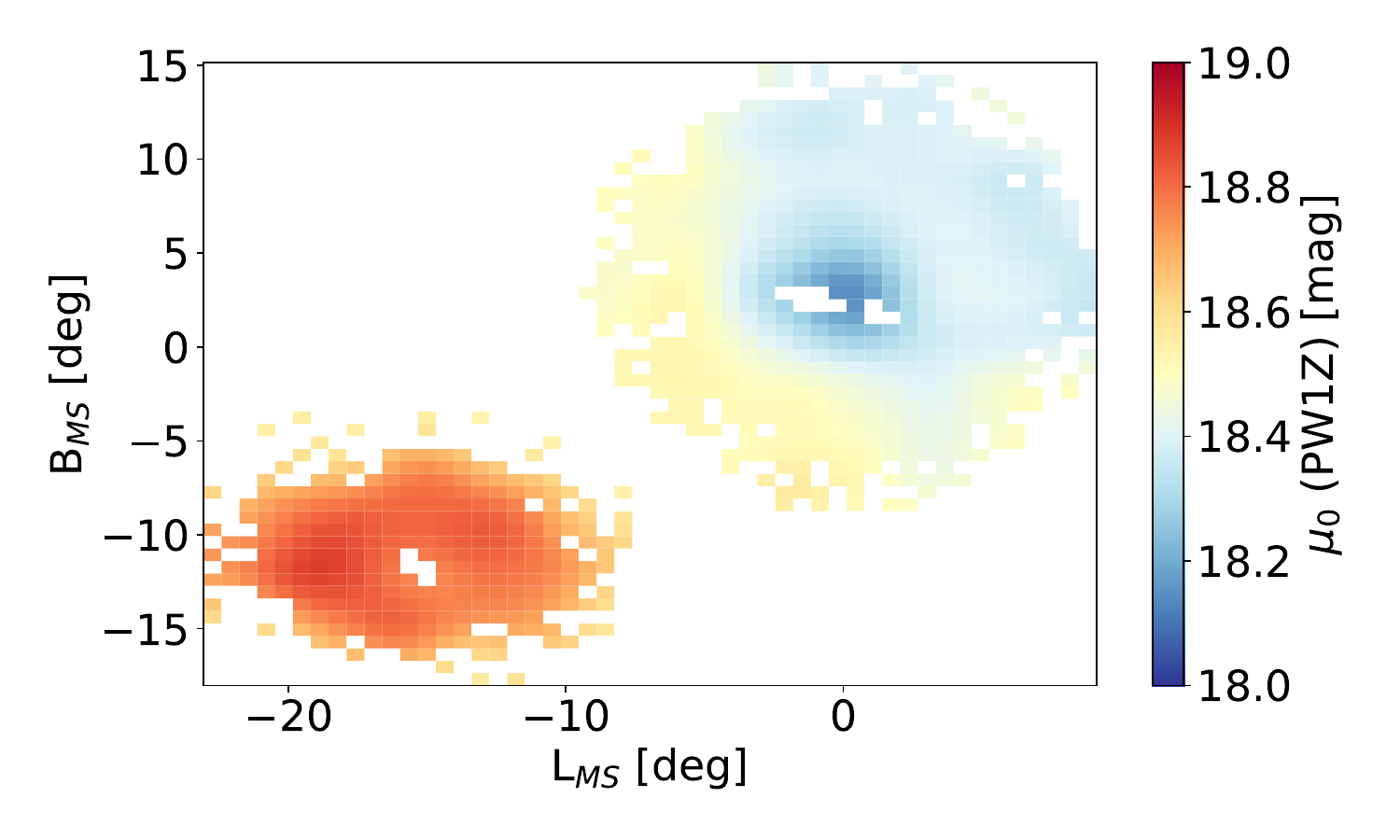}
    \includegraphics[width=0.49\linewidth]{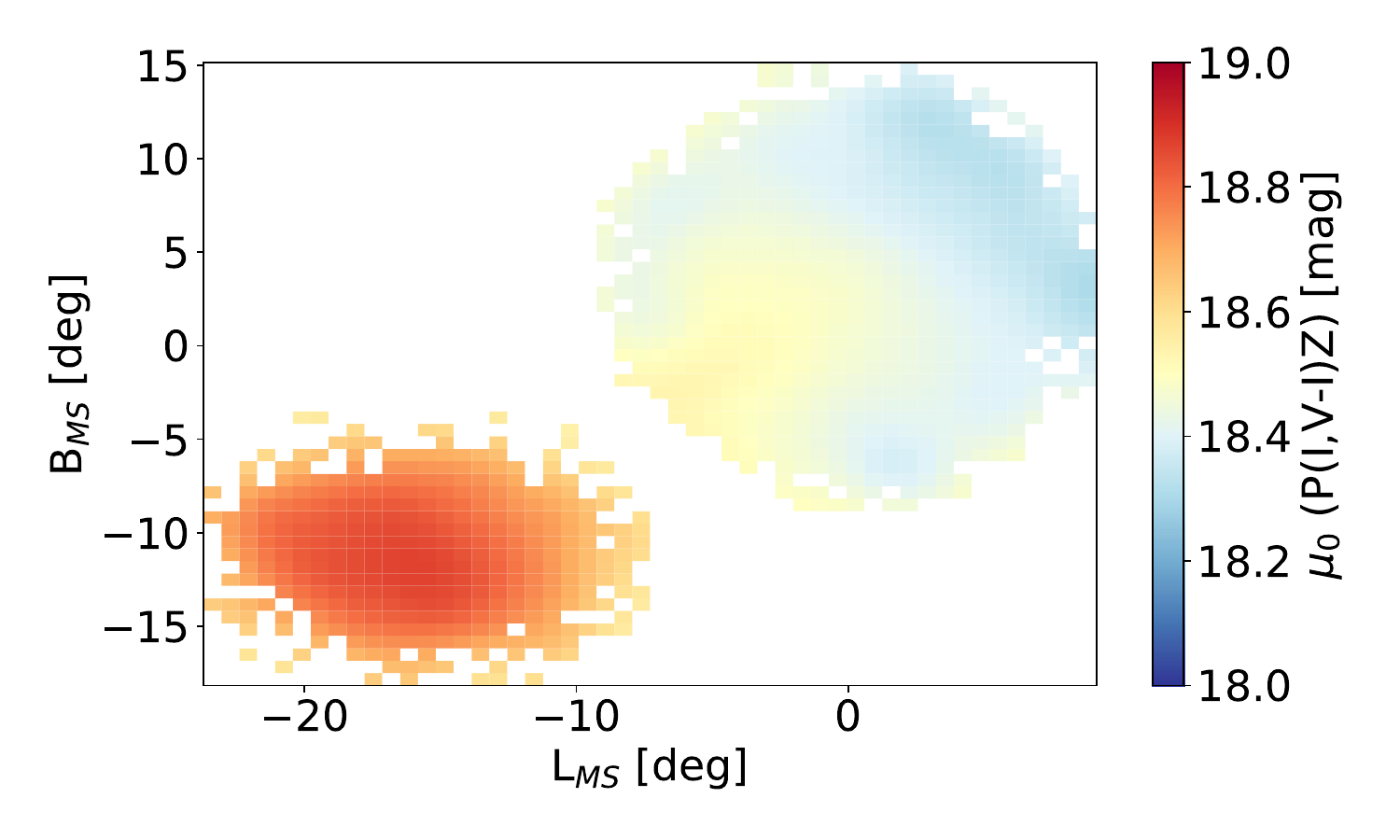}      
    \caption{Reddening (top-left) and distance moduli ($\mu_0$) maps (remaining plots) obtained using the  PLZ in $I$ and $W1$ band and the PWZ in $(I, V-I)$ in Table~\ref{tab:fits}. We use the Magellanic Stream reference frame to have a better perspective of the LMC and SMC. The binned maps are smoothed with a Gaussian kernel, with size 1 bin for the reddening map and 2 bins for the distance maps.}
    \label{fig:distance_maps}
\end{figure*}

\section{Distances to the LMC/SMC} \label{sec:distances}

Given that our PLZ and PWZ relations are anchored to Gaia DR3 parallaxes of Galactic RRL, we can use them to derive an independent distance of the LMC and SMC, to compare with reference literature values, such as the ones from eclipsing binaries \citep{Pietrzynski2019}. In fact, since our relations can be applied to each RRL in our sample, we can actually probe the tri-dimensional structure of these galaxies by measuring their individual distances. This is shown in Figure~\ref{fig:distance_maps}.

The maps are derived by binning the RRL projected on the plane of the sky, in the Magellanic Stream reference frame, using a regular $50\times50$ bins grid smoothed with a Gaussian kernel with a 2 bins width. The color-scale reports the median distance modulus in each bin.

The two panels on the right are constructed using only the Optical catalog. As such, they are the most complete of the set, since they do not require MIR CatWISE photometry, which is only available for a subset of the LMC and SMC stars. In the case of the LMC a distance gradient of $\sim 0.2$~mag is clearly detected, with the same orientation of the LMC disk as derived by \cite{vanderMarel2001} and \cite{Inno2016}. This is remarkable since RRL are expected to have a spheroidal distribution, not necessarily aligned with the disk. Yet, the disk alignment is preserved in the overall orientation of the old population spheroid, traced by the RRL. We also note how the central region of the galaxy, where the LMC bar is located, appears to be at a slightly larger distance than the surrounding area. This is likely a consequence of the enhanced extinction in this region (top-left panel in the figure), which is not fully corrected by the procedure described in section~\ref{sec:reddening}. In the case of the SMC, on the other hand, no clear distance gradient is visible. This is consistent with the orientation of this galaxy, characterized by a major axis almost parallel to the line of sight.

The distance map derived from the mid-infrared photometry is shown in the bottom-left panel of Figure~\ref{fig:distance_maps}. Note the empty bins located at the center of the two galaxies: this is due to an insufficient number of RRL detected in these areas. The culprit here is crowding, having an outsize effect on the CatWISE photometry, due to the large ($\sim 5$~arcsec) PSF of the WISE telescope. In the LMC, outside the overcrowded bins, the RRL appear however to trace the bar of this galaxy, as suggested by their median distance being closer than in the surrounding disk. This could be another consequence of the larger crowding along this structure, biasing the photometry towards brighter magnitudes. We do not see effects of extinction in the infrared maps, as expected at these wavelengths.

Since the two galaxies are spatially resolved in all three dimensions, a direct comparison between their distance measured with different indicators (e.g. RRL vs. eclipsing binaries) should take into account the relative spatial distribution of the distance indicators themselves. Figure~2 in \citet{Pietrzynski2019} shows that the eclipsing binary sample used to measure the LMC distance with 2.6\% accuracy belongs to a disk population, symmetrically distributed around the nominal center of the galaxy. The same cannot be said for RRLs, which are part of the LMC galactic spheroid. The distance moduli maps in Figure~\ref{fig:distance_maps}, however, show that the RRL still trace the inclination of the disk, and are also uniformly distributed around the center. The distance they provide, hence, should be directly comparable with the eclipsing binary distance, as long as the RRL distribution is not biased in the line-of-sight direction. While this is a fair assumption for the optical (OGLE) sample, the central region crowding breaks this assumption for the infrared sample, and any distance derived with PLZ/PWZ relations involving the $W1$ or $W2$ bands needs to be corrected.

We designate the average LMC distance obtained with the $W(I, V-I)$ PWZ relation as a reference value, since it is derived from the most complete catalog and is least subject to extinction, being based on Wesenheit magnitudes. We then re-derived the average distance in the same Wesenheit band, but restricting the sample to the stars that also have infrared photometry. We found that this second sample provides a smaller distance modulus (with $\Delta \mu_{LMC} \simeq 0.04$~mag). A similar procedure yields the same offset $\Delta \mu_{SMC} \simeq 0.04$~mag for the SMC.

\begin{figure}
    \centering
    \includegraphics[width=1.0\linewidth]{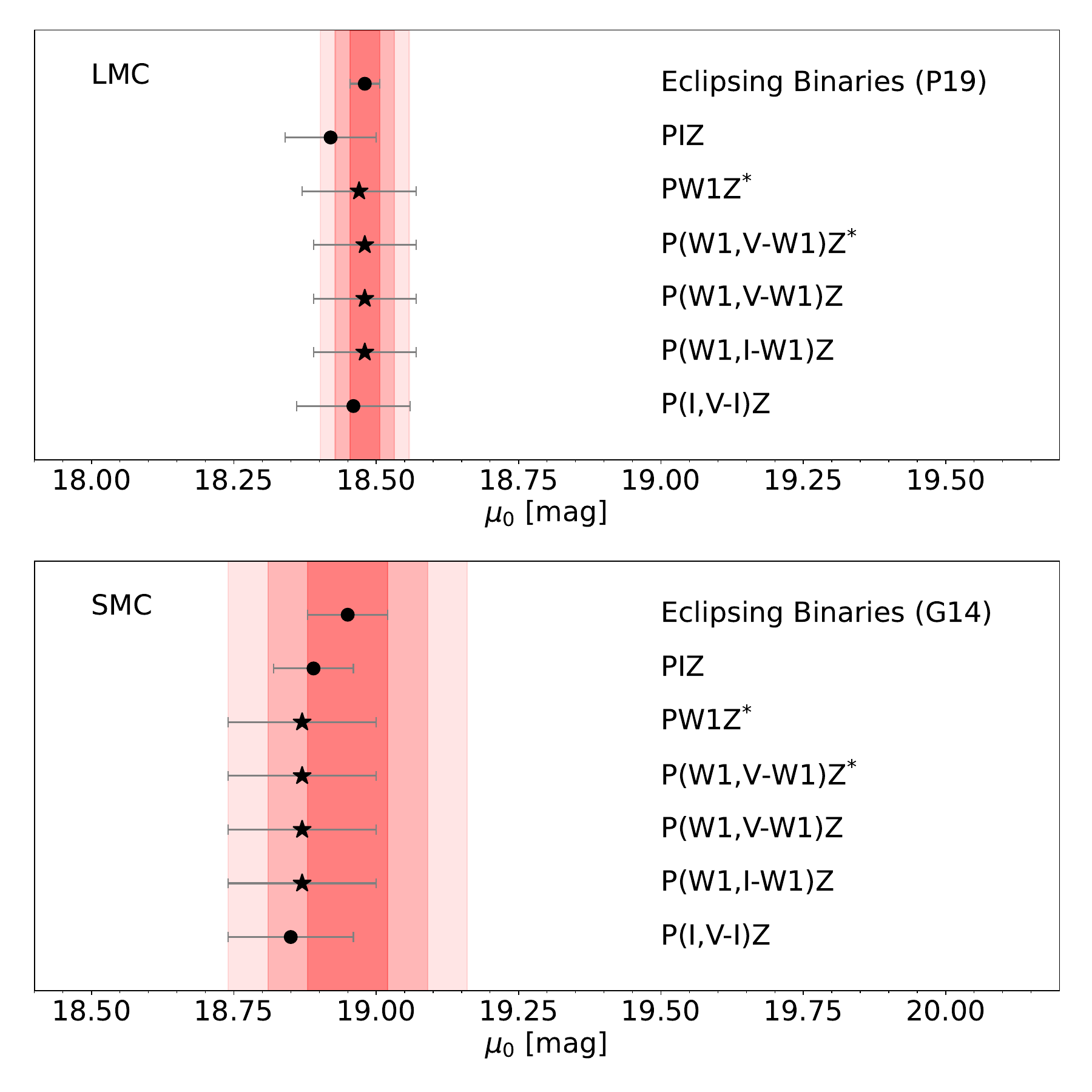}
    \caption{Comparison between the LMC and SMC distances obtained in this work and the most precise distance available to the date coming from the eclipsing binaries in the LMC \citep[P19; ][]{Pietrzynski2019} and SMC \citep[G14;][]{Graczyk2014}. Star symbols represent those values where an 0.04 mag was applied to fix the difference between the optical and the MIR sample to account for sample selection (see text). Red color stripes mark the locus of 1, 2, 3\,$\sigma$ (from darker to lighter) of the eclipsing binary distance.}
    \label{fig:dist_comp}
\end{figure}

Table~\ref{tab:mc_distances} and Figure~\ref{fig:dist_comp} present the distance moduli, and their statistical ($\sigma_{stat}$) and systematic ($\sigma_{sys}$) uncertainties, derived with the relations provided in Section~\ref{sec:plz}, for both LMC and SMC. We calculate $\sigma_{stat}$ by propagation of PLZ/PWZ errors considering the photometric uncertainties of the mean magnitudes in $V, I, W1$, the nominal uncertainties of 0.4 dex in individual photometry metallicities \citep{Mullen2021, Mullen2022} and an uncertainty of 10$^{-4}$ days for the period. In the case of the distances calculated from a PLZ, we also take into account the uncertainty resulting from the reddening value, which is considered to be 10\%. For $\sigma_{sys}$ we used the uncertainties coming from the propagation of errors of the PLZ/PWZ, but only considering the coefficients in the relationships (see Table~\ref{tab:fits}) as the source of error.

Figure~\ref{fig:dist_comp} shows how the value of the LMC and SMC distances obtained in this work compared to the most precise measurements derived from double eclipsing binaries: $\mu_{0, LMC} = 18.477 \pm 0.026 $~mag \citep{Pietrzynski2019} and $\mu_{0, SMC} = 18.97 \pm 0.07 $~mag \citep{Graczyk2014}. For the LMC, except for the distance from the PIZ, they are all within 1-$\sigma$ with respect to the distance obtained by \cite{Pietrzynski2019}. For the SMC, they are all within 2-$\sigma$ of agreement with the distance derived by \cite{Graczyk2014}.

Note finally how the scatter between the different measurements derived in this work is still much less than the error bars would suggest (only $\sigma_{stat}$ are displayed), if the uncertainty were purely statistical and the measurements were independent. But in fact, these are not independent measurements since we only have three independent bands, some of the relations assume the same metallicity slope, and there are likely still some systematic errors unaccounted for.

\begin{table}[]
\caption{Distance moduli for the LMC and SMC derived using our newly derived PLZ/PWZ relations.}
\hspace{-1.5cm}
\begin{tabular}{clccc}
\hline
System &  Method & $\mu_0$ & $\sigma_{stat}$ & $\sigma_{sys}$ \\
       &         &   (mag) &       (mag)     &      (mag)     \\
\hline
LMC    & P${I}$Z                   &  18.42  & 0.08 & 0.03 \\
LMC    & P${W1}$Z$^{*}$            &  18.47  & 0.10 & 0.04 \\
LMC    & P${W1, (V-W1)}$Z$^{*}$    &  18.48  & 0.09 & 0.06 \\
LMC    & P${W1, (V-W1)}$Z          &  18.48  & 0.09 & 0.01 \\
LMC    & P${W1, (I-W1)}$Z          &  18.48  & 0.09 & 0.01 \\
LMC    & P${I, (V-I)}$Z            &  18.46  & 0.10 & 0.03 \\
LMC    & Average$^{*}$   & \bf{18.47}  & \bf{0.07} & \bf{0.05} \\
\hline
SMC    & P${I}$Z                   &  18.89  & 0.07 & 0.02 \\
SMC    & P${W1}$Z$^{*}$            &  18.87  & 0.13 & 0.05 \\
SMC    & P${W1, (V-W1)}$Z$^{*}$    &  18.87  & 0.13 & 0.07 \\
SMC    & P${W1, (V-W1)}$Z          &  18.87  & 0.13 & 0.01 \\
SMC    & P${W1, (I-W1)}$Z          &  18.87  & 0.13 & 0.01 \\
SMC    & P${I, (V-I)}$Z            &  18.85  & 0.11 & 0.03 \\
SMC    & Average$^{*}$  & \bf{18.87} & \bf{0.09} & \bf{0.06} \\
\hline
\end{tabular}
\label{tab:mc_distances}
\begin{tablenotes}
    \item $^{*}$Relations derived with the Iterative method described in \S~\ref{sec:plz_iterative}. The average distance moduli are calculated from the P$W1$Z and P${W1, (V-W1)}Z$ derived with the Iterative method.
    \item An offset of 0.04~mag (see text for details) is applied to the PLZ/PWZ engaging the $W1$ band.
\end{tablenotes}
\end{table}

\section{Chemical enrichment in the Magellanic Clouds} \label{sec:enrichment}

\begin{figure}
    \centering
    \includegraphics[width=1.\linewidth]{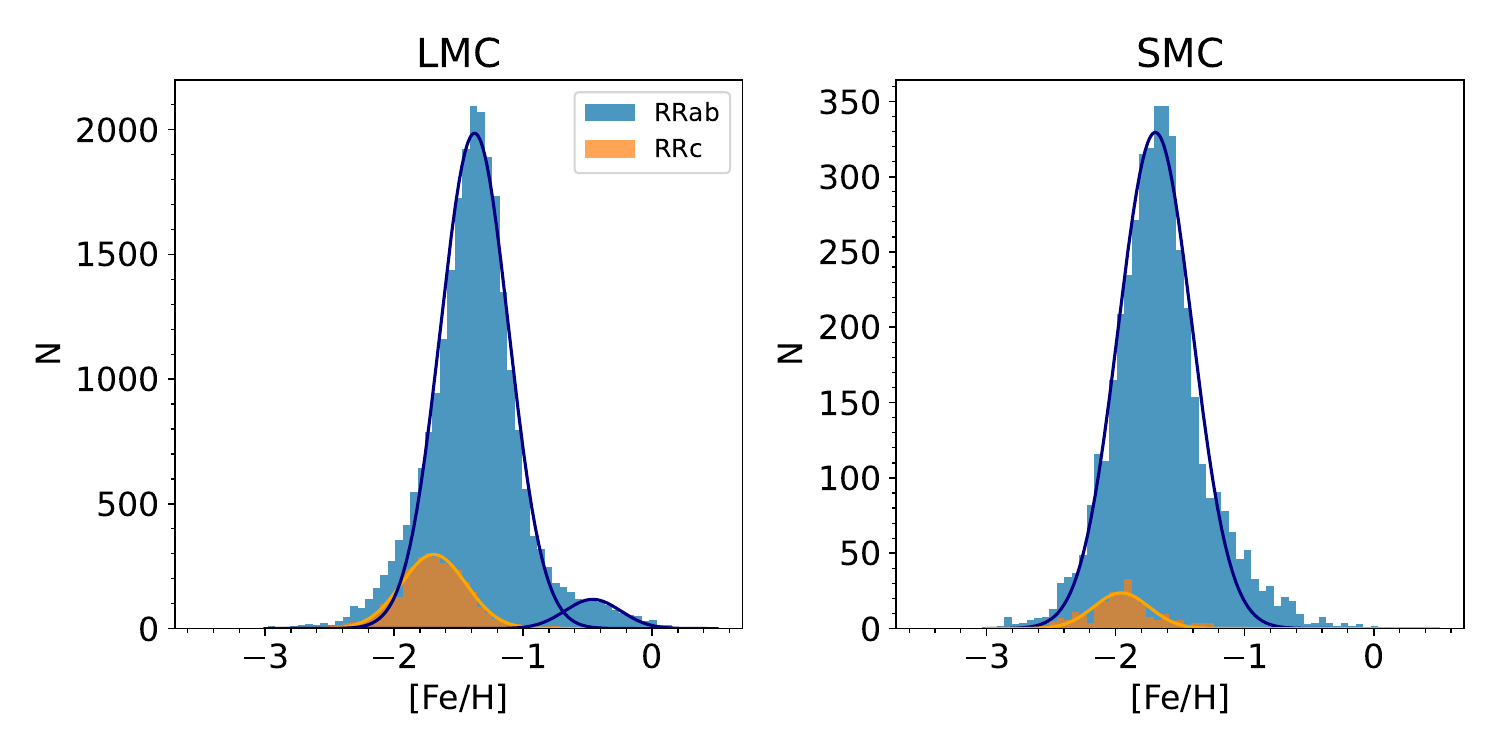}
    \caption{Metallicity distribution for the RRL in the LMC (left) and the SMC (right). The distribution of RRab (blue) and RRc (orange) are shown and fitted independently. Single Gaussian fitting was performed to the RRc. To interpret the RRab distribution we fit two Gaussians to the LMC and one Gaussian to the SMC. Individual Gaussian fits are display by solid blue lines in the case of the RRab star distributions, whereas the RRc distributions are represented by solid orange lines.}
    \label{fig:mdf}
\end{figure}

\begin{table}[]
\setlength{\tabcolsep}{3pt} 
\centering
\caption{Gaussian fit parameters of the metallicity and luminosity distribution in the LMC and SMC}
\begin{tabular}{ccccccc}
    System  & Set  & $\langle{\rm{[Fe/H]}}\rangle$ & $\sigma_{\rm[Fe/H]}$  & N$_{{\langle{\rm{[Fe/H]}}\rangle}\pm\sigma}$ & $\langle{V_0}\rangle$ & $\sigma_{V_0}$ \\
            &      & (dex) & (dex) &   & (mag) \\

    \hline
            & RRab & -1.38 & 0.26 & 15369 & 19.10 & 0.13 \\
    LMC     &      & -0.46 & 0.22 &  841  & 19.32 & 0.16 \\
            \cline{3-7}
            & RRc  & -1.69 & 0.25 & 2222 & 19.05 & 0.14 \\          
    \hline
    SMC    & RRab & -1.69 & 0.29 & 2833 & 19.53 & 0.15 \\
            \cline{3-7}
            & RRc  & -1.96 & 0.22 & 152 & 19.48 & 0.15 \\
  
    \hline
    \end{tabular}
    \label{tab:mdf_gaussian_fit}
\end{table}

Figure~\ref{fig:mdf} shows the metallicity distribution of the LMC (left) and SMC (right) RRL stars, as derived in section~\ref{sec:metallicity}. The RRc and RRab are plotted separately.

In both the LMC and SMC, the RRc population fits well to a single Gaussian peaked at $\mathrm{[Fe/H]} = -1.69$~dex ($\sigma$ = 0.25~dex) and $\mathrm{[Fe/H]} = -1.96$~dex ($\sigma$ = 0.22~dex), respectively. The RRab population in the LMC is however best fit as the sum of two distinct Gaussian distributions, the first peaking at $\mathrm{[Fe/H]} = -1.38$~dex, and the second at $-0.46$~dex. The two distributions have a similar spread with $\sigma = 0.26$~dex and 0.22~dex, respectively. The RRab population in the SMC is instead best represented by a single Gaussian peaking at $\mathrm{[Fe/H]} = -1.69$~dex (with dispersion $\sigma = 0.29$~dex). While the SMC RRab stars do not show a separate, peak as in the LMC, we notice an extended high-metallicity wing. This wing cannot be fit by a secondary Gaussian (doing so would return a broad distribution with $\sigma = 0.48$~dex, and with a peak coincident with the low-metallicity Gaussian), nor by a Lorentzian (that would instead over-fit the metal-poor regime). Therefore, even if we do not find a perfect fit for the metal rich end, we prefer to keep a single Gaussian fit for the SMC RRab, noting the presence of a $\sim 6$\% excess for $\mathrm{[Fe/H]} > -1.2$~dex above the best fit Gaussian. The mean and sigma values of the final fits to the metallicity distributions of the RRab and RRc stars in the LMC and SMC are listed in Table ~\ref{tab:mdf_gaussian_fit}.

\begin{figure}
    \centering
    \includegraphics[width=1.\linewidth]{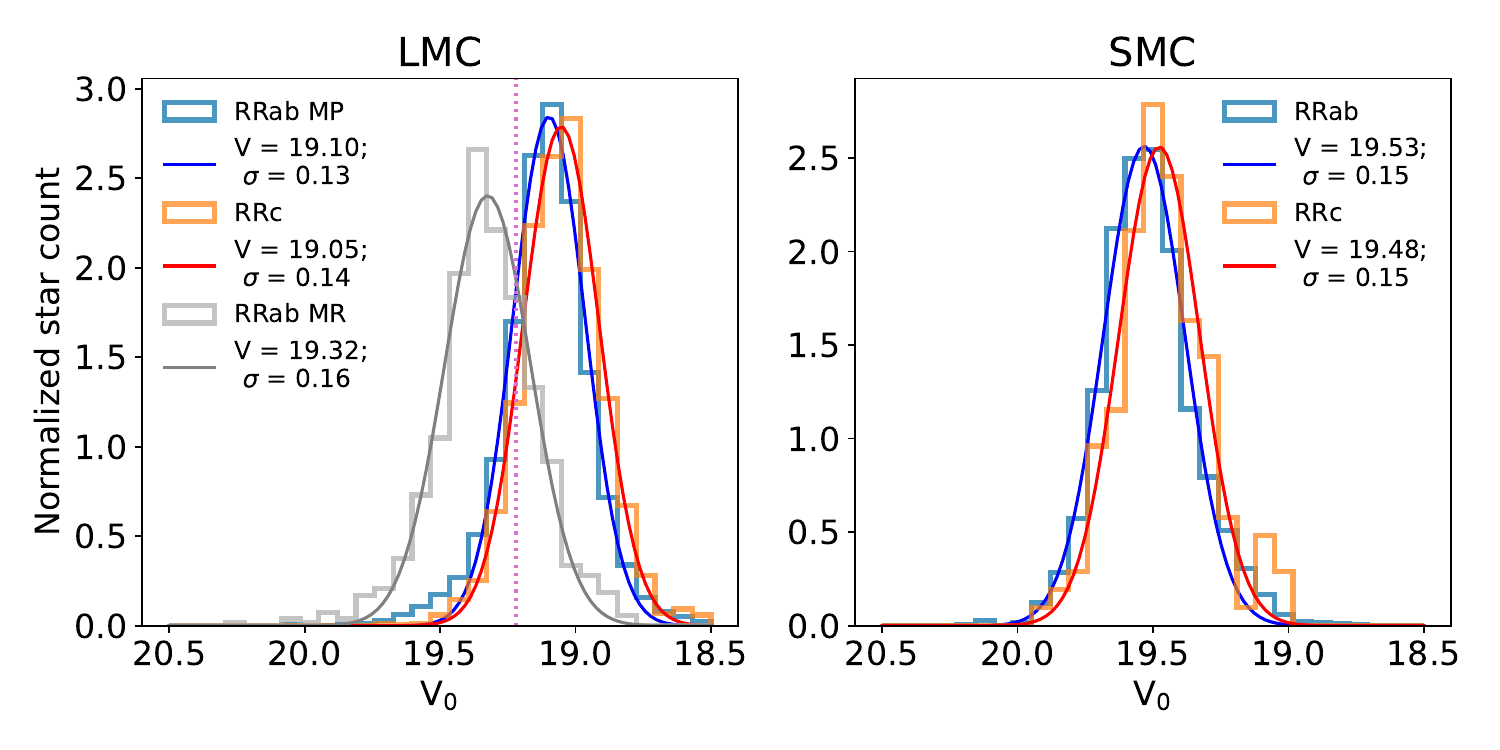}
    \caption{Reddening-corrected luminosity function in the $V$ band for the RRL in the LMC (left) and the SMC (right). Each distribution represents the stars embedded within 1$\sigma$ of the peak of their respective metallicity distribution (see Table~\ref{tab:fits}). In the labels we use MP (metal-poor) and MR (metal-richer) to differentiate between the two populations of RRab stars in the LMC. There is a clear split of these two populations also in their $V$ luminosities. The intersection between the two RRab distributions in the LMC is at $V_0 = 19.22$~mag (dotted purple line).}
    \label{fig:LFV}
\end{figure}

We note that in both galaxies the RRab and RRc populations seem to have a different peak metallicity. In the case of the LMC the RRc stars are, on average, more metal-poor than the RRab by $\Delta \mathrm{[Fe/H]} \simeq 0.31$~dex. A similar offset $\Delta \mathrm{[Fe/H]} \simeq 0.28$~dex is found for the SMC. The intrinsic dispersions of the two distributions are comparable; therefore, although the peaks are offset, the two metallicity distributions are not cleanly separated and show substantial overlap. To quantify the significance of the offset between the two peak metallicities, we estimate the uncertainty on the difference as

\begin{equation}
\sigma_{\Delta} = \sqrt{\frac{\sigma_{\rm[Fe/H], RRab}^{2}}{N_{\rm RRab}} + \frac{\sigma_{\rm[Fe/H], RRc}^{2}}{N_{\rm RRc}}}
\label{eqn:sigma_delta}
\end{equation}

\noindent
where $N_{\rm RRab}$ and $N_{\rm RRc}$ are the number of stars in each sample. The corresponding significance is then $\Delta \langle\rm{[Fe/H]\rangle} / \sigma_{\Delta} \simeq 67$ for the LMC, and 18 for the SMC. Thus, while the RRab and RRc metallicity distributions overlap substantially, the difference between their peak metallicities is statistically significant, with the RRc population shifted toward lower metallicities relative to the RRab population.

A similar difference in the metallicity distribution between RRab and RRc stars has already been found in Galactic RRL \citep{Fabrizio2021, Crestani2021}, and fully justified by evolutionary properties. Indeed, RRc stars are mainly found on the blue side of the instability strip, which is largely populated by lower metallicity stars (based on the well-known correlation between the blue extension of the HB and lower metallicity). As a consequence, we can expect RRc stars to draw from a population with lower metallicity than the their RRab counterparts.

However, the interpretation of this offset should be treated with caution as the RRc sample may be also affected by a metallicity-dependent selection bias. From the expected RRL population, RRc stars should represent approximately $20-30\%$ of the total sample (see Table 6 in \citealt{MartinezVazquez2017}), but only about half of this expected fraction is retrieved in the \textit{Optical} catalog. If the missing RRc stars are preferentially among the fainter objects, the observed RRc metallicity distribution could be biased. In particular, according to PLZ relation, more metal-rich RRL stars are expected to be fainter at fixed period. Therefore, an incompleteness affecting the faint end of the RRc sample would preferentially remove more metal-rich RRc stars from the observed distribution. This would naturally skew the recovered RRc metallicity distribution toward slightly lower metallicities. Consequently, the lower metallicity peak of the RRc sample may in part reflect an observational bias. 

The fact that there are two distinct populations of RRab stars in the LMC and a single population in the SMC is supported by their respective $V$-band luminosity function\footnote{We use the $V$ magnitude to facilitate the analysis, given that the horizontal-branch is completely flat in this band.}. Figure~\ref{fig:LFV} shows the reddening-corrected distribution of the $V$ band magnitudes for the RRL in the LMC (left) and the SMC (right), where each distribution is a subsample of the RRab and RRc stars within 1$\sigma$ of the peak of their respective metallicity distribution (see Table~\ref{tab:fits}). In the LMC there is a clear distinction of two differentiated peaks in the $V$ distribution for the metal-poorer (MP) and metal-richer (MR) RRab subpopulations, where the MP RRab and RRc subsample follows almost the same distribution. For the SMC, however, there is no indication of two different populations, and the RRab and RRc sub-samples follow a very similar distribution. This points to a complex star formation early in the history of the LMC, but not in the SMC. This is consistent with leading star-formation history studies of the LMC and the SMC (see \citealt{Massana2022} and references therein). These studies suggest that the SMC had a steady formation in its first 3 Gyrs (when all the RRL progenitors were formed). For the LMC, however, they support an early peak of active formation followed by intermittent star formation between 10 and 13.7 Gry ago. This scenario is consistent with the presence of distinct populations of RRab in the LMC but not in the SMC (note that a similar signal in the RRc is likely lost due to the much smaller number of the overtone pulsators).

\begin{figure}
    \centering
    \includegraphics[width=1.\linewidth]{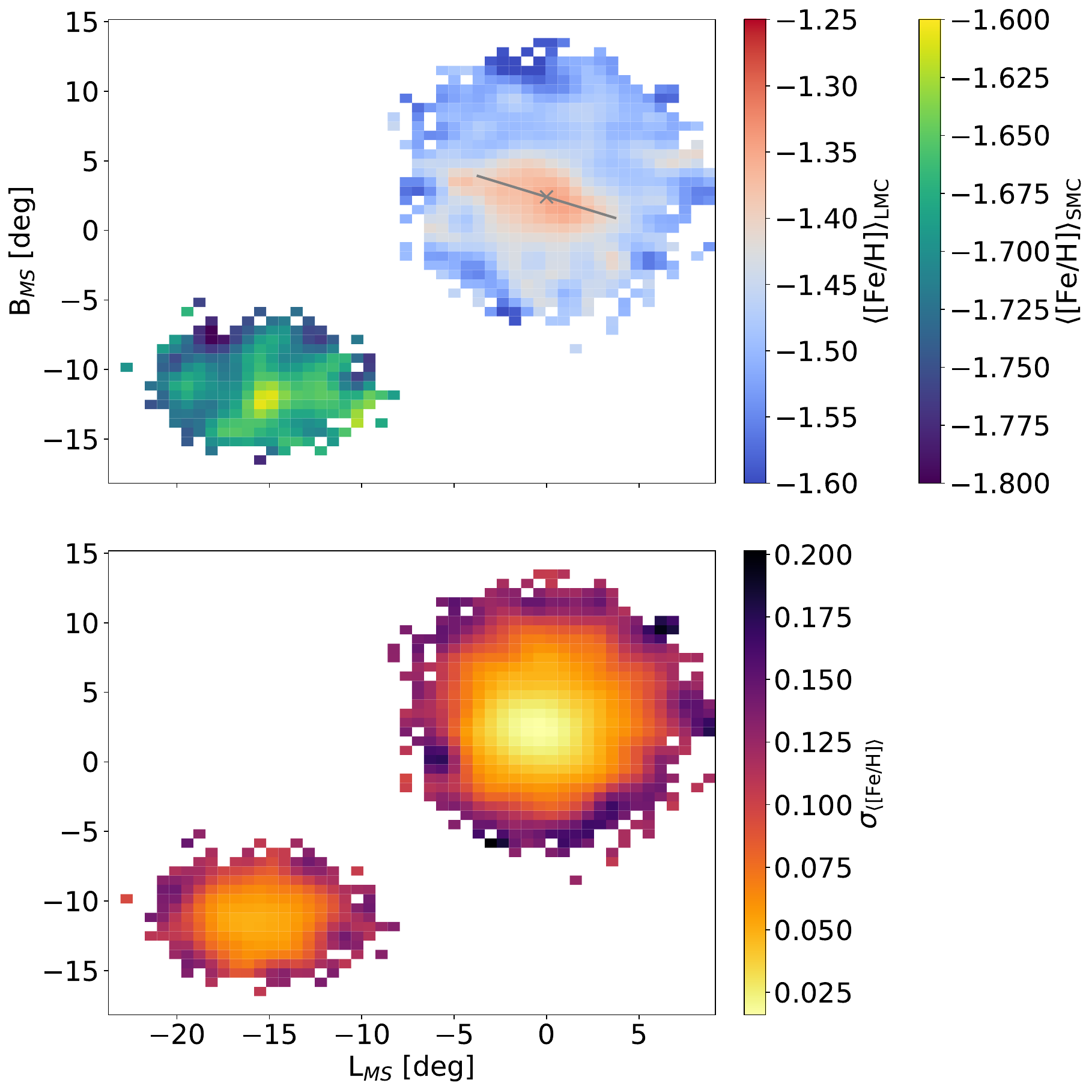}
    \caption{Metallicity map (top) and standard error of the mean metallicity (bottom) for the RRL in the SMC and LMC, in Magellanic Stream coordinates. The top panel shows the 5$\sigma$-clipped mean $\langle{\rm [Fe/H]}\rangle$, for the LMC and SMC samples computed on a common spatial grid and smoothed with a Gaussian kernel of $\sigma=1$ pixel. Only bins with more than 2 measurements (after clipping) are retained, and the LMC and SMC maps are displayed with independent color scales. The bottom panel shows the corresponding standard error on the clipped mean metallicity, also smoothed with a Gaussian kernel of $\sigma=1$ pixel and displayed with a common color scale. Cross and solid line indicate the adopted center and directions of the LMC bar from \cite{vanderMarel2014}.}
    \label{fig:feh_map}
\end{figure}

\citet{Gratton2004} derived the metallicity of 26 RRc and 62 RRab stars in the bar of the LMC, using their version of the $\Delta S$ method. They found $\textrm{[Fe/H]}_{RRab} = -1.51 \pm 0.03$~dex and $\textrm{[Fe/H]}_{RRc} = -1.42 \pm 0.07$~dex, in a scale they describe as 0.06~dex more metal-rich than \citet{Zinn1984}. Adopting the quadratic formula in \citet{Carretta2009}, and then equation~\ref{eq:C09toC21}, these values correspond to $\textrm{[Fe/H]}_{RRab} \simeq -1.41$~dex and $\textrm{[Fe/H]}_{RRc} \simeq -1.24$~dex in the metallicity scale used in this work. The value for the RRab is in remarkable agreement with the result reported in Table~\ref{tab:mdf_gaussian_fit} for the metal-poor component of the RRab stars. The metallicity of the RRc stars reported by \citet{Gratton2004} is however $\sim 1.4 \sigma$ more metal-rich than the average value we found. This difference could be explained by the relatively small number or RRc stars included in the \citet{Gratton2004} analysis, sampling a small region in the bar of the LMC (our analysis instead provides the average value of the entire population of RRc stars in the galaxy).

\subsection{Metallicity maps and gradients} \label{subsec:gradients}

Figure~\ref{fig:feh_map} (top panel) shows the projected metallicity distribution, in Magellanic Stream coordinates, of the RRL in the MCs. We constructed the map by binning the RRL of the two galaxies, and deriving their $5 \sigma$-clipped mean [Fe/H] abundance in each bin. Bins with less than 3 stars have been masked-out, and the surviving bins have been smoothed with a Gaussian kernel of $\sigma = 1$~bin. The map in the bottom panel presents the standard error $\sigma_{\langle\rm[Fe/H]\rangle}$ of the clipped mean metallicity in each bin.

The LMC shows a concentration of RRL with higher metallicity in the central region, while the outskirts of the galaxy appear to be more metal poor. This pattern is commonly found in other dwarf galaxies (e.g., Sculptor or Eridanus~II, \citealt{MartinezVazquez2016a, MartinezVazquez2021b}). Interestingly, the high metallicity region in the LMC central area follows the same orientation of the LMC bar (as determined by \citealt{vanderMarel2014}). This feature appears to be statistically significant: the mean metallicity in the bar region is $\sim -1.35$~dex while in the surrounding region is $\sim -1.50$~dex (a difference $\Delta \mathrm{[Fe/H]} \simeq 0.15$~dex), with an uncertainty $\sigma_{\langle\rm[Fe/H]\rangle} \lesssim 0.05$~dex in the same general area.

Since RRL stars trace a population older than approximately 10 Gyr, the metal-enhanced structure aligned with the LMC bar should not be interpreted as evidence that these stars formed during the assembly of the present-day bar. Instead, if the alignment is dynamical, it suggests that a chemically structured ancient population participates in the bar. This result is consistent with the scenario proposed by \citet{Monteagudo2018}, who found similar temporal patterns in the star-formation histories of the bar and inner disk and argued that the bar formed through a redistribution of pre-existing disk material rather than through a distinct episode of star formation. In this picture, the metallicity enhancement could arise from the preferential trapping of relatively metal-rich stars from the inner disk, transforming part of a pre-existing radial metallicity gradient into a non-axisymmetric chemical structure.

The SMC map suggests a similar pattern, with a central region having metallicity higher than the surrounding area by $\simeq 0.07$~dex. This feature, however, is not statistically significant, since the local uncertainty $\sigma_{\langle\rm[Fe/H]\rangle} \sim 0.07$~dex is comparable to the signal.

\begin{figure}
    \centering
    \includegraphics[width=1.\linewidth]{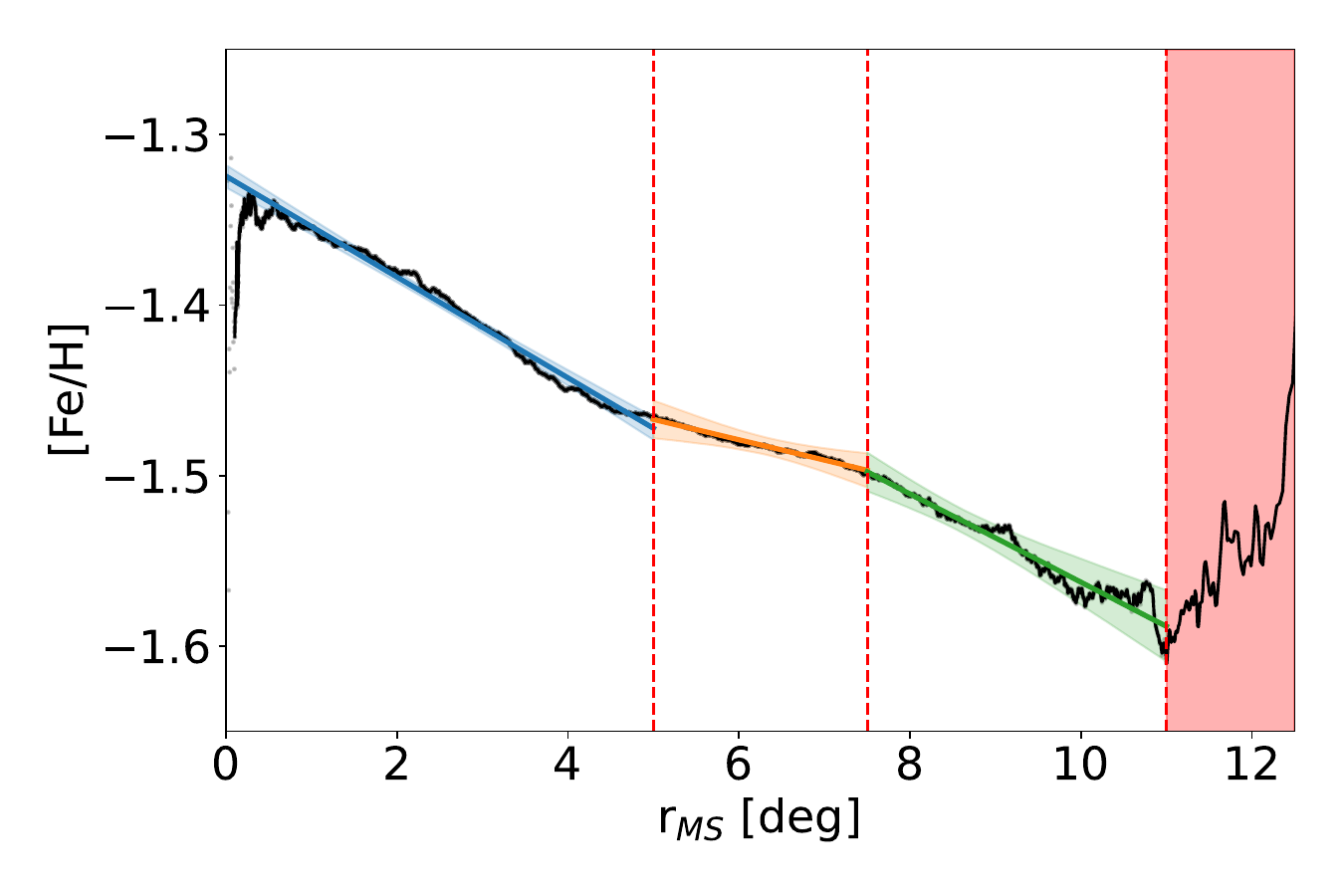}
    \caption{Metallicity gradient for the LMC as a function of the radial distance obtained in the Magellanic Stream reference frame ($r_{MS}$). Red dotted lines identify three intervals at $r_{MS}<5$, $5<r_{MS}<7.5$, and $7.5<r_{MS}<11$. Blue, orange, and green lines are the linear fit to each of the regions (see text for details). The red shaded area was masked since it is not significant and shows a lot of fluctuation.}
    \label{fig:feh_gradient}
\end{figure}

Figure~\ref{fig:feh_gradient} shows the [Fe/H] radial gradient in the LMC, with the radius derived in Magellanic Streams coordinates. The black solid line was derived by computing a running average using variable-size windows that increase toward the center of the sample and decrease toward the edges. For each window, we calculate the mean radial distance and the mean value of the metallicity. This produces a radial profile of the measured quantity as a function of distance. The resulting mean metallicity profile is then smoothed using a Savitzky--Golay filter with a window length of 5 and a third-order polynomial.

We can identify different trends in three separate radial intervals, defined as $r_{MS}<5$, $5<r_{MS}<7.5$, and $7.5<r_{MS}<11$. In order to characterize the trend across each region, we fit linear relations in the form $\rm{[Fe/H]} = a \cdot (r_{MS} - r_{MS, pivot}) + b $, where $r_{MS,pivot}$ was set to the median metallicity value of the data points contained within that region. This choice reduces the covariance between the slope, $a$, and the fitted normalization, $b$, and provides a direct estimate of the value of the relation at the central location of each subsample. The uncertainties on the fitted parameters were estimated through a Monte Carlo approach. In each realization, the observed [Fe/H] values were perturbed assuming Gaussian uncertainties with dispersion $\sigma=0.4$~dex \citep{Mullen2021, Mullen2022}, and the linear fit was repeated independently for each region. The final best-fit relation in each interval was taken from the median of the resulting Monte Carlo distributions, while the 16th and 84th percentiles were used to define the corresponding $1\sigma$ confidence interval. In Figure~\ref{fig:feh_gradient}, the solid lines in color show the median-fitted relation for each region, and the shaded bands represent the Monte Carlo uncertainty envelope. The parameters of these relations are listed in Table~\ref{tab:feh_gradients}.

\begin{table}
\centering
\caption{Linear-fit parameters to the metallicity gradients derived from the RRL stars in the LMC for the three fitted regions. The fit is written as $\rm{[Fe/H]} = a \cdot (r_{MS} - r_{MS, pivot}) + b$. Uncertainties are the standard deviations of the Monte Carlo distributions.}
\label{tab:feh_gradients}
\begin{tabular}{lccc}
\hline
LMC Region  & $r_{MS, pivot}$ & $a$ & $b$ \\
  (deg)     &  (deg)          & (deg dex$^{-1}$) & (dex) \\
\hline
$r_{MS} \leq 5$         & 2.315 & $-0.030 \pm 0.002$ & $-1.393 \pm 0.003$ \\
$5 < r_{MS} \leq 7.5$   & 6.460 & $-0.012 \pm 0.007$ & $-1.485 \pm 0.005$ \\
$7.5 < r_{MS} \leq 11$  & 8.412 & $-0.026 \pm 0.008$ & $-1.521 \pm 0.007$ \\
\hline
\end{tabular}
\end{table}

\subsection{Clusters}

As a final application of the metallicities we derived in section~\ref{sec:metallicity}, we performed a search for RRL stars around the old globular clusters ($t > 10$~Gyr) of the LMC and SMC\footnote{We note that Reticulum and NGC~1841 were not included here because they fall outside the Optical catalog footprint.} This allows us to compare the metallicities we obtained photometrically for the individual cluster RRL, with those obtained spectroscopically for the clusters. The age, spectroscopic metallicities, distance moduli and morphological parameters of the clusters used in this analysis were mainly obtained from \cite{Pace2025} database -- except for the case of NGC 121 where the [Fe/H] value ($-1.33 \pm 0.12$~dex) was obtained from \cite{Suntzeff1999}, and NGC~1786, NGC~1928 and NGC~1939 where the apparent angular sizes (NGC 1786: 2.0~arcmin , NGC~1928: 1.2~arcmin, NGC 1939: 1.4~arcmin) are from \citet{Bonatto2010}.

\begin{figure*}
    \centering
    \includegraphics[width=1.\linewidth]{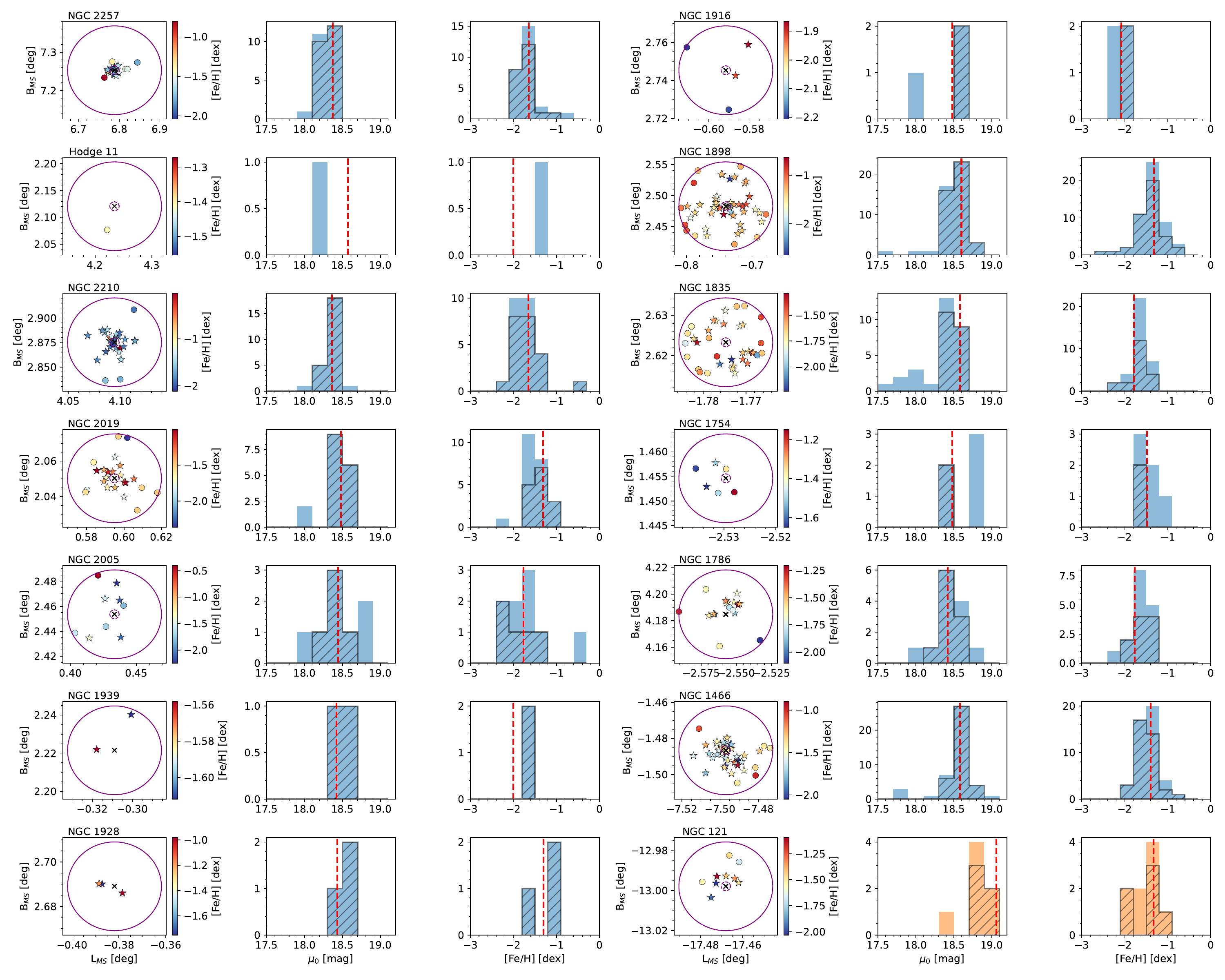}
    \caption{Comparison of photometric metallicities versus spectroscopic metallicities in the neighborhood of LMC and SMC old ($>10$~Gyr) globular clusters. The plot is a $6\times7$ grid, where each row represents two clusters (i.e., the first 3 columns belong to a cluster while the last three columns belong to another). For each cluster, we present the spatial distribution of the RRL around the cluster together with its distance and the metallicity distribution. The purple circles represent the half-light radii (dashed) and 10$\times$half-light radii (solid lines). Red dashed lines in the histograms are the mean values of the distance and spectroscopic metallicities reported for those clusters in the literature. The hashed histograms were obtained by filtering the spatially selected RRL stars by the mean distance modulus of the cluster $\pm$0.25 mag. Star symbols in the leftmost panels of each cluster show where these stars are located.}
    \label{fig:clusters}
\end{figure*}

Figure~\ref{fig:clusters} shows a grid of plots where each row represents two globular clusters (3 panels each). For each cluster, the first panel represents the spatial distribution of the RRL in the Optical catalog around the cluster while the second and third panels are the distance moduli (derived using the PW(I, V-I)Z relation) and photometric metallicity distributions of the RRL within 10$\times$ the half-light radius of each cluster. We additionally did a selection of RRL based on their individual distance moduli, excluding stars outside an interval $\Delta_{\mu_0} = \pm 0.25$~mag. The retained stars are represented by star symbols in the spatial plot and by hatched histograms in the $\mu_0$ and [Fe/H] distribution plots. The dashed vertical lines mark the locus of the mean spectroscopic metallicities and the distance moduli reported for those clusters. In general, we see that the mean spectroscopic metallicities fall close to the middle of each RRL photometric metallicity distribution, and the distance moduli distribution is also consistent with the distance of the clusters from the literature. This is an important further validation of the reliability of both the Fourier-metallicity relations and the PLZ/PWZ derived in this work. 

\section{Conclusions} \label{sec:Conclusions}

In this paper we provide new empirical PLZs in the $I$ and the $W1$ bands and PWZs in the $W(W1, V-W1)$, $W(W1, I-W1)$ and $W(I, V-I)$ bands. These relations are anchored to Gaia DR3 parallaxes and spectroscopic metallicities of Galactic RRL, combined with OGLE and WISE data of LMC and SMC variables. We introduce a new methodology designed to minimize the correlation between the period slope and the metallicity slope. This new approach allowed us to test for systematic biases in RRL distances anchored on Gaia parallaxes, which we previously found to be systematically smaller ($\sim$2–3$\sigma$) than other literature values \citep[see][]{Mullen2021}. Testing our new relations on nearby systems (Reticulum, M4 and Sculptor) indicate that Gaia DR3-based RRL distances still tend to be smaller than other methods, but the bias, if present, is reduced to less than 1 - 2$\sigma$.

For a more robust comparison we then apply our new relations to a large sample of  RRL in the LMC and SMC, to determine their individual distances, as well as the overall distance moduli of the galaxies. We find that our newly derived distances are in excellent agreement with the most precise and accurate distances obtained in the MCs, which are provided by eclipsing binaries \citep{Pietrzynski2019, Graczyk2014}. The average true distance moduli we found for the LMC and SMC are 18.47$\pm$0.07(stat)$\pm$0.05(syst) and 18.87$\pm$0.09(stat)$\pm$0.06(syst), respectively, in both cases well within $1\sigma$ from the eclipsing binary determinations, when taking into account the respective uncertainties. In the case of the LMC, which is an anchor for the Cosmological Distance Scale \citep{Riess2019}, the agreement is in fact better than 0.01~mag (which is less than 1\%). This result suggests that the Gaia DR3-based RRL distances are in fact accurate when compared to \emph{robust} independent determinations.

We also used the individual RRL distances determined with our PWZ relations to trace the tri-dimensional distribution of the old-stellar population in the two galaxies. We found a gradient in the LMC distance modulus of $\sim 0.2$~mag whose orientation is broadly consistent with that expected from the inclined LMC disk. This is noteworthy because RRL stars trace an old, dynamically hot population that is usually associated with an extended halo or spheroidal/ellipsoidal component, rather than with the young, thin disk. However, previous studies have shown that the inner LMC RR Lyrae population can display disk-like signatures, including a bar-like projected density distribution and viewing angles comparable to those of the disk \citep[e.g.,][]{Subramaniam2006,Subramaniam2009,Haschke2012}. Our result therefore suggests that, although the RRL population is spatially more extended than young disk tracers, its large-scale distance structure still retains an imprint of the global orientation of the LMC.

To the best of our knowledge, this is the first time $W1$ is used in the MC RRL to obtain distances. We note the similarities between the $W1$ band and the JWST/NIRCam F335M and F356W filters, as well as between the $I$ band and the JWST/NIRCam F090W filter. A suitable set of color transformations, tuned to approximate NIRCam flux-averaged magnitudes for stars with RRLs spectral types, may enable extending the relations derived in this work to key NIRCam bands. This in turn will allow obtaining very precise distances to already mapped areas with JWST, extending the reach of this technique to very crowded regions within the entirety of the Local Group of galaxies.

Although eclipsing binaries provide one of the most direct and precise routes to extragalactic distance determinations, reaching percent-level accuracy in the Magellanic Clouds \citep[e.g.,][]{Pietrzynski2019,Graczyk2014}, their application is observationally expensive and limited to carefully selected systems. In particular, accurate eclipsing-binary distances require both well-sampled light curves and spectroscopic radial-velocity curves, which can demand substantial telescope time for each target \citep[e.g.,][]{Harries2003}. RRL stars provide a more scalable alternative for mapping distances in resolved galaxies. They are numerous in old stellar populations, have been detected in nearly all nearby galaxies searched for them, and can be used to derive distances from time-series photometry alone through calibrated PLZ or PWZ relations. These relations have been increasingly validated using Gaia parallaxes and through comparisons with independent distances to Galactic globular clusters and the Magellanic Clouds \citep[e.g.,][]{Muraveva2018_carnegie,Neeley2019, Mullen2023}. This work, presenting an independent distance of the LMC agreeing within 1\% with the state-of-the-art distance derived with eclipsing binaries, is an important step in this direction and to establish RRLs as precision distance indicators within the Local Group.

PLZ and PWZ relations, to be effective, need a robust estimate of the metallicity of each star. As shown in this paper, individual metallicities can also be derived with photometric techniques, using period--Fourier--metallicity relations \citep[e.g.,][]{Mullen2021, Mullen2022} applied to sufficiently sampled light curves. With this technique, we have obtained detailed metallicity maps of the LMC and SMC, and spatially resolved the metallicity and [Fe/H] radial gradients of their old population. We found two distinct populations of fundamental mode RRL stars (RRab) in the LMC. The RRL with higher metallicity appear to be concentrated in the central region, with a distribution roughly aligned with the bar of the galaxy. The RRLs in the outskirts of the LMC tend instead to be markedly more metal poor. As shown in Figure~\ref{fig:feh_gradient}, this feature gives rise to a steep radial metallicity gradient in the galaxy. We have identified three distinct regions in this profile: a steep gradient within the inner 5~deg from the center (co-located with the bar), an intermediate annulus with $5 < r_{MS} \leq 7.5$, characterized by a shallower slope, and a somewhat steeper slope in the external part of the LMC, for $7.5 < r_{MS} \leq 11$~deg. The overall decrease in metallicity along this profile is as much as $\sim$0.3~dex, along a radius of $\sim 10$~deg in the sky. The SMC, on the other hand, does not show any significant metallicity gradient.

This work is a step forward to further promote RRL stars as precision probes of metallicity and distance in resolved stellar populations. While not as accurate as other distance markers like Classical Cepheids, they have the advantage of existing in larger numbers, and in a variety of environments where Classical Cepheids are not found. They also trace a different population, whose old age ($\ga 10$~Gyr) makes it an important probe for the stars preserving the record of the initial assembly process of galaxies. The ability of providing reliable metallicity distributions without the need of spectroscopy make the techniques explored in this work an important tool-set to assess the properties of old stellar populations, providing essential constrains for modeling the star formation and chemical history of the Local Group of galaxies. This is especially important in the current era of large time-domain surveys (such as the Vera C. Rubin Observatory Legacy Survey of Space and Time, LSST), which are mapping the local universe at unprecedented speed and scale.

\begin{acknowledgments}

C.~E.~Mart\'inez-V\'azquez dedicates this paper to the memory of Antonio Sollima, a great colleague and astronomer who unfortunately left us too soon. She thanks P. Massana for the suggestion of using Magellanic Stream coordinates rather than Equatorial coordinates in this study, and J. P. Blakeslee, V. M. Placco and P. Martín Ravelo for helpful discussions. 

The work of C.~E.~Mart\'inez-V\'azquez is supported by NOIRLab, which is managed by the Association of Universities for Research in Astronomy (AURA) under a cooperative agreement with the U.S. National Science Foundation. This work was also partially supported by the international Gemini Observatory, a program of NSF NOIRLab, which is managed by AURA under a cooperative agreement with the U.S. National Science Foundation, on behalf of the Gemini partnership of Argentina, Brazil, Canada, Chile, the Republic of Korea, and the United States of America. M.~Marengo is supported by the National Science Foundation under Grant No. AST-2407965. G.~Bono acknowledges partial support from the INFN InDARK project. M.~Monelli acknowledges support from the project ``ASTRA: Across Space and Time: Relics \& Archaeology'' (P.I. M. Marconi) funded by INAF-Instituto Nazionale di Astrofisica.

This publication makes use of data products from WISE, which is a joint project of the University of California, Los Angeles, and the Jet Propulsion Laboratory (JPL)/California Institute of Technology (Caltech), funded by the National Aeronautics and Space Administration (NASA), and from NEOWISE, which is a JPL/Caltech project funded by NASA’s Planetary Science Division. 

This work has made use of data from the European Space Agency (ESA) mission {\it Gaia} (\url{https://www.cosmos.esa.int/gaia}), processed by the {\it Gaia} Data Processing and Analysis Consortium (DPAC, \url{https://www.cosmos.esa.int/web/gaia/dpac/consortium}). Funding for the DPAC has been provided by national institutions, in particular the institutions participating in the {\it Gaia} Multilateral Agreement.

\end{acknowledgments}

\begin{contribution}

CEMV, JPM, MM collaborated together in the analysis and interpretation of this work and were responsible for writing and submitting the manuscript. The remaining authors contributed to scientific discussion. All authors reviewed and approved the manuscript.

\end{contribution}

\vspace{5mm}
\facilities{WISE, OGLE, ASAS-SN, \textit{Gaia}}
\software{Astropy \citep{2013A&A...558A..33A}, SciPy \citep{2020SciPy-NMeth}}

\appendix

\section{Ensuring Homogeneity between spectroscopic and photometric metallicity scales}\label{ap:metallicity_check}

To verify the reliability of the photometric metallicities used in this work and quantify possible offsets between the metallicities derived for RRab and RRc stars, we applied the \citet{Mullen2021, Mullen2022} \rm{P}--$\phi_{31}-\rm{[Fe/H]}$ relations to individual RRL in a sample of single-metallicity globular clusters, and compared them with their spectroscopic abundances. We chose a sample of eight Galactic Globular Clusters (GGCs) with homogeneous $V$-band photometry, sufficient epochs coverage to perform a Fourier decomposition of the light curves of their individual stars, and metallicities spanning [Fe/H] = $-1.0$ to $-2.3$~dex. Table~\ref{tab:clusters} shows the list of clusters used in this analysis. The table provides two different [Fe/H] abundances for each cluster. Column~2 reports the metallicity measured for each cluster by \citet{Carretta2009}, hereafter C09. Column~3 instead reports the metallicity transformed into the scale adopted by \citet{Crestani2021} when calibrating their $\Delta S$ method (C21 hereafter). As noted in C21 (see bottom panel in their Figure~10), metallicities derived by the $\Delta S$ method are related to the metallicities in the C09 scale by an offset of 0.08~dex, such that:

\begin{equation}
    \mathrm{[Fe/H]_{C21} = \mathrm{[Fe/H]}_{C09} + 0.08}\label{eq:C09toC21}
\end{equation}

\begin{figure*}
    \centering
    \includegraphics[width=0.5\linewidth]{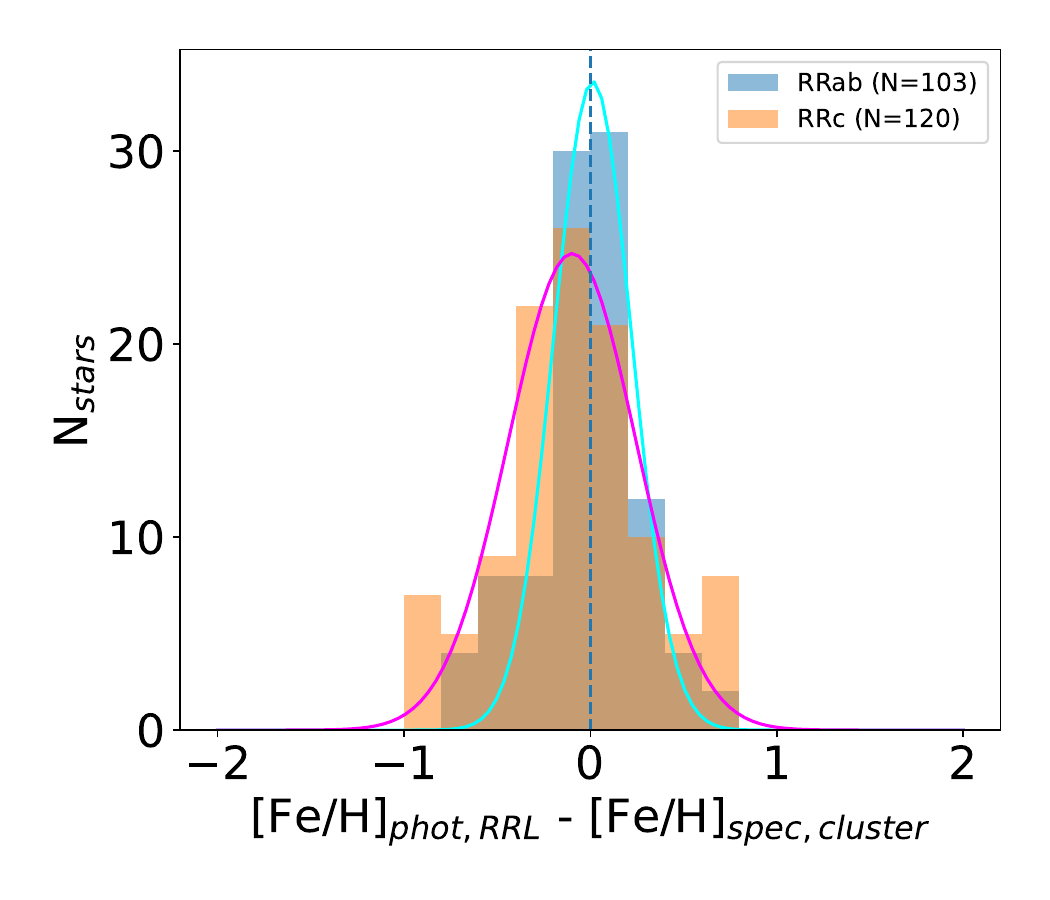}
    \caption{Distribution of the difference between the individual photometric metallicity derived for each RRL star in the eight GGCs and the spectroscopic metallicity of each cluster in C21 scale, separately for RRab (blue histogram) and RRc (orange histogram). The cyan and magenta lines represent a Gaussian fit of the two distributions.}
    \label{fig:metal_recalibration}
\end{figure*}

For each star in each cluster we have derived its photometric metallicity using Mullen's relation for the corresponding pulsation mode, based on their $V$-band Fourier parameters (photometric time series have been obtained from the homogeneous data set of P. B. Stetson\footnote{\url{https://www.canfar.net/storage/list/STETSON/homogeneous/Latest_photometry_for_targets_with_at_least_BVI}} for all clusters, except for NGC~3201 which comes from \citealt{Piersimoni2002}). In total, we ended up with a sample of 103 RRab stars and 120 RRc stars. The difference between the photometric metallicity of each star and the spectroscopic metallicity [Fe/H]$_{C21}$ of the cluster they belong to, is shown in Figure~\ref{fig:metal_recalibration}, separately for the RRab and RRc stars.

\begin{table}[!t]

    \begin{center}
    \caption{Globular clusters and their spectroscopic metallicities in C09 and C21 scale.}\label{tab:clusters}
    \begin{tabular}{lcc}
    \tableline
    \tableline
    Cluster & [Fe/H]$_{C09}$ & [Fe/H]$_{C21}$ \\
    \tableline
    NGC~7078 (M15) & $-2.33\pm0.02$ & $-2.25\pm0.02$ \\
    NGC~4590 (M68) & $-2.27\pm0.04$ & $-2.19\pm0.04$ \\
    NGC~4833       & $-1.89\pm0.05$ & $-1.81\pm0.05$ \\
    NGC~5286       & $-1.70\pm0.07$ & $-1.62\pm0.07$ \\
    NGC~3201       & $-1.51\pm0.02$ & $-1.43\pm0.02$ \\
    NGC~5272 (M3)  & $-1.50\pm0.05$ & $-1.42\pm0.05$ \\
    NGC~5904 (M5)  & $-1.33\pm0.02$ & $-1.25\pm0.02$ \\
    NGC~6362       & $-1.07\pm0.04$ & $-0.99\pm0.07$ \\
    \tableline
    \end{tabular}  
    \end{center}
\end{table}

The distribution of RRab stars metallicities (blue histogram and cyan smoothed curve) is centered around zero. This confirms that the photometric metallicities derived from Mullen's relations provide accurate and unbiased [Fe/H] abundances in the C21 scale. In the case of the RRc (orange histogram and magenta line) we instead observe an offset, which we measure at the level of $-0.098$~dex. Note that this offset is much smaller than the nominal accuracy of the photometric metallicity for individual stars ($\sigma_{[Fe/H]} \simeq 0.3$-0.4~dex), but can be detected in the larger ensemble of all stars combined from all clusters. The presence of a similar offset had already been noted in \citet{Mullen2022}, and attributed to small unaccounted residuals in the $\Delta S$ calibration of the \citet{Crestani2021} HR sample (see their Figure 9) and the fact that \citet{Mullen2021, Mullen2022} used a combination of HR+$\Delta S$ metallicities, also potentially subjected to small calibration offsets. This offset needs to be included in Mullen's relation for RRc stars, in order to have photometric metallicities fully consistent with the $\Delta S$ metallicities we use in the iterative method (section~\ref{sec:plz_iterative}) to calibrate the PLZ and PWZ relations presented in this section~\ref{sec:plz}.

For clarity and convenience we list below the final version of $V$-band Mullen's $\rm{P}-\phi_{31}-\rm{[Fe/H]}$ relations, corrected for bias, that we use in this work:

\begin{eqnarray}
\mathrm{[Fe/H]}_{C21}^V &=& -1.22\pm0.01 - (7.60\pm0.24) \cdot (P - 0.58) + (1.42\pm0.05) \cdot (\phi_{31}^{s,V} - 5.25) \ \ \ \textrm{for RRab} \\
\mathrm{[Fe/H]}_{C21}^V &=& -1.52\pm0.01 - (7.60\pm0.24) \cdot (P_F - 0.43) + (0.30\pm0.02) \cdot (\phi_{31}^{c,V} - 3.20) \ \ \ \textrm{for RRc} 
\end{eqnarray}

The RRab relation is unchanged from \citet{Mullen2021}. The RRc relation, instead, includes the offset identified from the analysis described above and uses the \emph{fundamentalized} period $P_F$, derived from the measured period using equation~\ref{eq:fundamentalization}. Following \citet{Mullen2021}, the $\phi_{31}$ term in the RRab relation is derived from the sine form of the Fourier expansion. Conversely, the $\phi_{31}$ term in the RRc relation uses the Fourier expansion in the cosine form. Converting one form into the other is as simple as applying the transformation $\phi_{31}^c = \phi_{31}^s - \pi$. As noted in \citet{Mullen2021}, due to the $2\pi$ ambiguity in the $\phi_{31}$ coefficients, the $\phi_{31}$ parameters of some stars require adding or subtracting $2 \pi$ to their phase, in order to lie close to the pivot value in the relations above (5.25 for RRab stars and 3.2 for RRc).

Both relations provide the [Fe/H] abundance in the C21 scale, since Mullen's relations have been calibrated using $\Delta S$ metallicities from the same source. These metallicities are appropriate for the analysis presented in this work, since the metallicity term of PLZ and PWZ relations rely on the $\Delta S$ metallicities for the same sample of calibrators (and is also consistent with the [Fe/H] parameters in theoretical works such as \citealt{2015ApJ...808...50M}).  Equation~\ref{eq:C09toC21} can be used to convert the metallicities from the above equations to the C09 scale and from there to any other empirical scale.

\begin{thebibliography}{}
	\expandafter\ifx\csname natexlab\endcsname\relax\def\natexlab#1{#1}\fi
	\providecommand{\url}[1]{\href{#1}{#1}}
	\providecommand{\dodoi}[1]{doi:~\href{http://doi.org/#1}{\nolinkurl{#1}}}
	\providecommand{\doeprint}[1]{\href{http://ascl.net/#1}{\nolinkurl{http://ascl.net/#1}}}
	\providecommand{\doarXiv}[1]{\href{https://arxiv.org/abs/#1}{\nolinkurl{https://arxiv.org/abs/#1}}}
	
	\bibitem[{ {Astropy Collaboration} {et~al.}(2013){Astropy Collaboration},
		{Robitaille}, {Tollerud}, {Greenfield}, {Droettboom}, {Bray}, {Aldcroft},
		{Davis}, {Ginsburg}, {Price-Whelan}, {Kerzendorf}, {Conley}, {Crighton},
		{Barbary}, {Muna}, {Ferguson}, {Grollier}, {Parikh}, {Nair}, {Unther},
		{Deil}, {Woillez}, {Conseil}, {Kramer}, {Turner}, {Singer}, {Fox}, {Weaver},
		{Zabalza}, {Edwards}, {Azalee Bostroem}, {Burke}, {Casey}, {Crawford},
		{Dencheva}, {Ely}, {Jenness}, {Labrie}, {Lim}, {Pierfederici}, {Pontzen},
		{Ptak}, {Refsdal}, {Servillat}, \& {Streicher}}]{2013A&A...558A..33A}
	{Astropy Collaboration}, {Robitaille}, T.~P., {Tollerud}, E.~J., {et~al.} 2013,
	\bibinfo{title}{{Astropy: A community Python package for astronomy},} \aap,
	558, A33, \dodoi{10.1051/0004-6361/201322068}
	
	\bibitem[{P.~T. Boggs {et~al.}(1989)Boggs, Donaldson, Byrd, \&
		Schnabel}]{Boggs}
	Boggs, P.~T., Donaldson, J.~R., Byrd, R.~h., \& Schnabel, R.~B. 1989,
	\bibinfo{title}{Algorithm 676: ODRPACK: Software for Weighted Orthogonal
		Distance Regression,} ACM Trans. Math. Softw., 15, 348–364,
	\dodoi{10.1145/76909.76913}
	
	\bibitem[{C. {Bonatto} \& E. {Bica}(2010){Bonatto} \& {Bica}}]{Bonatto2010}
	{Bonatto}, C., \& {Bica}, E. 2010, \bibinfo{title}{{Hierarchical structures in
			the Large and Small Magellanic Clouds},} \mnras, 403, 996,
	\dodoi{10.1111/j.1365-2966.2009.16177.x}
	
	\bibitem[{V.~F. {Braga} {et~al.}(2016){Braga}, {Stetson}, {Bono}, {Dall'Ora},
		{Ferraro}, {Fiorentino}, {Freyhammer}, {Iannicola}, {Marengo}, {Neeley},
		{Valenti}, {Buonanno}, {Calamida}, {Castellani}, {da Silva},
		{Degl'Innocenti}, {Di Cecco}, {Fabrizio}, {Freedman}, {Giuffrida}, {Lub},
		{Madore}, {Marconi}, {Marinoni}, {Matsunaga}, {Monelli}, {Persson},
		{Piersimoni}, {Pietrinferni}, {Prada-Moroni}, {Pulone}, {Stellingwerf},
		{Tognelli}, \& {Walker}}]{Braga2016}
	{Braga}, V.~F., {Stetson}, P.~B., {Bono}, G., {et~al.} 2016,
	\bibinfo{title}{{On the RR Lyrae Stars in Globulars. IV.
			{\ensuremath{\omega}} Centauri Optical UBVRI Photometry},} \aj, 152, 170,
	\dodoi{10.3847/0004-6256/152/6/170}
	
	\bibitem[{V.~F. {Braga} {et~al.}(2022){Braga}, {Fiorentino}, {Bono}, {Stetson},
		{Mart{\'\i}nez-V{\'a}zquez}, {Kwak}, {Dall'Ora}, {Di Criscienzo}, {Fabrizio},
		{Marengo}, {Marinoni}, {Marrese}, {Monelli}, \&
		{Tantalo}}]{2022MNRAS.tmp.2792B}
	{Braga}, V.~F., {Fiorentino}, G., {Bono}, G., {et~al.} 2022,
	\bibinfo{title}{{On the use of field RR lyrae as galactic probes. VI. Mixed
			mode RR Lyrae variables in Fornax and in nearby dwarf galaxies},} \mnras,
	\dodoi{10.1093/mnras/stac2813}
	
	\bibitem[{J.~A. {Cardelli} {et~al.}(1989){Cardelli}, {Clayton}, \&
		{Mathis}}]{Cardelli1989}
	{Cardelli}, J.~A., {Clayton}, G.~C., \& {Mathis}, J.~S. 1989,
	\bibinfo{title}{{The Relationship between Infrared, Optical, and Ultraviolet
			Extinction},} \apj, 345, 245, \dodoi{10.1086/167900}
	
	\bibitem[{E. {Carretta} {et~al.}(2009){Carretta}, {Bragaglia}, {Gratton},
		{D'Orazi}, \& {Lucatello}}]{Carretta2009}
	{Carretta}, E., {Bragaglia}, A., {Gratton}, R., {D'Orazi}, V., \& {Lucatello},
	S. 2009, \bibinfo{title}{{Intrinsic iron spread and a new metallicity scale
			for globular clusters},} \aap, 508, 695, \dodoi{10.1051/0004-6361/200913003}
	
	\bibitem[{M.~R.~L. Cioni {et~al.}(2011)Cioni, Clementini, Girardi, Guandalini,
		Gullieuszik, Miszalski, Moretti, Ripepi, Rubele, Bagheri, Bekki, Cross,
		de~Blok, de~Grijs, Emerson, Evans, Gibson, Gonzales-Solares, Groenewegen,
		Irwin, Ivanov, Lewis, Marconi, Marquette, Mastropietro, Moore, Napiwotzki,
		Naylor, Oliveira, Read, Sutorius, van Loon, Wilkinson, \& Wood}]{Cioni2011}
	Cioni, M. R.~L., Clementini, G., Girardi, L., {et~al.} 2011,
	\bibinfo{title}{The VMC survey. I. Strategy and first data,} Astronomy and
	Astrophysics, 527, A116, \dodoi{10.1051/0004-6361/201016137}
	
	\bibitem[{J. {Crestani} {et~al.}(2021){Crestani}, {Fabrizio}, {Braga},
		{Sneden}, {Preston}, {Ferraro}, {Iannicola}, {Bono}, {Alves-Brito}, {Nonino},
		{D'Orazi}, {Inno}, {Monelli}, {Storm}, {Altavilla}, {Chaboyer}, {Dall'Ora},
		{Fiorentino}, {Gilligan}, {Grebel}, {Lala}, {Lemasle}, {Marengo}, {Marinoni},
		{Marrese}, {Mart{\'\i}nez-V{\'a}zquez}, {Matsunaga}, {Mullen}, {Neeley},
		{Prudil}, {da Silva}, {Stetson}, {Th{\'e}venin}, {Valenti}, {Walker}, \&
		{Zoccali}}]{Crestani2021}
	{Crestani}, J., {Fabrizio}, M., {Braga}, V.~F., {et~al.} 2021,
	\bibinfo{title}{{On the Use of Field RR Lyrae as Galactic Probes. II. A New
			{\ensuremath{\Delta}}S Calibration to Estimate Their Metallicity},} \apj,
	908, 20, \dodoi{10.3847/1538-4357/abd183}
	
	\bibitem[{F. Cusano {et~al.}(2021)Cusano, Moretti, Clementini, Ripepi, Marconi,
		Cioni, Rubele, Garofalo, de~Grijs, Groenewegen, Oliveira, Subramanian, Sun,
		\& van Loon}]{Cusano2021}
	Cusano, F., Moretti, M.~I., Clementini, G., {et~al.} 2021, \bibinfo{title}{The
		VMC Survey - XLII. Near-infrared period-luminosity relations for RR Lyrae
		stars and the structure of the Large Magellanic Cloud,} Monthly Notices of
	the Royal Astronomical Society, 504, 1–15, \dodoi{10.1093/mnras/stab901}
	
	\bibitem[{S. {Deb} \& H.~P. {Singh}(2010){Deb} \&
		{Singh}}]{2010MNRAS.402..691D}
	{Deb}, S., \& {Singh}, H.~P. 2010, \bibinfo{title}{{Physical parameters of the
			Small Magellanic Cloud RR Lyrae stars and the distance scale},} \mnras, 402,
	691, \dodoi{10.1111/j.1365-2966.2009.15927.x}
	
	\bibitem[{I. {D{\'e}k{\'a}ny} {et~al.}(2021){D{\'e}k{\'a}ny}, {Grebel}, \&
		{Pojma{\'n}ski}}]{2021ApJ...920...33D}
	{D{\'e}k{\'a}ny}, I., {Grebel}, E.~K., \& {Pojma{\'n}ski}, G. 2021,
	\bibinfo{title}{{Metallicity Estimation of RR Lyrae Stars From Their I-Band
			Light Curves},} \apj, 920, 33, \dodoi{10.3847/1538-4357/ac106f}
	
	\bibitem[{L.-C. {Deng} {et~al.}(2012){Deng}, {Newberg}, {Liu}, {Carlin},
		{Beers}, {Chen}, {Chen}, {Christlieb}, {Grillmair}, {Guhathakurta}, {Han},
		{Hou}, {Lee}, {L{\'e}pine}, {Li}, {Liu}, {Pan}, {Sellwood}, {Wang}, {Wang},
		{Yang}, {Yanny}, {Zhang}, {Zhang}, {Zheng}, \& {Zhu}}]{2012RAA....12..735D}
	{Deng}, L.-C., {Newberg}, H.~J., {Liu}, C., {et~al.} 2012,
	\bibinfo{title}{{LAMOST Experiment for Galactic Understanding and Exploration
			(LEGUE) {\textemdash} The survey's science plan},} Research in Astronomy and
	Astrophysics, 12, 735, \dodoi{10.1088/1674-4527/12/7/003}
	
	\bibitem[{E.~A. {Dorfi} \& M.~U. {Feuchtinger}(1999){Dorfi} \&
		{Feuchtinger}}]{1999A&A...348..815D}
	{Dorfi}, E.~A., \& {Feuchtinger}, M.~U. 1999, \bibinfo{title}{{Theoretical UBVI
			light curves of RR Lyrae stars},} \aap, 348, 815
	
	\bibitem[{P.~R.~M. {Eisenhardt} {et~al.}(2020){Eisenhardt}, {Marocco},
		{Fowler}, {Meisner}, {Kirkpatrick}, {Garcia}, {Jarrett}, {Koontz},
		{Marchese}, {Stanford}, {Caselden}, {Cushing}, {Cutri}, {Faherty}, {Gelino},
		{Gonzalez}, {Mainzer}, {Mobasher}, {Schlegel}, {Stern}, {Teplitz}, \&
		{Wright}}]{Eisenhardt2020}
	{Eisenhardt}, P. R.~M., {Marocco}, F., {Fowler}, J.~W., {et~al.} 2020,
	\bibinfo{title}{{The CatWISE Preliminary Catalog: Motions from WISE and
			NEOWISE Data},} \apjs, 247, 69, \dodoi{10.3847/1538-4365/ab7f2a}
	
	\bibitem[{D. {El Youssoufi} {et~al.}(2019){El Youssoufi}, {Cioni}, {Bell},
		{Rubele}, {Bekki}, {de Grijs}, {Girardi}, {Ivanov}, {Matijevic},
		{Niederhofer}, {Oliveira}, {Ripepi}, {Subramanian}, \& {van
			Loon}}]{ElYoussoufi2019}
	{El Youssoufi}, D., {Cioni}, M.-R.~L., {Bell}, C. P.~M., {et~al.} 2019,
	\bibinfo{title}{{The VMC survey - XXXIV. Morphology of stellar populations in
			the Magellanic Clouds},} \mnras, 490, 1076, \dodoi{10.1093/mnras/stz2400}
	
	\bibitem[{M. {Fabrizio} {et~al.}(2021){Fabrizio}, {Braga}, {Crestani}, {Bono},
		{Ferraro}, {Fiorentino}, {Iannicola}, {Preston}, {Sneden}, {Th{\'e}venin},
		{Altavilla}, {Chaboyer}, {Dall'Ora}, {da Silva}, {Grebel}, {Gilligan},
		{Lala}, {Lemasle}, {Magurno}, {Marengo}, {Marinoni}, {Marrese},
		{Mart{\'\i}nez-V{\'a}zquez}, {Matsunaga}, {Monelli}, {Mullen}, {Neeley},
		{Nonino}, {Prudil}, {Salaris}, {Stetson}, {Valenti}, \&
		{Zoccali}}]{Fabrizio2021}
	{Fabrizio}, M., {Braga}, V.~F., {Crestani}, J., {et~al.} 2021,
	\bibinfo{title}{{On the Use of Field RR Lyrae As Galactic Probes: IV. New
			Insights Into and Around the Oosterhoff Dichotomy},} \apj, 919, 118,
	\dodoi{10.3847/1538-4357/ac1115}
	
	\bibitem[{M.~W. {Feast} \& A.~R. {Walker}(1987){Feast} \&
		{Walker}}]{FeastWalter1987}
	{Feast}, M.~W., \& {Walker}, A.~R. 1987, \bibinfo{title}{{Cepheids as distance
			indicators.},} \araa, 25, 345, \dodoi{10.1146/annurev.aa.25.090187.002021}
	
	\bibitem[{W.~L. {Freedman}(2021){Freedman}}]{Freedman2021}
	{Freedman}, W.~L. 2021, \bibinfo{title}{{Measurements of the Hubble Constant:
			Tensions in Perspective},} \apj, 919, 16, \dodoi{10.3847/1538-4357/ac0e95}
	
	\bibitem[{W.~L. {Freedman} {et~al.}(2012){Freedman}, {Madore}, {Scowcroft},
		{Burns}, {Monson}, {Persson}, {Seibert}, \& {Rigby}}]{Freedman2012}
	{Freedman}, W.~L., {Madore}, B.~F., {Scowcroft}, V., {et~al.} 2012,
	\bibinfo{title}{{Carnegie Hubble Program: A Mid-infrared Calibration of the
			Hubble Constant},} \apj, 758, 24, \dodoi{10.1088/0004-637X/758/1/24}
	
	\bibitem[{W.~L. {Freedman} {et~al.}(2009){Freedman}, {Rigby}, {Madore},
		{Persson}, {Sturch}, \& {Mager}}]{Freedman2009}
	{Freedman}, W.~L., {Rigby}, J., {Madore}, B.~F., {et~al.} 2009,
	\bibinfo{title}{{The Cepheid Period-Luminosity Relation (The Leavitt Law) at
			Mid-Infrared Wavelengths. IV. Cepheids in IC 1613},} \apj, 695, 996,
	\dodoi{10.1088/0004-637X/695/2/996}
	
	\bibitem[{ {Gaia Collaboration} {et~al.}(2023){Gaia Collaboration},
		{Vallenari}, {Brown}, {Prusti}, {de Bruijne}, {Arenou}, {Babusiaux},
		{Biermann}, {Creevey}, {Ducourant}, {Evans}, {Eyer}, {Guerra}, {Hutton},
		{Jordi}, {Klioner}, {Lammers}, {Lindegren}, {Luri}, {Mignard}, {Panem},
		{Pourbaix}, {Randich}, {Sartoretti}, {Soubiran}, {Tanga}, {Walton},
		{Bailer-Jones}, {Bastian}, {Drimmel}, {Jansen}, {Katz}, {Lattanzi}, {van
			Leeuwen}, {Bakker}, {Cacciari}, {Casta{\~n}eda}, {De Angeli}, {Fabricius},
		{Fouesneau}, {Fr{\'e}mat}, {Galluccio}, {Guerrier}, {Heiter}, {Masana},
		{Messineo}, {Mowlavi}, {Nicolas}, {Nienartowicz}, {Pailler}, {Panuzzo},
		{Riclet}, {Roux}, {Seabroke}, {Sordo}, {Th{\'e}venin}, {Gracia-Abril},
		{Portell}, {Teyssier}, {Altmann}, {Andrae}, {Audard}, {Bellas-Velidis},
		{Benson}, {Berthier}, {Blomme}, {Burgess}, {Busonero}, {Busso},
		{C{\'a}novas}, {Carry}, {Cellino}, {Cheek}, {Clementini}, {Damerdji},
		{Davidson}, {de Teodoro}, {Nu{\~n}ez Campos}, {Delchambre}, {Dell'Oro},
		{Esquej}, {Fern{\'a}ndez-Hern{\'a}ndez}, {Fraile}, {Garabato},
		{Garc{\'\i}a-Lario}, {Gosset}, {Haigron}, {Halbwachs}, {Hambly}, {Harrison},
		{Hern{\'a}ndez}, {Hestroffer}, {Hodgkin}, {Holl}, {Jan{\ss}en}, {Jevardat de
			Fombelle}, {Jordan}, {Krone-Martins}, {Lanzafame}, {L{\"o}ffler}, {Marchal},
		{Marrese}, {Moitinho}, {Muinonen}, {Osborne}, {Pancino}, {Pauwels},
		{Recio-Blanco}, {Reyl{\'e}}, {Riello}, {Rimoldini}, {Roegiers}, {Rybizki},
		{Sarro}, {Siopis}, {Smith}, {Sozzetti}, {Utrilla}, {van Leeuwen}, {Abbas},
		{{\'A}brah{\'a}m}, {Abreu Aramburu}, {Aerts}, {Aguado}, {Ajaj},
		{Aldea-Montero}, {Altavilla}, {{\'A}lvarez}, {Alves}, {Anders}, {Anderson},
		{Anglada Varela}, {Antoja}, {Baines}, {Baker}, {Balaguer-N{\'u}{\~n}ez},
		{Balbinot}, {Balog}, {Barache}, {Barbato}, {Barros}, {Barstow},
		{Bartolom{\'e}}, {Bassilana}, {Bauchet}, {Becciani}, {Bellazzini},
		{Berihuete}, {Bernet}, {Bertone}, {Bianchi}, {Binnenfeld}, {Blanco-Cuaresma},
		{Blazere}, {Boch}, {Bombrun}, {Bossini}, {Bouquillon}, {Bragaglia},
		{Bramante}, {Breedt}, {Bressan}, {Brouillet}, {Brugaletta}, {Bucciarelli},
		{Burlacu}, {Butkevich}, {Buzzi}, {Caffau}, {Cancelliere}, {Cantat-Gaudin},
		{Carballo}, {Carlucci}, {Carnerero}, {Carrasco}, {Casamiquela}, {Castellani},
		{Castro-Ginard}, {Chaoul}, {Charlot}, {Chemin}, {Chiaramida}, {Chiavassa},
		{Chornay}, {Comoretto}, {Contursi}, {Cooper}, {Cornez}, {Cowell}, {Crifo},
		{Cropper}, {Crosta}, {Crowley}, {Dafonte}, {Dapergolas}, {David}, {David},
		{de Laverny}, {De Luise}, \& {De March}}]{GaiaDR3}
	{Gaia Collaboration}, {Vallenari}, A., {Brown}, A.~G.~A., {et~al.} 2023,
	\bibinfo{title}{{Gaia Data Release 3. Summary of the content and survey
			properties},} \aap, 674, A1, \dodoi{10.1051/0004-6361/202243940}
	
	\bibitem[{A. {Garofalo} {et~al.}(2018){Garofalo}, {Scowcroft}, {Clementini},
		{Johnston}, {Cohen}, {Freedman}, {Madore}, {Majewski}, {Monson}, {Neeley},
		{Grillmair}, {Hendel}, {Kallivayalil}, {Marengo}, \& {van der
			Marel}}]{Garofalo2018}
	{Garofalo}, A., {Scowcroft}, V., {Clementini}, G., {et~al.} 2018,
	\bibinfo{title}{{SMHASH: a new mid-infrared RR Lyrae distance determination
			for the Local Group dwarf spheroidal galaxy Sculptor},} \mnras, 481, 578,
	\dodoi{10.1093/mnras/sty2222}
	
	\bibitem[{C.~K. {Gilligan} {et~al.}(2021){Gilligan}, {Chaboyer}, {Marengo},
		{Mullen}, {Bono}, {Braga}, {Crestani}, {Dall'Ora}, {Fiorentino}, {Monelli},
		{Neeley}, {Fabrizio}, {Mart{\'\i}nez-V{\'a}zquez}, {Th{\'e}venin}, \&
		{Sneden}}]{Gilligan2021}
	{Gilligan}, C.~K., {Chaboyer}, B., {Marengo}, M., {et~al.} 2021,
	\bibinfo{title}{{Metallicities from high-resolution spectra of 49 RR Lyrae
			variables},} \mnras, 503, 4719, \dodoi{10.1093/mnras/stab857}
	
	\bibitem[{D. Graczyk {et~al.}(2014)Graczyk, Pietrzyński, Thompson, Gieren,
		Pilecki, Konorski, Udalski, Soszyński, Villanova, Górski, Suchomska,
		Karczmarek, Kudritzki, Bresolin, \& Gallenne}]{Graczyk2014}
	Graczyk, D., Pietrzyński, G., Thompson, I.~B., {et~al.} 2014,
	\bibinfo{title}{The Araucaria Project. The Distance to the Small Magellanic
		Cloud from Late-type Eclipsing Binaries,} The Astrophysical Journal, 780, 59,
	\dodoi{10.1088/0004-637X/780/1/59}
	
	\bibitem[{R.~G. {Gratton} {et~al.}(2004){Gratton}, {Bragaglia}, {Clementini},
		{Carretta}, {Di Fabrizio}, {Maio}, \& {Taribello}}]{Gratton2004}
	{Gratton}, R.~G., {Bragaglia}, A., {Clementini}, G., {et~al.} 2004,
	\bibinfo{title}{{Metal abundances of RR Lyrae stars in the bar of the Large
			Magellanic Cloud},} \aap, 421, 937, \dodoi{10.1051/0004-6361:20035840}
	
	\bibitem[{T.~J. {Harries} {et~al.}(2003){Harries}, {Hilditch}, \&
		{Howarth}}]{Harries2003}
	{Harries}, T.~J., {Hilditch}, R.~W., \& {Howarth}, I.~D. 2003,
	\bibinfo{title}{{Ten eclipsing binaries in the Small Magellanic Cloud:
			fundamental parameters and Cloud distance},} \mnras, 339, 157,
	\dodoi{10.1046/j.1365-8711.2003.06169.x}
	
	\bibitem[{R. {Haschke} {et~al.}(2012){Haschke}, {Grebel}, \&
		{Duffau}}]{Haschke2012}
	{Haschke}, R., {Grebel}, E.~K., \& {Duffau}, S. 2012,
	\bibinfo{title}{{Three-dimensional Maps of the Magellanic Clouds using RR
			Lyrae Stars and Cepheids. I. The Large Magellanic Cloud},} \aj, 144, 106,
	\dodoi{10.1088/0004-6256/144/4/106}
	
	\bibitem[{T.~J. {Hoyt}(2023){Hoyt}}]{Hoyt2023}
	{Hoyt}, T.~J. 2023, \bibinfo{title}{{Sub-per-cent determination of the
			brightness at the tip of the red giant branch in the Magellanic Clouds},}
	Nature Astronomy, 7, 590, \dodoi{10.1038/s41550-023-01913-1}
	
	\bibitem[{J. {Iben} \& J. {Huchra}(1971){Iben} \&
		{Huchra}}]{1971A&A....14..293I}
	{Iben}, I., J., \& {Huchra}, J. 1971, \bibinfo{title}{{Comments on the
			Instability Strip for Halo Population Variables},} \aap, 14, 293
	
	\bibitem[{L. {Inno} {et~al.}(2016){Inno}, {Bono}, {Matsunaga}, {Fiorentino},
		{Marconi}, {Lemasle}, {da Silva}, {Soszy{\'n}ski}, {Udalski}, {Romaniello},
		\& {Rix}}]{Inno2016}
	{Inno}, L., {Bono}, G., {Matsunaga}, N., {et~al.} 2016, \bibinfo{title}{{The
			Panchromatic View of the Magellanic Clouds from Classical Cepheids. I.
			Distance, Reddening, and Geometry of the Large Magellanic Cloud Disk},} \apj,
	832, 176, \dodoi{10.3847/0004-637X/832/2/176}
	
	\bibitem[{G. {Iorio} \& V. {Belokurov}(2020){Iorio} \&
		{Belokurov}}]{2020arXiv200802280I}
	{Iorio}, G., \& {Belokurov}, V. 2020, \bibinfo{title}{{Chemo-kinematics of the
			$Gaia$ RR Lyrae: the halo and the disc},} arXiv e-prints, arXiv:2008.02280.
	\newblock \doarXiv{2008.02280}
	
	\bibitem[{{\v{Z}}. {Ivezi{\'c}} {et~al.}(2014){Ivezi{\'c}}, {Connolly},
		{VanderPlas}, \& {Gray}}]{Ivezic2014}
	{Ivezi{\'c}}, {\v{Z}}., {Connolly}, A.~J., {VanderPlas}, J.~T., \& {Gray}, A.
	2014, {Statistics, Data Mining, and Machine Learning in Astronomy: A
		Practical Python Guide for the Analysis of Survey Data},
	\dodoi{10.1515/9781400848911}
	
	\bibitem[{A.~M. Jacyszyn-Dobrzeniecka {et~al.}(2017)Jacyszyn-Dobrzeniecka,
		Skowron, Mróz, Soszyński, Udalski, Pietrukowicz, Skowron, Poleski,
		Kozłowski, Wyrzykowski, Pawlak, Szymański, \&
		Ulaczyk}]{JacyszynDobrzeniecka2017}
	Jacyszyn-Dobrzeniecka, A.~M., Skowron, D.~M., Mróz, P., {et~al.} 2017,
	\bibinfo{title}{OGLE-ing the Magellanic System: Three-Dimensional Structure
		of the Clouds and the Bridge using RR Lyrae Stars,} Acta Astronomica, 67,
	1–35, \dodoi{10.32023/0001-5237/67.1.1}
	
	\bibitem[{T. {Jayasinghe} {et~al.}(2018){Jayasinghe}, {Kochanek}, {Stanek},
		{Shappee}, {Holoien}, {Thompson}, {Prieto}, {Dong}, {Pawlak}, {Shields},
		{Pojmanski}, {Otero}, {Britt}, \& {Will}}]{2018MNRAS.477.3145J}
	{Jayasinghe}, T., {Kochanek}, C.~S., {Stanek}, K.~Z., {et~al.} 2018,
	\bibinfo{title}{{The ASAS-SN catalogue of variable stars I: The Serendipitous
			Survey},} \mnras, 477, 3145, \dodoi{10.1093/mnras/sty838}
	
	\bibitem[{J. {Jurcsik} \& G. {Kovacs}(1996){Jurcsik} \&
		{Kovacs}}]{1996A&A...312..111J}
	{Jurcsik}, J., \& {Kovacs}, G. 1996, \bibinfo{title}{{Determination of [Fe/H]
			from the light curves of RR Lyrae stars.},} \aap, 312, 111
	
	\bibitem[{A.~C. {Layden} {et~al.}(2019){Layden}, {Tiede}, {Chaboyer}, {Bunner},
		\& {Smitka}}]{Layden19}
	{Layden}, A.~C., {Tiede}, G.~P., {Chaboyer}, B., {Bunner}, C., \& {Smitka},
	M.~T. 2019, \bibinfo{title}{{Infrared K-band Photometry of Field RR Lyrae
			Variable Stars},} \aj, 158, 105, \dodoi{10.3847/1538-3881/ab2e10}
	
	\bibitem[{H.~S. {Leavitt} \& E.~C. {Pickering}(1912){Leavitt} \&
		{Pickering}}]{Leavitt1912}
	{Leavitt}, H.~S., \& {Pickering}, E.~C. 1912, \bibinfo{title}{{Periods of 25
			Variable Stars in the Small Magellanic Cloud.},} Harvard College Observatory
	Circular, 173, 1
	
	\bibitem[{X.-Y. {Li} {et~al.}(2023){Li}, {Huang}, {Liu}, {Beers}, \&
		{Zhang}}]{2023ApJ...944...88L}
	{Li}, X.-Y., {Huang}, Y., {Liu}, G.-C., {Beers}, T.~C., \& {Zhang}, H.-W. 2023,
	\bibinfo{title}{{Photometric Metallicity and Distance Estimates for 136,000
			RR Lyrae Stars from Gaia Data Release 3},} \apj, 944, 88,
	\dodoi{10.3847/1538-4357/acadd5}
	
	\bibitem[{L. {Lindegren} {et~al.}(2021){Lindegren}, {Bastian}, {Biermann},
		{Bombrun}, {de Torres}, {Gerlach}, {Geyer}, {Hern{\'a}ndez}, {Hilger},
		{Hobbs}, {Klioner}, {Lammers}, {McMillan}, {Ramos-Lerate},
		{Steidelm{\"u}ller}, {Stephenson}, \& {van Leeuwen}}]{EDR3bias}
	{Lindegren}, L., {Bastian}, U., {Biermann}, M., {et~al.} 2021,
	\bibinfo{title}{{Gaia Early Data Release 3. Parallax bias versus magnitude,
			colour, and position},} \aap, 649, A4, \dodoi{10.1051/0004-6361/202039653}
	
	\bibitem[{X.~W. {Liu} {et~al.}(2014){Liu}, {Yuan}, {Huo}, {Deng}, {Hou},
		{Zhao}, {Zhao}, {Shi}, {Luo}, {Xiang}, {Zhang}, {Huang}, \&
		{Zhang}}]{2014IAUS..298..310L}
	{Liu}, X.~W., {Yuan}, H.~B., {Huo}, Z.~Y., {et~al.} 2014, in IAU Symposium,
	Vol. 298, Setting the scene for Gaia and LAMOST, ed. S.~{Feltzing},
	G.~{Zhao}, N.~A. {Walton}, \& P.~{Whitelock}, 310--321,
	\dodoi{10.1017/S1743921313006510}
	
	\bibitem[{A. {Mainzer} {et~al.}(2011){Mainzer}, {Bauer}, {Grav}, {Masiero},
		{Cutri}, {Dailey}, {Eisenhardt}, {McMillan}, {Wright}, {Walker}, {Jedicke},
		{Spahr}, {Tholen}, {Alles}, {Beck}, {Brand enburg}, {Conrow}, {Evans},
		{Fowler}, {Jarrett}, {Marsh}, {Masci}, {McCallon}, {Wheelock}, {Wittman},
		{Wyatt}, {DeBaun}, {Elliott}, {Elsbury}, {Gautier}, {Gomillion}, {Leisawitz},
		{Maleszewski}, {Micheli}, \& {Wilkins}}]{2011ApJ...731...53M}
	{Mainzer}, A., {Bauer}, J., {Grav}, T., {et~al.} 2011,
	\bibinfo{title}{{Preliminary Results from NEOWISE: An Enhancement to the
			Wide-field Infrared Survey Explorer for Solar System Science},} \apj, 731,
	53, \dodoi{10.1088/0004-637X/731/1/53}
	
	\bibitem[{M. {Marconi} {et~al.}(2015){Marconi}, {Coppola}, {Bono}, {Braga},
		{Pietrinferni}, {Buonanno}, {Castellani}, {Musella}, {Ripepi}, \&
		{Stellingwerf}}]{2015ApJ...808...50M}
	{Marconi}, M., {Coppola}, G., {Bono}, G., {et~al.} 2015, \bibinfo{title}{{On a
			New Theoretical Framework for RR Lyrae Stars. I. The Metallicity
			Dependence},} \apj, 808, 50, \dodoi{10.1088/0004-637X/808/1/50}
	
	\bibitem[{F. Marocco {et~al.}(2021)Marocco, Eisenhardt, Fowler, Kirkpatrick,
		Meisner, Schlafly, Stanford, Garcia, Caselden, Cushing, Cutri, Faherty,
		Gelino, Gonzalez, Jarrett, Koontz, Mainzer, Marchese, Mobasher, Schlegel,
		Stern, Teplitz, \& Wright}]{Marocco2021}
	Marocco, F., Eisenhardt, P. R.~M., Fowler, J.~W., {et~al.} 2021,
	\bibinfo{title}{The CatWISE2020 Catalog,} The Astrophysical Journal
	Supplement Series, 253, 8, \dodoi{10.3847/1538-4365/abd805}
	
	\bibitem[{P.~M. {Marrese} {et~al.}(2022){Marrese}, {Marinoni}, {Fabrizio}, \&
		{Altavilla}}]{Marrese2022}
	{Marrese}, P.~M., {Marinoni}, S., {Fabrizio}, M., \& {Altavilla}, G. 2022,
	\bibinfo{title}{{Gaia DR3 documentation Chapter 15: Cross-match with external
			catalogues},}, Gaia DR3 documentation, European Space Agency; Gaia Data
	Processing and Analysis Consortium. Online at
	https://gea.esac.esa.int/archive/documentation/GDR3/index.html, id. 15
	
	\bibitem[{C.~E. {Mart\'inez-V\'azquez} {et~al.}(2016){Mart\'inez-V\'azquez},
		{Monelli}, {Bono}, {Stetson}, {Gallart}, {Bernard}, {Fiorentino}, \&
		{Dall'Ora}}]{Martinez-Vazquez2016}
	{Mart\'inez-V\'azquez}, C.~E., {Monelli}, M., {Bono}, G., {et~al.} 2016,
	\bibinfo{title}{{A new Phi\_31-period-metallicity relation for RR Lyrae
			stars},} Commmunications of the Konkoly Observatory Hungary, 105, 53
	
	\bibitem[{C.~E. {Mart{\'{\i}}nez-V{\'a}zquez}
		{et~al.}(2016){Mart{\'{\i}}nez-V{\'a}zquez}, {Monelli}, {Gallart}, {Bono},
		{Bernard}, {Stetson}, {Ferraro}, {Walker}, {Dall'Ora}, {Fiorentino}, \&
		{Iannicola}}]{MartinezVazquez2016a}
	{Mart{\'{\i}}nez-V{\'a}zquez}, C.~E., {Monelli}, M., {Gallart}, C., {et~al.}
	2016, \bibinfo{title}{{Probing the early chemical evolution of the Sculptor
			dSph with purely old stellar tracers},} \mnras, 461, L41,
	\dodoi{10.1093/mnrasl/slw093}
	
	\bibitem[{C.~E. {Mart{\'{\i}}nez-V{\'a}zquez}
		{et~al.}(2017){Mart{\'{\i}}nez-V{\'a}zquez}, {Monelli}, {Bernard}, {Gallart},
		{Stetson}, {Skillman}, {Bono}, {Cassisi}, {Fiorentino}, {McQuinn}, {Cole},
		{McConnachie}, {Martin}, {Dolphin}, {Boylan-Kolchin}, {Aparicio}, {Hidalgo},
		\& {Weisz}}]{MartinezVazquez2017}
	{Mart{\'{\i}}nez-V{\'a}zquez}, C.~E., {Monelli}, M., {Bernard}, E.~J., {et~al.}
	2017, \bibinfo{title}{{The ISLAnds Project. III. Variable Stars in Six
			Andromeda Dwarf Spheroidal Galaxies},} \apj, 850, 137,
	\dodoi{10.3847/1538-4357/aa9381}
	
	\bibitem[{C.~E. {Mart{\'\i}nez-V{\'a}zquez}
		{et~al.}(2021){Mart{\'\i}nez-V{\'a}zquez}, {Monelli}, {Cassisi}, {Taibi},
		{Gallart}, {Vivas}, {Walker}, {Mart{\'\i}n-Ravelo}, {Zenteno}, {Battaglia},
		{Bono}, {Calamida}, {Carollo}, {Cicu{\'e}ndez}, {Fiorentino}, {Marconi},
		{Salvadori}, {Balbinot}, {Bernard}, {Dall'Ora}, \&
		{Stetson}}]{MartinezVazquez2021b}
	{Mart{\'\i}nez-V{\'a}zquez}, C.~E., {Monelli}, M., {Cassisi}, S., {et~al.}
	2021, \bibinfo{title}{{Variable stars in Local Group galaxies - V. The fast
			and early evolution of the low-mass Eridanus II dSph galaxy},} \mnras, 508,
	1064, \dodoi{10.1093/mnras/stab2493}
	
	\bibitem[{P. {Massana} {et~al.}(2022){Massana}, {Ruiz-Lara}, {No{\"e}l},
		{Gallart}, {Nidever}, {Choi}, {Sakowska}, {Besla}, {Olsen}, {Monelli},
		{Dorta}, {Stringfellow}, {Cassisi}, {Bernard}, {Zaritsky}, {Cioni},
		{Monachesi}, {van der Marel}, {de Boer}, \& {Walker}}]{Massana2022}
	{Massana}, P., {Ruiz-Lara}, T., {No{\"e}l}, N.~E.~D., {et~al.} 2022,
	\bibinfo{title}{{The synchronized dance of the magellanic clouds' star
			formation history},} \mnras, 513, L40, \dodoi{10.1093/mnrasl/slac030}
	
	\bibitem[{L. {Monteagudo} {et~al.}(2018){Monteagudo}, {Gallart}, {Monelli},
		{Bernard}, \& {Stetson}}]{Monteagudo2018}
	{Monteagudo}, L., {Gallart}, C., {Monelli}, M., {Bernard}, E.~J., \& {Stetson},
	P.~B. 2018, \bibinfo{title}{{The origin of the LMC stellar bar: clues from
			the SFH of the bar and inner disc},} \mnras, 473, L16,
	\dodoi{10.1093/mnrasl/slx158}
	
	\bibitem[{M.~I. Moretti {et~al.}(2014)Moretti, Clementini, Muraveva, Ripepi,
		Marquette, Cioni, Marconi, Girardi, Rubele, Tisserand, de~Grijs, Groenewegen,
		Guandalini, Ivanov, \& van Loon}]{Moretti2014}
	Moretti, M.~I., Clementini, G., Muraveva, T., {et~al.} 2014,
	\bibinfo{title}{The VMC Survey - X. Cepheids, RR Lyrae stars and binaries as
		probes of the Magellanic System’s structure,} Monthly Notices of the Royal
	Astronomical Society, 437, 2702–2719, \dodoi{10.1093/mnras/stt2081}
	
	\bibitem[{S. {Morgan}(2014){Morgan}}]{2014IAUS..301..461M}
	{Morgan}, S. 2014, in Precision Asteroseismology, ed. J.~A. {Guzik}, W.~J.
	{Chaplin}, G.~{Handler}, \& A.~{Pigulski}, Vol. 301, 461--462,
	\dodoi{10.1017/S1743921313015056}
	
	\bibitem[{S.~M. {Morgan} {et~al.}(2007){Morgan}, {Wahl}, \&
		{Wieckhorst}}]{2007MNRAS.374.1421M}
	{Morgan}, S.~M., {Wahl}, J.~N., \& {Wieckhorst}, R.~M. 2007,
	\bibinfo{title}{{[Fe/H] relations for c-type RR Lyrae variables based upon
			Fourier coefficients},} \mnras, 374, 1421,
	\dodoi{10.1111/j.1365-2966.2006.11247.x}
	
	\bibitem[{J.~P. Mullen {et~al.}(2021)Mullen, Marengo, Martínez-Vázquez,
		Neeley, Bono, Dall’Ora, Chaboyer, Thévenin, Braga, Crestani, Fabrizio,
		Fiorentino, Gilligan, Monelli, \& Stetson}]{Mullen2021}
	Mullen, J.~P., Marengo, M., Martínez-Vázquez, C.~E., {et~al.} 2021,
	\bibinfo{title}{Metallicity of Galactic RR Lyrae from Optical and Infrared
		Light Curves. I. Period-Fourier-Metallicity Relations for Fundamental-mode RR
		Lyrae,} The Astrophysical Journal, 912, 144, \dodoi{10.3847/1538-4357/abefd4}
	
	\bibitem[{J.~P. Mullen {et~al.}(2022)Mullen, Marengo, Martínez-Vázquez, Bono,
		Braga, Chaboyer, Crestani, Dall’Ora, Fabrizio, Fiorentino, Monelli, Neeley,
		Stetson, \& Thévenin}]{Mullen2022}
	Mullen, J.~P., Marengo, M., Martínez-Vázquez, C.~E., {et~al.} 2022,
	\bibinfo{title}{Metallicity of Galactic RR Lyrae from Optical and Infrared
		Light Curves. II. Period-Fourier-Metallicity Relations for First Overtone RR
		Lyrae,} The Astrophysical Journal, 931, 131, \dodoi{10.3847/1538-4357/ac67ee}
	
	\bibitem[{J.~P. Mullen {et~al.}(2023)Mullen, Marengo, Martínez-Vázquez,
		Chaboyer, Bono, Braga, Dall’Ora, D’Orazi, Fabrizio, Monelli, \&
		Thévenin}]{Mullen2023}
	Mullen, J.~P., Marengo, M., Martínez-Vázquez, C.~E., {et~al.} 2023,
	\bibinfo{title}{RR Lyrae Mid-infrared Period-Luminosity-Metallicity and
		Period-Wesenheit-Metallicity Relations Based on Gaia DR3 Parallaxes,} The
	Astrophysical Journal, 945, 83, \dodoi{10.3847/1538-4357/acb20a}
	
	\bibitem[{T. {Muraveva} {et~al.}(2018){Muraveva}, {Garofalo}, {Scowcroft},
		{Clementini}, {Freedman}, {Madore}, \& {Monson}}]{Muraveva2018_carnegie}
	{Muraveva}, T., {Garofalo}, A., {Scowcroft}, V., {et~al.} 2018,
	\bibinfo{title}{{The Carnegie RR Lyrae Program: mid-infrared
			period-luminosity relations of RR Lyrae stars in Reticulum},} \mnras, 480,
	4138, \dodoi{10.1093/mnras/sty1959}
	
	\bibitem[{T. {Muraveva} {et~al.}(2025){Muraveva}, {Giannetti}, {Clementini},
		{Garofalo}, \& {Monti}}]{2025MNRAS.536.2749M}
	{Muraveva}, T., {Giannetti}, A., {Clementini}, G., {Garofalo}, A., \& {Monti},
	L. 2025, \bibinfo{title}{{Metallicity of RR Lyrae stars from the Gaia Data
			Release 3 catalogue computed with Machine Learning algorithms},} \mnras, 536,
	2749, \dodoi{10.1093/mnras/stae2679}
	
	\bibitem[{T. Muraveva {et~al.}(2018)Muraveva, Subramanian, Clementini, Cioni,
		Palmer, van Loon, Moretti, de~Grijs, Molinaro, Ripepi, Marconi, Emerson, \&
		Ivanov}]{Muraveva2018_vmc}
	Muraveva, T., Subramanian, S., Clementini, G., {et~al.} 2018,
	\bibinfo{title}{The VMC survey - XXVI. Structure of the Small Magellanic
		Cloud from RR Lyrae stars,} Monthly Notices of the Royal Astronomical
	Society, 473, 3131–3146, \dodoi{10.1093/mnras/stx2514}
	
	\bibitem[{J.~R. {Neeley} {et~al.}(2017){Neeley}, {Marengo}, {Bono}, {Braga},
		{Dall'Ora}, {Magurno}, {Marconi}, {Trueba}, {Tognelli}, {Prada Moroni},
		{Beaton}, {Freedman}, {Madore}, {Monson}, {Scowcroft}, {Seibert}, \&
		{Stetson}}]{Neeley2017}
	{Neeley}, J.~R., {Marengo}, M., {Bono}, G., {et~al.} 2017, \bibinfo{title}{{On
			a New Theoretical Framework for RR Lyrae Stars. II. Mid-infrared
			Period-Luminosity-Metallicity Relations},} \apj, 841, 84,
	\dodoi{10.3847/1538-4357/aa713d}
	
	\bibitem[{J.~R. Neeley {et~al.}(2019)Neeley, Marengo, Freedman, Madore, Beaton,
		Hatt, Hoyt, Monson, Rich, Sarajedini, Seibert, \& Scowcroft}]{Neeley2019}
	Neeley, J.~R., Marengo, M., Freedman, W.~L., {et~al.} 2019,
	\bibinfo{title}{Standard Galactic field RR Lyrae II: a Gaia DR2 calibration
		of the period-Wesenheit-metallicity relation,} Monthly Notices of the Royal
	Astronomical Society, 490, 4254–4270, \dodoi{10.1093/mnras/stz2814}
	
	\bibitem[{J.~M. {Nemec} {et~al.}(2013){Nemec}, {Cohen}, {Ripepi}, {Derekas},
		{Moskalik}, {Sesar}, {Chadid}, \& {Bruntt}}]{2013ApJ...773..181N}
	{Nemec}, J.~M., {Cohen}, J.~G., {Ripepi}, V., {et~al.} 2013,
	\bibinfo{title}{{Metal Abundances, Radial Velocities, and Other Physical
			Characteristics for the RR Lyrae Stars in The Kepler Field},} \apj, 773, 181,
	\dodoi{10.1088/0004-637X/773/2/181}
	
	\bibitem[{C.-C. {Ngeow} {et~al.}(2016){Ngeow}, {Yu}, {Bellm}, {Yang}, {Chang},
		{Miller}, {Laher}, {Surace}, \& {Ip}}]{2016ApJS..227...30N}
	{Ngeow}, C.-C., {Yu}, P.-C., {Bellm}, E., {et~al.} 2016, \bibinfo{title}{{The
			Palomar Transient Factory and RR Lyrae: The Metallicity-Light Curve Relation
			Based on ab-type RR Lyrae in the Kepler Field},} \apjs, 227, 30,
	\dodoi{10.3847/1538-4365/227/2/30}
	
	\bibitem[{D.~L. {Nidever}(2024){Nidever}}]{Nidever2024}
	{Nidever}, D.~L. 2024, \bibinfo{title}{{Discovery of a split stellar stream in
			the periphery of the Small Magellanic Cloud},} \mnras, 533, 3238,
	\dodoi{10.1093/mnras/stae1783}
	
	\bibitem[{D.~L. {Nidever} {et~al.}(2008){Nidever}, {Majewski}, \& {Butler
			Burton}}]{Nidever2008}
	{Nidever}, D.~L., {Majewski}, S.~R., \& {Butler Burton}, W. 2008,
	\bibinfo{title}{{The Origin of the Magellanic Stream and Its Leading Arm},}
	\apj, 679, 432, \dodoi{10.1086/587042}
	
	\bibitem[{D.~L. {Nidever} {et~al.}(2017){Nidever}, {Olsen}, {Walker}, {Vivas},
		{Blum}, {Kaleida}, {Choi}, {Conn}, {Gruendl}, {Bell}, {Besla}, {Mu{\~n}oz},
		{Gallart}, {Martin}, {Olszewski}, {Saha}, {Monachesi}, {Monelli}, {de Boer},
		{Johnson}, {Zaritsky}, {Stringfellow}, {van der Marel}, {Cioni}, {Jin},
		{Majewski}, {Martinez-Delgado}, {Monteagudo}, {No{\"e}l}, {Bernard},
		{Kunder}, {Chu}, {Bell}, {Santana}, {Frechem}, {Medina}, {Parkash},
		{Navarrete}, \& {Hayes}}]{Nidever2017}
	{Nidever}, D.~L., {Olsen}, K., {Walker}, A.~R., {et~al.} 2017,
	\bibinfo{title}{{SMASH: Survey of the MAgellanic Stellar History},} \aj, 154,
	199, \dodoi{10.3847/1538-3881/aa8d1c}
	
	\bibitem[{A.~B. {Pace}(2025){Pace}}]{Pace2025}
	{Pace}, A.~B. 2025, \bibinfo{title}{{The Local Volume Database: a library of
			the observed properties of nearby dwarf galaxies and star clusters},} The
	Open Journal of Astrophysics, 8, 142, \dodoi{10.33232/001c.144859}
	
	\bibitem[{A.~M. {Piersimoni} {et~al.}(2002){Piersimoni}, {Bono}, \&
		{Ripepi}}]{Piersimoni2002}
	{Piersimoni}, A.~M., {Bono}, G., \& {Ripepi}, V. 2002, \bibinfo{title}{{BVI
			Time-Series Data of the Galactic Globular Cluster NGC 3201. I. RR Lyrae
			Stars},} \aj, 124, 1528, \dodoi{10.1086/341821}
	
	\bibitem[{G. Pietrzyński {et~al.}(2013)Pietrzyński, Graczyk, Gieren,
		Thompson, Pilecki, Udalski, Soszyński, Kozłowski, Konorski, Suchomska,
		Bono, Moroni, Villanova, Nardetto, Bresolin, Kudritzki, Storm, Gallenne,
		Smolec, Minniti, Kubiak, Szymański, Poleski, Wyrzykowski, Ulaczyk,
		Pietrukowicz, Górski, \& Karczmarek}]{Pietrzynski2013}
	Pietrzyński, G., Graczyk, D., Gieren, W., {et~al.} 2013, \bibinfo{title}{An
		eclipsing-binary distance to the Large Magellanic Cloud accurate to two per
		cent,} Nature, 495, 76–79, \dodoi{10.1038/nature11878}
	
	\bibitem[{G. Pietrzyński {et~al.}(2019)Pietrzyński, Graczyk, Gallenne,
		Gieren, Thompson, Pilecki, Karczmarek, Górski, Suchomska, Taormina, Zgirski,
		Wielgórski, Kołaczkowski, Konorski, Villanova, Nardetto, Kervella,
		Bresolin, Kudritzki, Storm, Smolec, \& Narloch}]{Pietrzynski2019}
	Pietrzyński, G., Graczyk, D., Gallenne, A., {et~al.} 2019, \bibinfo{title}{A
		distance to the Large Magellanic Cloud that is precise to one per cent,}
	Nature, 567, 200–203, \dodoi{10.1038/s41586-019-0999-4}
	
	\bibitem[{G.~W. {Preston}(1959){Preston}}]{1959ApJ...130..507P}
	{Preston}, G.~W. 1959, \bibinfo{title}{{A Spectroscopic Study of the RR Lyrae
			Stars.},} \apj, 130, 507, \dodoi{10.1086/146743}
	
	\bibitem[{A. Price-Whelan {et~al.}(2025)Price-Whelan, Garrison, Souchereau,
		Wagg, Sipőcz, Starkman, Chen, Lenz, Greco, Hart, AlexKurek, Robert,
		Foreman-Mackey, HNLala, Lim, Oh, Koposov, Lilleengen, \&
		Li}]{Price-Whelan/gala-soft}
	Price-Whelan, A., Garrison, L., Souchereau, H., {et~al.} 2025,
	\bibinfo{title}{adrn/gala: v1.10.1,}, v1.10.1 Zenodo,
	\dodoi{10.5281/zenodo.16923466}
	
	\bibitem[{A.~M. Price-Whelan(2017)Price-Whelan}]{Price-Whelan/gala}
	Price-Whelan, A.~M. 2017, \bibinfo{title}{Gala: A Python package for galactic
		dynamics,} The Journal of Open Source Software, 2,
	\dodoi{10.21105/joss.00388}
	
	\bibitem[{A. {Rest} {et~al.}(2005){Rest}, {Stubbs}, {Becker}, {Miknaitis},
		{Miceli}, {Covarrubias}, {Hawley}, {Smith}, {Suntzeff}, {Olsen}, {Prieto},
		{Hiriart}, {Welch}, {Cook}, {Nikolaev}, {Huber}, {Prochtor}, {Clocchiatti},
		{Minniti}, {Garg}, {Challis}, {Keller}, \& {Schmidt}}]{Rest2005}
	{Rest}, A., {Stubbs}, C., {Becker}, A.~C., {et~al.} 2005,
	\bibinfo{title}{{Testing LMC Microlensing Scenarios: The Discrimination Power
			of the SuperMACHO Microlensing Survey},} \apj, 634, 1103,
	\dodoi{10.1086/497060}
	
	\bibitem[{M.~J. {Rieke} {et~al.}(2023){Rieke}, {Kelly}, {Misselt},
		{Stansberry}, {Boyer}, {Beatty}, {Egami}, {Florian}, {Greene}, {Hainline},
		{Leisenring}, {Roellig}, {Schlawin}, {Sun}, {Tinnin}, {Williams}, {Willmer},
		{Wilson}, {Clark}, {Rohrbach}, {Brooks}, {Canipe}, {Correnti}, {DiFelice},
		{Gennaro}, {Girard}, {Hartig}, {Hilbert}, {Koekemoer}, {Nikolov}, {Pirzkal},
		{Rest}, {Robberto}, {Sunnquist}, {Telfer}, {Wu}, {Ferry}, {Lewis}, {Baum},
		{Beichman}, {Doyon}, {Dressler}, {Eisenstein}, {Ferrarese}, {Hodapp},
		{Horner}, {Jaffe}, {Johnstone}, {Krist}, {Martin}, {McCarthy}, {Meyer},
		{Rieke}, {Trauger}, \& {Young}}]{Rieke2023}
	{Rieke}, M.~J., {Kelly}, D.~M., {Misselt}, K., {et~al.} 2023,
	\bibinfo{title}{{Performance of NIRCam on JWST in Flight},} \pasp, 135,
	028001, \dodoi{10.1088/1538-3873/acac53}
	
	\bibitem[{A.~G. {Riess} {et~al.}(2019){Riess}, {Casertano}, {Yuan}, {Macri}, \&
		{Scolnic}}]{Riess2019}
	{Riess}, A.~G., {Casertano}, S., {Yuan}, W., {Macri}, L.~M., \& {Scolnic}, D.
	2019, \bibinfo{title}{{Large Magellanic Cloud Cepheid Standards Provide a 1\%
			Foundation for the Determination of the Hubble Constant and Stronger Evidence
			for Physics beyond {\ensuremath{\Lambda}}CDM},} \apj, 876, 85,
	\dodoi{10.3847/1538-4357/ab1422}
	
	\bibitem[{A.~G. {Riess} {et~al.}(2023){Riess}, {Anand}, {Yuan}, {Casertano},
		{Dolphin}, {Macri}, {Breuval}, {Scolnic}, {Perrin}, \&
		{Anderson}}]{Riess2023}
	{Riess}, A.~G., {Anand}, G.~S., {Yuan}, W., {et~al.} 2023,
	\bibinfo{title}{{Crowded No More: The Accuracy of the Hubble Constant Tested
			with High-resolution Observations of Cepheids by JWST},} \apjl, 956, L18,
	\dodoi{10.3847/2041-8213/acf769}
	
	\bibitem[{A.~G. {Riess} {et~al.}(2024){Riess}, {Scolnic}, {Anand}, {Breuval},
		{Casertano}, {Macri}, {Li}, {Yuan}, {Huang}, {Jha}, {Murakami}, {Beaton},
		{Brout}, {Wu}, {Addison}, {Bennett}, {Anderson}, {Filippenko}, \&
		{Carr}}]{Riess2024}
	{Riess}, A.~G., {Scolnic}, D., {Anand}, G.~S., {et~al.} 2024,
	\bibinfo{title}{{JWST Validates HST Distance Measurements: Selection of
			Supernova Subsample Explains Differences in JWST Estimates of Local H
			$_{0}$},} \apj, 977, 120, \dodoi{10.3847/1538-4357/ad8c21}
	
	\bibitem[{V. Ripepi {et~al.}(2017)Ripepi, Cioni, Moretti, Marconi, Bekki,
		Clementini, de~Grijs, Emerson, Groenewegen, Ivanov, Molinaro, Muraveva,
		Oliveira, Piatti, Subramanian, \& van Loon}]{Ripepi2017}
	Ripepi, V., Cioni, M.-R.~L., Moretti, M.~I., {et~al.} 2017, \bibinfo{title}{The
		VMC survey - XXV. The 3D structure of the Small Magellanic Cloud from
		Classical Cepheids,} Monthly Notices of the Royal Astronomical Society, 472,
	808–827, \dodoi{10.1093/mnras/stx2096}
	
	\bibitem[{E.~F. {Schlafly} \& D.~P. {Finkbeiner}(2011){Schlafly} \&
		{Finkbeiner}}]{Schlafly2011}
	{Schlafly}, E.~F., \& {Finkbeiner}, D.~P. 2011, \bibinfo{title}{{Measuring
			Reddening with Sloan Digital Sky Survey Stellar Spectra and Recalibrating
			SFD},} \apj, 737, 103, \dodoi{10.1088/0004-637X/737/2/103}
	
	\bibitem[{D.~J. {Schlegel} {et~al.}(1998){Schlegel}, {Finkbeiner}, \&
		{Davis}}]{Schlegel1998}
	{Schlegel}, D.~J., {Finkbeiner}, D.~P., \& {Davis}, M. 1998,
	\bibinfo{title}{{Maps of Dust Infrared Emission for Use in Estimation of
			Reddening and Cosmic Microwave Background Radiation Foregrounds},} \apj, 500,
	525, \dodoi{10.1086/305772}
	
	\bibitem[{B.~J. {Shappee} {et~al.}(2014){Shappee}, {Prieto}, {Grupe},
		{Kochanek}, {Stanek}, {De Rosa}, {Mathur}, {Zu}, {Peterson}, {Pogge},
		{Komossa}, {Im}, {Jencson}, {Holoien}, {Basu}, {Beacom}, {Szczygie{\l}},
		{Brimacombe}, {Adams}, {Campillay}, {Choi}, {Contreras}, {Dietrich},
		{Dubberley}, {Elphick}, {Foale}, {Giustini}, {Gonzalez}, {Hawkins}, {Howell},
		{Hsiao}, {Koss}, {Leighly}, {Morrell}, {Mudd}, {Mullins}, {Nugent},
		{Parrent}, {Phillips}, {Pojmanski}, {Rosing}, {Ross}, {Sand}, {Terndrup},
		{Valenti}, {Walker}, \& {Yoon}}]{2014ApJ...788...48S}
	{Shappee}, B.~J., {Prieto}, J.~L., {Grupe}, D., {et~al.} 2014,
	\bibinfo{title}{{The Man behind the Curtain: X-Rays Drive the UV through NIR
			Variability in the 2013 Active Galactic Nucleus Outburst in NGC 2617},} \apj,
	788, 48, \dodoi{10.1088/0004-637X/788/1/48}
	
	\bibitem[{T. {Sicignano} {et~al.}(2024){Sicignano}, {Ripepi}, {Marconi},
		{Molinaro}, {Bhardwaj}, {Cioni}, {de Grijs}, {Storm}, {Groenewegen},
		{Ivanov}, \& {De Somma}}]{Sicignano2024}
	{Sicignano}, T., {Ripepi}, V., {Marconi}, M., {et~al.} 2024,
	\bibinfo{title}{{The VMC survey. L. Type II Cepheids in the Magellanic
			Clouds: Period-luminosity relations in the near-infrared bands},} \aap, 685,
	A41, \dodoi{10.1051/0004-6361/202348650}
	
	\bibitem[{D.~M. {Skowron} {et~al.}(2016){Skowron}, {Soszy{\'n}ski}, {Udalski},
		{Szyma{\'n}ski}, {Pietrukowicz}, {Skowron}, {Poleski}, {Wyrzykowski},
		{Ulaczyk}, {Koz{\l}owski}, {Mr{\'o}z}, \& {Pawlak}}]{2016AcA....66..269S}
	{Skowron}, D.~M., {Soszy{\'n}ski}, I., {Udalski}, A., {et~al.} 2016,
	\bibinfo{title}{{OGLE-ing the Magellanic System: Photometric Metallicity from
			Fundamental Mode RR Lyrae Stars},} \actaa, 66, 269,
	\dodoi{10.48550/arXiv.1608.00013}
	
	\bibitem[{D.~M. Skowron {et~al.}(2021)Skowron, Skowron, Udalski, Szymański,
		Soszyński, Wyrzykowski, Ulaczyk, Poleski, Kozłowski, Pietrukowicz, Mróz,
		Rybicki, Iwanek, Wrona, \& Gromadzki}]{Skowron2021}
	Skowron, D.~M., Skowron, J., Udalski, A., {et~al.} 2021,
	\bibinfo{title}{OGLE-ing the Magellanic System: Optical Reddening Maps of the
		Large and Small Magellanic Clouds from Red Clump Stars,} The Astrophysical
	Journal Supplement Series, 252, 23, \dodoi{10.3847/1538-4365/abcb81}
	
	\bibitem[{J. {Skowron} {et~al.}(2016){Skowron}, {Udalski}, {Koz{\l}owski},
		{Szyma{\'n}ski}, {Mr{\'o}z}, {Wyrzykowski}, {Poleski}, {Pietrukowicz},
		{Ulaczyk}, {Pawlak}, \& {Soszy{\'n}ski}}]{2016AcA....66....1S}
	{Skowron}, J., {Udalski}, A., {Koz{\l}owski}, S., {et~al.} 2016,
	\bibinfo{title}{{Analysis of Photometric Uncertainties in the OGLE-IV
			Galactic Bulge Microlensing Survey Data},} \actaa, 66, 1,
	\dodoi{10.48550/arXiv.1604.01966}
	
	\bibitem[{R. {Smolec}(2005){Smolec}}]{2005AcA....55...59S}
	{Smolec}, R. 2005, \bibinfo{title}{{Metallicity Dependence of the Blazhko
			Effect},} \actaa, 55, 59.
	\newblock \doarXiv{astro-ph/0503614}
	
	\bibitem[{I. {Soszy{\'n}ski} {et~al.}(2016){Soszy{\'n}ski}, {Udalski},
		{Szyma{\'n}ski}, {Wyrzykowski}, {Ulaczyk}, {Poleski}, {Pietrukowicz},
		{Koz{\l}owski}, {Skowron}, {Skowron}, {Mr{\'o}z}, \&
		{Pawlak}}]{Soszynski2016}
	{Soszy{\'n}ski}, I., {Udalski}, A., {Szyma{\'n}ski}, M.~K., {et~al.} 2016,
	\bibinfo{title}{{The OGLE Collection of Variable Stars. Over 45 000 RR Lyrae
			Stars in the Magellanic System},} \actaa, 66, 131,
	\dodoi{10.48550/arXiv.1606.02727}
	
	\bibitem[{A. {Subramaniam}(2006){Subramaniam}}]{Subramaniam2006}
	{Subramaniam}, A. 2006, \bibinfo{title}{{RR Lyrae stars in the inner Large
			Magellanic Cloud: halo-like location with a disk-like distribution},} \aap,
	449, 101, \dodoi{10.1051/0004-6361:20054214}
	
	\bibitem[{A. {Subramaniam} \& S. {Subramanian}(2009){Subramaniam} \&
		{Subramanian}}]{Subramaniam2009}
	{Subramaniam}, A., \& {Subramanian}, S. 2009, \bibinfo{title}{{RR Lyrae stars
			in the inner LMC: Where did they form?},} \aap, 503, L9,
	\dodoi{10.1051/0004-6361/200912694}
	
	\bibitem[{N.~B. {Suntzeff} {et~al.}(1999){Suntzeff}, {Walker}, {Smith},
		{Kraft}, {Klemola}, \& {Stetson}}]{Suntzeff1999}
	{Suntzeff}, N.~B., {Walker}, A.~R., {Smith}, V.~V., {et~al.} 1999, in IAU
	Symposium, Vol. 190, New Views of the Magellanic Clouds, ed. Y.-H. {Chu},
	N.~{Suntzeff}, J.~{Hesser}, \& D.~{Bohlender}, 393,
	\dodoi{10.48550/arXiv.astro-ph/9809358}
	
	\bibitem[{A. {Udalski} {et~al.}(1992){Udalski}, {Szymanski}, {Kaluzny},
		{Kubiak}, \& {Mateo}}]{Udalski1992}
	{Udalski}, A., {Szymanski}, M., {Kaluzny}, J., {Kubiak}, M., \& {Mateo}, M.
	1992, \bibinfo{title}{{The Optical Gravitational Lensing Experiment},}
	\actaa, 42, 253
	
	\bibitem[{A. {Udalski} {et~al.}(2015){Udalski}, {Szyma{\'n}ski}, \&
		{Szyma{\'n}ski}}]{Udalski2015}
	{Udalski}, A., {Szyma{\'n}ski}, M.~K., \& {Szyma{\'n}ski}, G. 2015,
	\bibinfo{title}{{OGLE-IV: Fourth Phase of the Optical Gravitational Lensing
			Experiment},} \actaa, 65, 1, \dodoi{10.48550/arXiv.1504.05966}
	
	\bibitem[{A. {Udalski} {et~al.}(2025){Udalski}, {Skowron}, {Skowron},
		{Szyma{\'n}ski}, {Soszy{\'n}ski}, {Pietrukowicz}, {Mr{\'o}z}, {Poleski},
		{Koz{\l}owski}, {Ulaczyk}, {Rybicki}, {Iwanek}, {Gromadzki}, {Wrona},
		{Ratajczak}, \& {Mr{\'o}z}}]{Udalski2025}
	{Udalski}, A., {Skowron}, D.~M., {Skowron}, J., {et~al.} 2025,
	\bibinfo{title}{{The Ultimate I-band Calibration of the TRGB Standard
			Candle},} \actaa, 75, 1, \dodoi{10.32023/0001-5237/75.1.1}
	
	\bibitem[{R.~P. {van der Marel} \& M.-R.~L. {Cioni}(2001){van der Marel} \&
		{Cioni}}]{vanderMarel2001}
	{van der Marel}, R.~P., \& {Cioni}, M.-R.~L. 2001, \bibinfo{title}{{Magellanic
			Cloud Structure from Near-Infrared Surveys. I. The Viewing Angles of the
			Large Magellanic Cloud},} \aj, 122, 1807, \dodoi{10.1086/323099}
	
	\bibitem[{R.~P. {van der Marel} \& N. {Kallivayalil}(2014){van der Marel} \&
		{Kallivayalil}}]{vanderMarel2014}
	{van der Marel}, R.~P., \& {Kallivayalil}, N. 2014,
	\bibinfo{title}{{Third-epoch Magellanic Cloud Proper Motions. II. The Large
			Magellanic Cloud Rotation Field in Three Dimensions},} \apj, 781, 121,
	\dodoi{10.1088/0004-637X/781/2/121}
	
	\bibitem[{P. {Virtanen} {et~al.}(2020){Virtanen}, {Gommers}, {Oliphant},
		{Haberland}, {Reddy}, {Cournapeau}, {Burovski}, {Peterson}, {Weckesser},
		{Bright}, {van der Walt}, {Brett}, {Wilson}, {Jarrod Millman}, {Mayorov},
		{Nelson}, {Jones}, {Kern}, {Larson}, {Carey}, {Polat}, {Feng}, {Moore}, {Vand
			erPlas}, {Laxalde}, {Perktold}, {Cimrman}, {Henriksen}, {Quintero}, {Harris},
		{Archibald}, {Ribeiro}, {Pedregosa}, {van Mulbregt}, \&
		{Contributors}}]{2020SciPy-NMeth}
	{Virtanen}, P., {Gommers}, R., {Oliphant}, T.~E., {et~al.} 2020,
	\bibinfo{title}{{SciPy 1.0: Fundamental Algorithms for Scientific Computing
			in Python},} Nature Methods, 17, 261,
	\dodoi{https://doi.org/10.1038/s41592-019-0686-2}
	
	\bibitem[{A.~R. {Walker}(2012){Walker}}]{Walker2012}
	{Walker}, A.~R. 2012, \bibinfo{title}{{The Large Magellanic Cloud and the
			distance scale},} \apss, 341, 43, \dodoi{10.1007/s10509-011-0961-x}
	
	\bibitem[{E.~L. {Wright} {et~al.}(2010){Wright}, {Eisenhardt}, {Mainzer},
		{Ressler}, {Cutri}, {Jarrett}, {Kirkpatrick}, {Padgett}, {McMillan},
		{Skrutskie}, {Stanford}, {Cohen}, {Walker}, {Mather}, {Leisawitz}, {Gautier},
		{McLean}, {Benford}, {Lonsdale}, {Blain}, {Mendez}, {Irace}, {Duval}, {Liu},
		{Royer}, {Heinrichsen}, {Howard}, {Shannon}, {Kendall}, {Walsh}, {Larsen},
		{Cardon}, {Schick}, {Schwalm}, {Abid}, {Fabinsky}, {Naes}, \&
		{Tsai}}]{2010AJ....140.1868W}
	{Wright}, E.~L., {Eisenhardt}, P. R.~M., {Mainzer}, A.~K., {et~al.} 2010,
	\bibinfo{title}{{The Wide-field Infrared Survey Explorer (WISE): Mission
			Description and Initial On-orbit Performance},} \aj, 140, 1868,
	\dodoi{10.1088/0004-6256/140/6/1868}
	
	\bibitem[{B. {Yanny} {et~al.}(2009){Yanny}, {Rockosi}, {Newberg}, {Knapp},
		{Adelman-McCarthy}, {Alcorn}, {Allam}, {Allende Prieto}, {An}, {Anderson},
		{Anderson}, {Bailer-Jones}, {Bastian}, {Beers}, {Bell}, {Belokurov},
		{Bizyaev}, {Blythe}, {Bochanski}, {Boroski}, {Brinchmann}, {Brinkmann},
		{Brewington}, {Carey}, {Cudworth}, {Evans}, {Evans}, {Gates}, {G{\"a}nsicke},
		{Gillespie}, {Gilmore}, {Nebot Gomez-Moran}, {Grebel}, {Greenwell}, {Gunn},
		{Jordan}, {Jordan}, {Harding}, {Harris}, {Hendry}, {Holder}, {Ivans},
		{Ivezi{\v{c}}}, {Jester}, {Johnson}, {Kent}, {Kleinman}, {Kniazev},
		{Krzesinski}, {Kron}, {Kuropatkin}, {Lebedeva}, {Lee}, {French Leger},
		{L{\'e}pine}, {Levine}, {Lin}, {Long}, {Loomis}, {Lupton}, {Malanushenko},
		{Malanushenko}, {Margon}, {Martinez-Delgado}, {McGehee}, {Monet}, {Morrison},
		{Munn}, {Neilsen}, {Nitta}, {Norris}, {Oravetz}, {Owen}, {Padmanabhan},
		{Pan}, {Peterson}, {Pier}, {Platson}, {Re Fiorentin}, {Richards}, {Rix},
		{Schlegel}, {Schneider}, {Schreiber}, {Schwope}, {Sibley}, {Simmons},
		{Snedden}, {Allyn Smith}, {Stark}, {Stauffer}, {Steinmetz}, {Stoughton},
		{SubbaRao}, {Szalay}, {Szkody}, {Thakar}, {Sivarani}, {Tucker}, {Uomoto},
		{Vanden Berk}, {Vidrih}, {Wadadekar}, {Watters}, {Wilhelm}, {Wyse}, {Yarger},
		\& {Zucker}}]{2009AJ....137.4377Y}
	{Yanny}, B., {Rockosi}, C., {Newberg}, H.~J., {et~al.} 2009,
	\bibinfo{title}{{SEGUE: A Spectroscopic Survey of 240,000 Stars with g =
			14-20},} \aj, 137, 4377, \dodoi{10.1088/0004-6256/137/5/4377}
	
	\bibitem[{R. {Zinn} \& M.~J. {West}(1984){Zinn} \& {West}}]{Zinn1984}
	{Zinn}, R., \& {West}, M.~J. 1984, \bibinfo{title}{{The globular cluster system
			of the Galaxy. III. Measurements of radial velocity and metallicity for 60
			clusters and a compilation of metallicities for 121 clusters.},} \apjs, 55,
	45, \dodoi{10.1086/190947}
	
	\bibitem[{J.~W. Zwolak {et~al.}(2007)Zwolak, Boggs, \& Watson}]{Zwolak}
	Zwolak, J.~W., Boggs, P.~T., \& Watson, L.~T. 2007, \bibinfo{title}{Algorithm
		869: ODRPACK95: A weighted orthogonal distance regression code with bound
		constraints.,} ACM Trans. Math. Softw., 33, 27.
	\newblock
	\url{http://dblp.uni-trier.de/db/journals/toms/toms33.html#ZwolakBW07}
	
\end{thebibliography}
\bibliographystyle{aasjournal}

\end{document}